\documentclass{article}
\usepackage{amssymb}
\usepackage{url}
\usepackage[usenames]{color}
\usepackage{gastex}

\newtheorem{thm}{Theorem}[section]
\newtheorem{cor}[thm]{Corollary}
\newtheorem{lem}[thm]{Lemma}
\newtheorem{prop}[thm]{Proposition}
\newtheorem{fact}[thm]{Fact}

\newtheorem{pro-example}[thm]{Example}
\newenvironment{expl}{\begin{pro-example}\rm}{\cqfd\end{pro-example}}

\newtheorem{pro-remark}[thm]{Remark}
\newenvironment{remark}{\begin{pro-remark}\rm}{\cqfd\end{pro-remark}}

\newtheorem{pro-zoltan}{Zoltan!}

\newenvironment{Proof}{\rm \trivlist \item[\hskip \labelsep{\bf
Proof.}]}{\cqfd\endtrivlist}

\def\cqfd{\skip10=\parfillskip\parfillskip=0pt
\enspace\hfill\symbolecqfd\par\parfillskip=\skip10\par\medskip}
\def\symbolecqfd{\rlap{$\sqcap$}$\sqcup$}
\def\preuve{\begin{Proof}}
\def\proof{\begin{Proof}}
\def\eop{\end{Proof}}

\def\NV{\textsf{NV}}

\def\pgp{$pg$-pair}
\def\pgs{$pg$-pairs}
\def\0{{\bf 0}}
\def\1{{\bf 1}}

\def\os{{\oplus}}

\def\V{{\bf V}}

\def\n{{\bf n}}

\def\k{{\bf k}}

\def\K{{\bf K}}

\def\cK{\mathcal{K}}

\def\FO{\textsf{FO}}
\def\MOD{\textsf{MOD}}
\def\L{{\cal L}}
\def\Lin{{\bf Lind}}
\def\Linlg{{{\cal L}ind}}
\def\Suc{{\sf Succ }}
\def\false{{\sf false}}
\def\true{{\sf true}}
\def\rank{{\sf rank}}

\def\mod{{\rm mod}}

\def\Box{\hbox{\rlap{$\sqcap$}$\sqcup$}}
\def\block{\mathbin{\Box}}

\def\inv{^{-1}}
\let\phi\varphi

\def\V{\hbox{\bf V}}
\def\W{\hbox{\bf W}}

\def\sfor{{\sf or}}

\def\leftf{{\bf left}}
\def\rightf{{\bf right}}

\def\Im{\mathop{\mathcal{I}\textrm{m}}_\emptyset}

\def\var{{\sf var}}
\def\psv{{\sf psv}}
\def\cP{\mathcal{P}}

\def\EX{\textsf{EX}}
\def\EF{\textsf{EF}}
\def\path{{\sf path}}
\def\leftf{{\sf left}}
\def\rightf{{\sf right}}
\def\next{{\sf next}}
\def\precl{{\sf precl}}
\def\pgpairs{{\sf pgp}}

\def\p{{\bf p}}
\def\q{{\bf q}}
\def\r{{\bf r}}

\def\L{{\bf L}}

\def\X{{\bf X}}

\def\B{\mathbb{B}}
\def\T{\mathbb{T}}

\def\calK{\mathcal{K}}
\def\calV{\mathcal{V}}

\def\card{\textsf{card}}
\def\str{\textsf{str}}
\def\root{\textsf{root}}
\def\maxf{\textsf{max}}
\def\minf{\textsf{min}}

\def\V{\hbox{\bf V}}

\let\epsilon\varepsilon

\begin{document}

    \title{Algebraic characterization of logically defined tree
    languages\protect\footnote{The first author acknowledges partial
    support from grant MTM2007 63422 from the Ministry of Education
    and Science of Spain.  The second author acknowledges partial
    support from the French \textit{ANR} (projet \textsc{dots}) and the
    Indo-French project \textsc{Timed discoveri}.  Both authors
    acknowledge support from the European Science Foundation program
    \textsc{AutoMathA}.}}

\author{
Zolt\'an \'Esik
\\
\small{GRLMC, Rovira i Virgili University, Tarragona, Spain}
\\
\small{and}
\\
\small{Department of Computer Science, University of Szeged, Hungary}
\\
\null
\\
Pascal Weil
\protect\footnote{%
Corresponding author. \protect\url{pascal.weil@labri.fr}. Postal 
address: LaBRI, 351 cours de la Lib\'eration, 33405
Talence Cedex, France.} \\
\small{LaBRI, Universit\'e de Bordeaux, CNRS}
}

\date{}

\maketitle

\begin{abstract}
    We give an algebraic characterization of the tree languages that
    are defined by logical formulas using certain Lindstr\"om
    quantifiers.  An important instance of our result concerns
    first-order definable tree languages.  Our characterization relies
    on the usage of preclones, an algebraic structure introduced by
    the authors in a previous paper, and of the block product
    operation on preclones.  Our results generalize analogous results
    on finite word languages, but it must be noted that, as they
    stand, they do not yield an algorithm to decide whether a given
    regular tree language is first-order definable.
\end{abstract} 

\bigskip

\noindent\textbf{Classification}: ACM: F.4.3, F.4.1. MSC: 03B70, 68Q70, 68Q45

\bigskip

One of Bret Tilson's lasting contributions is the introduction (with
John Rhodes) of the notions of block product and two-sided semidirect
product, and their use in the structure theory of finite monoids.
This tool was initially introduced to derive iterated decompositions
of morphisms and to refine the wreath product-based Krohn-Rhodes
decomposition of finite monoids \cite{RhodesTilson}.  It quickly found
applications in formal language theory (see
\cite{mps1,mps2,jcss,AMSV,Straubing} among others).  One of the more
fruitful applications of this work has been in the investigation of
the logical aspects of automata theory (on finite words).  For
instance, the expressive power of first-order formulas with a certain
quantifier depth, can be captured by monoids which divide an iterated
block product of semilattices of the same length, see Straubing's book
\cite{Straubing}. 

Automata are particularly well suited to discuss the behavior of
terminating sequential systems (languages of finite words), and this
field of research has benefited from the start from the
well-established connection between automata and monoid theory.  There
is however also much interest in automata-theoretic descriptions of
languages of other structures than finite words, corresponding to
other natural ideas of implementation and other natural models of
computation (infinite, branching, concurrent, timed, etc).  This paper
is a contribution to the investigation of an important problem of this
sort: can we decide whether a regular tree language is first-order
definable?  Here trees are finite, ranked and ordered.  The latter
properties signify that the nodes of the trees are labeled with
symbols of a given arity (the rank of the node), and the children of a
node of rank $r$ form a totally ordered set of cardinality $r$.  A
tree language is said to be regular if it is accepted by a classical
(deterministic) bottom-up tree automaton.

The notion of automata recognizability for (finite) word languages is
easily translated to an algebraic notion of recognizability, expressed
in terms of monoids: the set of all words on a given alphabet $A$ is a
monoid (the free monoid $A^*$ over that alphabet), and one shows that
a language is recognized by a finite state automaton if and only if it
is the inverse image of a set, under a morphism from $A^*$ into a
\textit{finite} monoid.  Moreover, if a language is recognizable, then
there is a least finite monoid recognizing it, called its syntactic
monoid.  This point of view opens vast possibilities for the
classification and the discussion of the properties of recognizable
languages, which can be characterized in terms of the algebraic
properties of their syntactic monoid, see Eilenberg's variety theory
\cite{Eilenberg,Pin,Almeida}.  As the syntactic monoid of the language
accepted by a given automaton is computable, this can lead to
interesting decision algorithms.

It is well-known (B\"uchi, 1960), that a language is recognizable if
and only if it is definable in monadic second-order logic.  It was
also shown that a language is definable by a first-order formula if
and only if its syntactic monoid is aperiodic.  This statement is
actually the combination of two classical theorems due to
Sch\"utzenberger and to McNaughton and Papert.  It can also be proved
directly (as in \cite{Straubing}), using the Krohn-Rhodes decomposition
theorem, which implies that a monoid is aperiodic if and only if it
belongs to the least pseudovariety closed under block product and
containing $U_1 = \{1,0\}$.  As the syntactic monoid of a regular
language is computable and as aperiodicity is decidable, it is also
decidable whether a regular language is $\FO$-definable.

Considering logically defined sets of trees (or other discrete
combinatorial structures) is just as natural as for words.  But the
literature on these questions shows that classification and decision
results are much harder to reach, in part because we lack the
versatile and powerful algebraic tool provided in the word case by
finite monoid theory.  The weakness of our understanding of automata
theory for tree languages is highlighted by the fact that the
decidability of first-order definability is still an open problem.

For most discrete structures, there is no obvious algebraic structure
that can be used in lieu of monoids, or at least no algebraic
structure that gives rise to the same wealth of structure theorems and
variety characterizations, see \cite{WeilMFCS}.  In the tree case (for
finite, ranked and ordered trees), several propositions can be found
in the literature, see the work of Steinby, Salehi, Heuter, Podelski,
Wilke, \'Esik, etc
\cite{Steinby1,Steinby2,Salehi,SalehiSteinby,Heuter,Podelski,Wilke,Esik}.
Until recently none was very convincing in terms of its capacity to
characterize significant classes of languages, but there are recent
encouraging results expressed in terms of minimal tree automata, that
is, in terms of $\Sigma$-algebras by Benedikt and S\'egoufin
\cite{BenediktSegoufin}, Boja\'nczyk and Walukiewicz \cite{BW}, \'Esik
\cite{Esik TCS 2006} and \'Esik and Iv\'an \cite{ESz1,ESz2}\footnote{%
After our results were announced in 2003 \cite{FSTTCS} and while this
paper was in preparation or under refereeing, an interesting approach
of tree languages in terms of so-called \textit{forest algebras} was
introduced by Boja\'nczyk and Walukiewicz.  See the conclusion of this
paper for a brief discussion.}.  In a previous paper \cite{EsikWeil
TCS}, the authors introduce a new algebraic framework -- the so-called
\textit{preclones} -- to classify and discuss the properties of
recognizable tree languages.

It turns out that the setting of preclones makes it natural to discuss
not only the recognizable sets of trees, but also recognizable sets of
trees with variables.  Variables can be seen as unlabeled leaves of
the tree, and the rank of a tree is the number of such unlabeled
leaves.  Alternately, one can regard these leaves as labeled by
particular letters $\{v_1,v_2,\ldots\}$, in such a way that in a rank
$k$ tree, the variable leaves are labeled $v_1,\ldots,v_k$ from left
to right.

We verified in \cite{EsikWeil TCS} that the notion of recognizability 
induced by the algebraic structure of preclones coincides with the 
usual notion of recognizable tree languages, that the syntactic 
preclone of a recognizable tree language is completely determined by 
the minimal deterministic bottom-up automaton of the language (all 
very reassuring facts), and that these notions are robust enough to 
allow for an Eilenberg-like development in terms of varieties of 
languages and pseudovarieties of preclones.

In this paper, we use this algebraic framework to derive an algebraic
characterization of first-order definable tree languages, and more
generally, of the classes of tree languages determined by certain
families of Lindstr\"om quantifiers.  This characterization requires
the introduction of a block product operation on preclones, a complex
algebraic operation which generalizes Tilson's block product of
monoids.  Our main result implies that the first-order definable tree
languages are exactly those languages whose syntactic preclone sits in
the least pseudovariety of preclones closed under block product and
containing $T_\exists$, a very simple preclone whose properties were
discussed in \cite{EsikWeil TCS} and which can be viewed as an
analogue of the monoid $U_1$.  This result was announced without 
proof in the authors' communication at FST-TCS~\cite{FSTTCS}.

As it is, our result does not yield an algorithm to decide whether a
given recognizable tree language is first-order definable.  This
question is briefly discussed in the conclusion of the paper, but
whatever the case may be, such a decidability result remains one of
the main goals in this field.  Our result however suggests the
feasibility of an algebra-based solution.

The plan of this paper is as follows. Section~\ref{algebraic} 
summarizes the essential properties of preclones that are necessary 
for this study. Section~\ref{sec logically} introduces the logical 
apparatus we will use, including the Lindstr\"om quantifiers and the 
closure properties of the associated operators on classes of 
languages. Finally, in Section~\ref{sec alg charact}, we introduce 
the block product operation on preclones, and we prove our main 
results. The paper closes on a conclusion where we discuss certain 
questions raised by these results.

\paragraph{Notation}
Let $n\ge 0$. We denote by $[n]$ the set $\{1,\ldots,n\}$ if $n > 0$, 
the empty set if $n = 0$.

\section{The algebraic framework}\label{algebraic}

Throughout the paper, we will be discussing sets of finite ranked
trees, that is, trees in which the set of children of each inner node
is linearly ordered; $\Sigma$ designates a \textit{ranked alphabet},
that is, $\Sigma = (\Sigma_n)_{n\ge 0}$ where the $\Sigma_n$ are
pairwise disjoint sets and $\bigcup_n\Sigma_n$ is finite.  An element
of $\Sigma_n$ is said to have \textit{rank} $n$.

This section summarizes the main facts relative to the algebraic
framework, which we will use to establish our main theorem.  Most of
these results are taken from the authors' earlier paper \cite{EsikWeil
TCS} and are stated here without proof.

\subsection{Preclones}

A \textit{preclone} is a many-sorted algebra $S = ((S_n)_{n\ge 0},
\bullet, \1)$.  The elements of $S_n$ are said to have \textit{rank}
$n$, the element $\1$ belongs to $S_1$, and the composition operation
$\bullet$ associates with each $f\in S_n$ and $n$-tuple $g =
(g_1,\ldots,g_n)$ (with $g_i \in S_{m_i}$, $1\le i \le n$), an element
$f\bullet g \in S_m$ where $m = \sum_i m_i$.  Moreover, $\1$ and
$\bullet$ satisfy the axioms given below.

For convenience, a tuple $g$ as above is written $g = g_1 \oplus
\cdots \oplus g_n$, we say that $g$ has \textit{total rank} $m$,
written $\rank(g) = m$, and we let $S_{n,m}$ be the set of $n$-tuples
of total rank $m$.  Note that $S_{1,m} = S_m$ for all $m$.  We also
write $\n$ for the $\oplus$-sum of $n$ copies of $\1$, so that $\n \in
S_{n,n}$.  The axioms defining preclones are the following:

\begin{eqnarray*}
    \1\cdot f &=& f \enspace = \enspace f \cdot \n \enspace\textrm{
    for each $f\in S_n$, $n\ge 0$,} \\
    \textrm{and}\enspace (f\cdot g)\cdot h &=& f\cdot (g_1\cdot\bar
    h_1 \oplus \cdots \oplus g_n\cdot\bar h_n)
\end{eqnarray*}
where $f\in S_n$, $g = \bigoplus_{i=1}^n g_i$ with each $g_i \in
S_{m_i}$, $h = \bigoplus_{j=1}^m h_j$ with each $h_j \in S_{p_j}$, $m
= \sum_i m_i$, $\bar h_1$ equal to the $\oplus$-sum of the $m_1$ first
$h_j$'s, $\bar h_2$ equal to the $\oplus$-sum of the $m_2$ next
$h_j$'s, etc.

It is interesting to remark that $S_1$ is naturally equipped with a 
monoid structure.

\textit{Sub-preclones} of preclones are defined in the natural way.  A
\textit{morphism of preclones}, $\phi\colon S \rightarrow T$, is a
rank preserving map, which also preserves the unit element \1\ and the
composition operation.  Similarly, a congruence is an equivalence
relation, that relates only elements of equal rank, and which is
stable under the composition operation.  The quotient of a preclone by
a congruence is naturally endowed with a preclone structure, and the
projection map is an onto morphism.

The least sub-preclone of a preclone $S$, containing a given subset
$A$ is called \textit{the sub-preclone of $S$ generated by $A$}.  If
this preclone is $S$ itself, we say that $S$ is \textit{generated} by
$A$.  A preclone is \textit{finitely generated} if it admits a finite
set of generators.

A preclone $S$ is said to be \textit{finitary} if each $S_n$ is finite.
Observe that as soon as some $S_k$, $k\ge 2$ is non-empty, then
infinitely many $S_k$ are non-empty, and hence $S$ is not finite.

\subsection{Examples of preclones}\label{sec examples}

The following examples will be essential for our study.

\paragraph{Trees and free preclones}
Let $\Sigma$ be a ranked alphabet.  The free preclone generated by
$\Sigma$, written $\Sigma M$, can be described as follows (see
\cite[Section 2.2]{EsikWeil TCS}).  Let $(v_k)_{k\ge 1}$ be a sequence
of variable names: we let $\Sigma M_n$ be the set of finite trees,
whose inner nodes are labeled by elements of $\Sigma$ (where a rank
$k$ letter labels a node with $k$ children), whose leaves are labeled
by elements of $\Sigma_0 \cup \{v_1,\ldots,v_n\}$, and whose
\textit{frontier} (the left to right sequence of leaf labels) contains
exactly one occurrence of $v_1$, \dots, $v_n$, in that order (that is,
belongs to $\Sigma_0^*v_1\Sigma_0^* \cdots v_n\Sigma_0^*$).  The
elements of $\Sigma M_n$ are called \textit{trees of rank $n$} or
\textit{$n$-ary trees} over $\Sigma$.

If $t$ is such a tree, we let $\NV(t)$ be the set of nodes of $t$ with
a label in $\Sigma$ ($\NV$ stands for \textit{non-variable labeled}).
In the logical discussion to follow (Section~\ref{sec logically}), we
will give the nodes in $\NV(t)$ a particular r\^ole.

\begin{figure}[ht]
    \centering
    \begin{picture}(110,48)(0,-48)
	\drawline[AHnb=0](15.0,0.0)(0,-30)
	\drawline[AHnb=0](15,0)(30,-30)
	\drawline[AHnb=0](0,-30)(30,-30)
	\node[Nfill=y,fillcolor=Black,Nw=1.0,Nh=1.0,Nmr=1.0](n0)(5.0,-30){}
	\node[Nfill=y,fillcolor=Black,Nw=1.0,Nh=1.0,Nmr=1.0](n1)(25,-30){}
	\put(14,-17){$f$}
	\put(4,-34){$v_1$}
	\put(13,-33){\dots}
	\put(24,-34){$v_n$}
	\drawline[AHnb=0](55.0,0.0)(40,-30)
	\drawline[AHnb=0](55,0)(70,-30)
	\drawline[AHnb=0](40,-30)(70,-30)
	\node[Nfill=y,fillcolor=Black,Nw=1.0,Nh=1.0,Nmr=1.0](n0)(45.0,-30){}
	\node[Nfill=y,fillcolor=Black,Nw=1.0,Nh=1.0,Nmr=1.0](n1)(65,-30){}
	\put(54,-17){$f$}
	\put(54,-29){$n$}
	\drawline[AHnb=0](45,-30)(40,-45)
	\drawline[AHnb=0](45,-30)(50,-45)
	\drawline[AHnb=0](40,-45)(50,-45)
	\put(44,-40){$g_1$}
	\put(44,-48){$m_1$}
	\drawline[AHnb=0](65,-30)(60,-45)
	\drawline[AHnb=0](65,-30)(70,-45)
	\drawline[AHnb=0](60,-45)(70,-45)
	\put(64,-40){$g_n$}
	\put(64,-48){$m_n$}
	\drawline[AHnb=0](95.0,0.0)(80,-30)
	\drawline[AHnb=0](95,0)(110,-30)
	\drawline[AHnb=0](80,-30)(110,-30)
	\drawline[AHnb=0](80.0,-31)(75,-45)
	\drawline[AHnb=0](110,-31)(115,-45)
	\drawline[AHnb=0](80.0,-31)(110,-31)
	\drawline[AHnb=0](75,-45)(115,-45)
	\put(94,-17){$f$}
	\put(94,-29){$n$}
	\put(94,-39){$g$}
	\put(94,-48){$m$}
\end{picture}
\caption{$f \in \Sigma M_n$, $g = g_1 \oplus \cdots \oplus g_n \in
\Sigma M_{n,m}$ and two views of $f\cdot g$}
\label{composition of trees}
\end{figure}
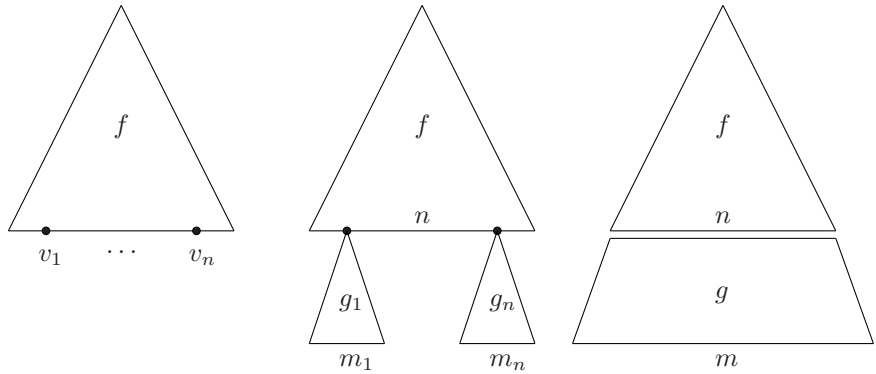

$\Sigma M$ is a preclone for the following operations: if $f\in \Sigma
M_n$ and $g_i \in \Sigma M_{m_i}$ ($1 \le i \le n$), then the
composite tree $f\cdot (g_1\oplus \cdots \oplus g_n)$ is obtained by
replacing the $v_i$-labeled leaf of $f$ by the root of $g_i$, and by
renumbering the variable labeled leaves of the resulting tree with
consecutively indexed $v_j$'s, see Figure~\ref{composition of trees}.
The unit element $\1$ is the graph in $\Sigma M_1$ consisting of a
single node labeled $v_1$.

Each element of $\Sigma$ can be identified with an element of $\Sigma
M$: the letter $\sigma \in \Sigma_n$ is identified with the tree with
$n+1$ nodes, consisting of the root, labeled $\sigma$, and the $n$
children of the root, labeled $v_1,\ldots,v_n$ in this order.

The elements of rank 0, $\Sigma M_0$, are the ordinary
$\Sigma$-labeled trees.

\begin{expl}\label{words as trees}
    As discussed in the introduction, our results can be seen as
    generalizations to trees of known results on recognizable word
    languages.  This meta-statement can be made precise in the
    following fashion: if $A$ is a finite (unranked) alphabet, we can
    view $A$ as a ranked alphabet, all of whose elements have rank 1.
    Then the elements of $AM_1$ can be seen as the words of the form
    $wv_1$, where $w\in A^*$.  In particular, the monoid $AM_1$ is
    isomorphic to, and will be identified with the free monoid $A^*$. 
    The sets $AM_n$ ($n\ne 1$) are empty.
\end{expl}    

\paragraph{Preclone of transformations, preclone of an automaton}
If $Q$ is a set, let $\T_n(Q)$ be the set of $n$-ary transformations 
of $Q$, that is, the set of mappings $Q^n \rightarrow Q$. Let also 
$\T(Q) = (\T_n(Q))_{n\ge 0}$. Composition of mappings endows 
$\T(Q)$ with a preclone structure in the following sense: if $f\in 
\T_n(Q)$, $g_i \in \T_{m_i}(Q)$ ($1\le i\le n$) and $m =  \sum_im_i$, 
then $f\cdot (g_1\oplus \cdots \oplus g_n)$ maps $(q_1,\ldots,q_m)$ to
$$f\big(g_1(q_1,\ldots,q_{m_1}),
g_2(q_{m_1+1},\ldots,q_{m_1+m_2}),\ldots,
g_n(q_{m-m_n+1},\ldots,q_m)\big).$$

If $\Sigma$ is a ranked alphabet and $Q$ is a $\Sigma$-algebra, each
element $\sigma \in \Sigma_n$ determines naturally an $n$-ary
transformation of $Q$.  The sub-preclone of $\T(Q)$ generated by
$\Sigma$ is called the \textit{preclone associated} with $Q$.

\begin{expl}\label{automata and transition monoids}
    Let $A$ be an unranked alphabet, viewed as a ranked alphabet as in
    Example~\ref{words as trees}.  An $A$-algebra $Q$ is simply a set,
    equipped with an action of $A$, that is, a deterministic complete
    automaton.  Each letter $a\in A$ then defines a mapping $Q
    \rightarrow Q$.  Thus the preclone associated with $Q$ has
    elements of rank 1 only, which form the usual transition monoid of
    the automaton, see \cite{Eilenberg,Pin}.
\end{expl}    

We note that $\Sigma$-algebras are natural objects in our context: a
deterministic complete bottom-up tree automaton accepting trees in
$\Sigma M_0$ (see \cite{tata}), with state set $Q$, can be
described as a finite $\Sigma$-algebra $Q$, equipped with a set
$F\subseteq Q$ of \textit{final states}.

\paragraph{The preclones $T_\exists$ and $T_p$}

Let $\B$ be the Boolean semiring $\B = \{\true,\false\}$, and let
$T_{\exists}$ be the subset of $\T(\B)$ whose rank $n$ elements are the
$n$-ary \sfor\ function and the $n$-ary constant $\true$, written
respectively $\sfor_{n}$ and $\true_{n}$ (by convention, $\sfor_0$ is
the nullary constant $\false_0$).  Then $T_{\exists}$ is a preclone,
which is generated by the binary $\sfor_{2}$ function and the nullary
constants $\true_{0}$ and $\false_{0}$.  The set $\B$ equipped with
these 3 generators can be viewed as a finite tree automaton, and
$T_{\exists}$ is the preclone associated with this automaton.
    
It is interesting to note that the rank 1 elements of $T_{\exists}$
form a 2-element monoid, isomorphic to the multiplicative monoid
$\{0,1\}$, and known as $U_{1}$ in the literature on monoid theory,
e.g. \cite{Pin}.

Similarly, if $p \ge 2$ is an integer and $\B_{p} =
\{0,1,\ldots,p-1\}$, let $T_{p}$ be the subset of $\T(\B_{p})$ whose
rank $n$ elements ($n\geq 0$) are the mappings
$f_{n,r}\colon(r_{1},\ldots,r_{n})\mapsto r_{1}+\cdots+r_{n}+r \bmod
p$ for $0\leq r < p$.  Again, $T_{p}$ is a finitely generated
preclone, generated by the nullary constant 0, the unary increment
function $f_{1,1}$ and the binary sum $f_{2,0}$.  Moreover, $T_{p}$
can be seen as the preclone associated with a $p$-element automaton,
and its rank 1 elements form a monoid isomorphic to the cyclic group
of order $p$.

\paragraph{Preclone-generator pairs}
If $S$ is a preclone and $A$ is a set of generators of $S$, we say
that $(S,A)$ is a \textit{preclone-generator pair}, or \textit{\pgp}.
A \pgp\ $(S,A)$ is said to be \textit{finitary} if $S$ is finitary and
$A$ is finite.  The notions of sub-\pgp\ and morphisms of \pgs\ are
defined naturally: $(S,A)$ is a \textit{sub-\pgp} of $(T,B)$ is $A
\subseteq B$ (so that $S$ is by construction a sub-preclone of $T$);
and a \textit{morphism of \pgs} $\phi\colon (S,A) \rightarrow (T,B)$
is a preclone morphism from $S$ to $T$ such that $\phi(A) \subseteq
B$.

\subsection{Syntactic preclones}\label{sec syntactic}

Let $S$ be a preclone and let $L \subseteq S_k$.  We say that $L$ is
\textit{recognizable} if there exists a morphism $\phi\colon S
\rightarrow T$ into a finitary preclone and a subset $P \subseteq T_k$
such that $L = \phi\inv(P)$.  Then we say that $L$ is
\textit{recognized by} $T$, and by the morphism $\phi$.  If $(S,A)$
and $(T,B)$ are \pgs\ and $\phi$ is a morphism between these \pgs, we
say that $L$ is recognized by $(T,B)$.

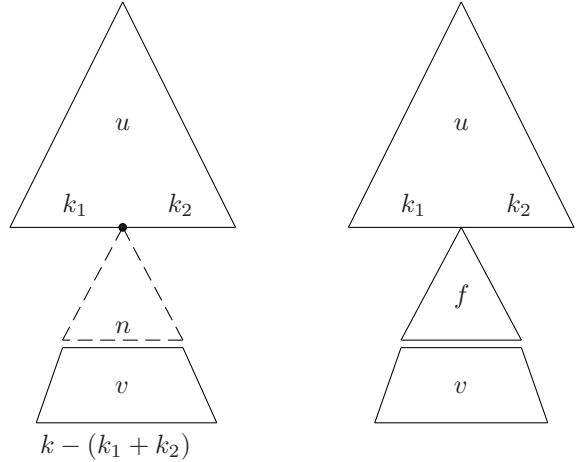
\begin{figure}[ht]
    \centering
    \begin{picture}(75,59)(0,-59)
	\drawline[AHnb=0](15.0,0.0)(0,-30)
	\drawline[AHnb=0](15,0)(30,-30)
	\drawline[AHnb=0](0,-30)(30,-30)
	\node[Nfill=y,fillcolor=Black,Nw=1.0,Nh=1.0,Nmr=1.0](n0)(15.0,-30){}
	\put(14,-17){$u$}
	\put(7,-28){$k_1$}
	\put(21,-28){$k_2$}
	\drawline[dash={2.0 1.0}{0.0},AHnb=0](15.0,-30.0)(7,-45)
	\drawline[dash={2.0 1.0}{0.0},AHnb=0](15,-30)(23,-45)
	\drawline[dash={2.0 1.0}{0.0},AHnb=0](7,-45)(23,-45)
	\drawline[AHnb=0](7.0,-46)(3.5,-56)
	\drawline[AHnb=0](23,-46)(27.5,-56)
	\drawline[AHnb=0](7.0,-46)(23,-46)
	\drawline[AHnb=0](3.5,-56)(27.5,-56)
	\put(14,-52){$v$}
	\put(14,-44){$n$}
	\put(4,-60){$k-(k_1+k_2)$}
	\drawline[AHnb=0](60.0,0.0)(45,-30)
	\drawline[AHnb=0](60,0)(75,-30)
	\drawline[AHnb=0](45,-30)(75,-30)
	\drawline[AHnb=0](60.0,-30.0)(52,-45)
	\drawline[AHnb=0](60,-30)(68,-45)
	\drawline[AHnb=0](52,-45)(68,-45)
	\put(59,-17){$u$}
	\put(52,-28){$k_1$}
	\put(66,-28){$k_2$}
	\put(59,-40){$f$}
	\drawline[AHnb=0](52.0,-46)(48.5,-56)
	\drawline[AHnb=0](68,-46)(71.5,-56)
	\drawline[AHnb=0](52.0,-46)(68,-46)
	\drawline[AHnb=0](48.5,-56)(71.5,-56)
	\put(59,-52){$v$}
\end{picture}
\caption{an $n$-ary context in $\Sigma M_k$; is it an $L$-context of 
$f$ ($f \in \Sigma_n$)?}
\label{contexts}
\end{figure}

Let $f\in S_n$.  A \textit{context of $f$ in $L$} is a pair $(u,v)$
where $u$ is an element of $S$ and $v$ is an $n$-tuple of elements of
$S$, such that $f$ can be inserted under $u$ and above $v$ to produce
an element of $L$, see Figure~\ref{contexts}.  We would like this
condition to read $u\cdot f\cdot v \in L$, but it has to be a little
more technical, to specify where precisely $f$ is attached under $u$.

Formally, for each $k \geq 0$, $n > 0$, let $I_{k,n}$ be the set of
\textit{$n$-ary contexts in $S_{k}$}, that is, the set of tuples of
the form $(u,k_1,v,k_2)$, where $k_1, k_2\ge 0$ and $k_1+k_2 \le k$,
$u\in S_{k_1+1+k_2}$ and $v\in S_{n,\ell}$ with $\ell = k-(k_1+k_2)$,
see Figure~\ref{contexts}.
If $L \subseteq S_k$, $f\in S_n$, we say that a context $(u,k_1,v,k_2)
\in I_{k,n}$ is an \textit{$L$-context of $f$} if $u \cdot (\k_1
\oplus f\cdot v \oplus \k_2) \in L$.  We also let the set of 0-ary
contexts in $S_k$ be the set $I_{k,0}$ of tuples $(u,k_1,\0,k_2)$
where $k_1,k_2\ge 0$ and $u\in S_{k_1+1+k_2}$ (the symbol $\0$ is
introduced here to preserve the uniformity of notation).  We say that
such a context is an $L$-context of $f\in S_0$ if $u \cdot (\k_1
\oplus f \oplus \k_2) = u \cdot (\k_1 \oplus f \cdot \0 \oplus \k_2)
\in L$.

Next we say that elements of $f,g$ are \textit{$L$-equivalent},
written $f \sim_L g$ if $f$ and $g$ have the same $L$-contexts.  The
relation $\sim_L$ is a congruence, called the \textit{syntactic
congruence} of $L$, the quotient preclone $S/\!\sim_L$ is called the
\textit{syntactic preclone} of $L$, and the projection morphism is the
\textit{syntactic morphism}.  Finally, if $A$ is a set of generators
of $S$, the \textit{syntactic \pgp} of $L$ is the pair $(T,B)$ where
$T = S /\!\sim_L$ and $B$ is the image of $A$ in the syntactic
morphism.  We note the following result, proved in \cite[Proposition
3.2]{EsikWeil TCS}.

\begin{prop}\label{prop syntactic}
    Let $S$ be a preclone, $k\ge 0$ and $L \subseteq S_k$. Then the 
    following statements hold.
    \begin{itemize}
	\item A morphism of preclones recognizes $L$ if and only if it
	it can be factored through the syntactic morphism of $L$.
	
	\item If $T$ is a sub-preclone or a quotient of $T'$, and if
	$T$ recognizes $L$, then so does $T'$.
	
	\item $L$ is recognizable if and only if its syntactic
	preclone is finitary.	
	
	\item The analogous statements hold for \pgs.
    \end{itemize}
\end{prop}    

We will be primarily concerned with the case where $S$ is a finitely
generated free preclone, $S = \Sigma M$ with $\Sigma$ a ranked
alphabet.  The subsets of each $\Sigma M_k$ are called \textit{tree
languages}.  In that case, the notion of recognizable tree languages
defined above coincides with the classical notion of recognizability,
and in particular, the syntactic preclone of a recognizable tree
language $L \subseteq \Sigma M_0$ coincides with the preclone of the
minimal automaton of $L$, see \cite[Section 3.2]{EsikWeil TCS}.  It is
interesting to note that the syntactic $\Sigma$-algebra of $L$
\cite{Steinby2} is exactly the rank 0 part of the syntactic preclone of $L$,
and that the syntactic tree monoid \cite{Podelski} of $L$ is the monoid of
rank 1 elements of its syntactic preclone.  In particular, if $L
\subseteq \Sigma M_0$, then the syntactic preclone of $L$ is finitary
if and only if its rank 0 part is finite.

We now consider two important examples.  In each, the alphabet is a
\textit{ranked Boolean alphabet} $\Delta$, that is, a ranked alphabet
such that whenever $\Delta_n \ne\emptyset$, then $\Delta_n = \{1_n,
0_n\}$, see \cite[Section 3.3]{EsikWeil TCS} for more details.

\begin{expl}\label{lg Texists}
    For $k\ge 0$, let $K_k(\exists)$ be the set of trees in $\Delta
    M_k$ containing at least one node labeled $1_{n}$ (for some
    $n$).  Then $K_{k}(\exists)$ is recognizable and its syntactic
    preclone is $T_{\exists}$ defined in Section~\ref{sec examples},
    see \cite[Section 3.3]{EsikWeil TCS}.  More generally, let
    $\phi\colon \Sigma M \rightarrow T_\exists$ be a morphism, with
    $\Sigma$ an arbitrary ranked alphabet.  Let
    $$\Sigma^{(0)} = \bigcup_n \{\sigma \in \Sigma_n \mid 
    \phi(\sigma) = \sfor_n\}$$
    and let $\Sigma^{(1)}$ be the complement of $\Sigma^{(0)}$ in
    $\Sigma$.  Then the subsets of $\Sigma M_k$ recognized by $\phi$
    are $\emptyset$, $\Sigma M_k$, $\phi\inv(\sfor_k) = \Sigma^{(0)}
    M_k$ and $\phi\inv(\true_k)$, the set of trees in $\Sigma M_k$
    with at least one occurrence of a letter in $\Sigma^{(1)}$.
    
    Similarly, if $p,r$ are integers with $0\leq r < p$ and if
    $K_{k}(\exists^r_p)$ consists of the trees in $\Delta M_k$ such
    that the number of nodes labeled $1_{n}$ (for some $n$) is
    congruent to $r$ modulo $p$, then $K_{k}(\exists^r_p)$ is
    recognizable and its syntactic preclone is $T_{p}$.  If
    $\phi\colon \Sigma M \rightarrow T_p$ is a morphism, then the
    subsets of $\Sigma M_k$ recognized by $\phi$ are the finite unions
    of the $\phi\inv(f_{n,r})$ ($0\le r < p$).  For each such $r$, let
    $$\Sigma^{(r)} = \bigcup_n \{\sigma \in \Sigma_n \mid 
    \phi(\sigma) = f_{n,r}\}.$$
    For each $t\in \Sigma M_k$, let $w_r(t)$ be the number of nodes in
    $\NV(t)$ labeled by a letter in $\Sigma^{(r)}$, and let $w(t) =
    \sum_r r\,w_r(t)$.  Then $\phi\inv(f_{n,r})$ is the set of all
    $t\in\Sigma M_k$ such that $w(t) = r \pmod p$.
\end{expl}

\subsection{Varieties of tree languages and pseudovarieties}\label{sec var}

A \textit{pseudovariety of preclones} is a class of finitary preclones
which is closed under taking finite direct products, sub-preclones,
quotients, finitary unions of $\omega$-chains and finitary inverse
limits of $\omega$-diagrams, see \cite[Section 4]{EsikWeil TCS}.
Here, we say that a union $T = \bigcup_{n\ge 0} T^{(n)}$ is
\textit{finitary} if $T$ is finitary.  \textit{Finitary inverse
limits} of an $\omega$-diagram of the form $\phi^{(n)} \colon
T^{(n+1)} \rightarrow T^{(n)}$ are defined similarly.

The definition of a \textit{pseudovariety of \pgs} is similar: it is a
class of finitary \pgs\ which is closed under taking finite direct
products, sub-preclones, quotients, and finitary inverse limits of
$\omega$-diagrams (there is no need to consider unions of
$\omega$-diagrams, see \cite[Section 4.4]{EsikWeil TCS}).

We note the following proposition \cite[Corollary 4.22]{EsikWeil TCS}.

\begin{prop}\label{423 TCS}
    Let \V\ be a pseudovariety of preclones and let $S$ be a finitary
    preclone such that, for all $s\ne t \in S$, there exists a
    morphism $\phi$ from $S$ into a preclone in \V\ with $\phi(s) \ne
    \phi(t)$.  Then $S \in \V$.
\end{prop}    

An analogous statement for pseudovarieties of \pgs\ also holds.

\begin{prop}\label{424 TCS}
    Let \V\ be a pseudovariety of \pgs\ and let $(S,A)$ be a finitary
    \pgp\ such that, for all $s\ne t \in S$, there exists a morphism
    $\phi$ from $(S,A)$ into a \pgp\ in \V\ with $\phi(s) \ne
    \phi(t)$.  Then $(S,A) \in \V$.
\end{prop}    

\proof
This statement is proved in the same fashion as Proposition~\ref{423
TCS} (see \cite{EsikWeil TCS}), using also \cite[Proposition
4.23]{EsikWeil TCS}.
\eop

If $\K$ is a class of finitary preclones (resp.  \pgs), there exists a
least pseudovariety containing $\K$, which is said to be
\textit{generated by $\K$} and is denoted by $\langle\K\rangle$, see
\cite[Section 4.2]{EsikWeil TCS}.  We record in particular the
following results, which follow from \cite[Propositions 3.3 and 4.16,
Corollary 4.8]{EsikWeil TCS}.

\begin{prop}\label{prop psv gen}
    Let \K\ be a class of finitary preclones (resp.  \pgs) and let
    $\V$ be the pseudovariety generated by \K. The syntactic preclone
    (resp.  \pgp) of a recognizable tree language belongs to \V\ if
    and only if it is a quotient of a sub-preclone (resp.  sub-\pgp)
    of a direct product of elements of \K.
\end{prop}    

\begin{prop}\label{cor 48}
  A pseudovariety of preclones (resp.  \pgs) is entirely determined
  by the syntactic preclones (resp.  \pgs) it contains.
\end{prop}    

Note that pseudovarieties of preclones can be seen as particular
examples of pseudovarieties of \pgs, in the sense of
Proposition~\ref{psv pgs or preclones} below\footnote{The notion of
pseudovarieties of \pgs\ could be generalised to the notion of
\textit{varieties of stamps} as is done for word languages, see
\cite{StraubingStamps,PinStraubingStamps}.}.
If \K\ is a class of \pgs, we let $\precl(\K)$ be the class of
preclones $S$ such that $(S,A) \in \K$ for some set $A$.  Conversely,
if $\L$ is a class of preclones, we let $\pgpairs(\L)$ be the class of
finitary \pgs\ $(S,A)$ such that $S\in \L$.  Let us say that a class
$\K$ of \pgs\ is \textit{full} if membership of a \pgp\ $(S,A)$ in
$\K$ depends only on $S$; that is, $\K = \pgpairs(\precl(\K))$.

\begin{prop}\label{psv pgs or preclones}    
    A pseudovariety \V\ of \pgs\ is full if and only if
    there exists a pseudovariety \W\ of preclones such that $\V = 
    \pgpairs(\W)$ and in that case, \W\ is the pseudovariety 
    generated by $\precl(\V)$.
    
    Moreover, if $\K$ is a full class of \pgs\ and \V\ is the
    pseudovariety generated by \K, then \V\ is full as well and
    $\precl(\K)$ and $\precl(\V)$ generate the same pseudovariety of
    preclones.
\end{prop}

\proof
Let \W\ be a pseudovariety of preclones and let $\V = \pgpairs(\W)$.
The class \V\ is full by definition.  Let us first verify that it is
closed under taking sub-\pgs, quotients, finite direct products and
finitary inverse limits of $\omega$-diagrams.  Suppose for instance
that $(S,A)$ is a sub-\pgp\ of $(T,B)$ with $T\in \W$.  Then $S$ is a
sub-preclone of $T$, so $S \in \W$ and $(S,A) \in \V$.  The
verification is equally routine for quotients and finite direct
products.  As for inverse limits of $\omega$-diagrams, it was shown
\cite[Proposition 4.23]{EsikWeil TCS} that if $(S,A) =
\lim_n(S^{(n)},A^{(n)})$, then $S = \lim_n S^{(n)}$.  Thus, if the
$(S^{(n)},A^{(n)})$ are in \V, then the $S^{(n)}$ are in \W\ and hence
$S \in \W$ and $(S,A) \in \V$.

Now let us show that if \L\ is a class of finitely generated finitary
preclones, then $\pgpairs(\langle \L \rangle) \subseteq
\langle\pgpairs(\L)\rangle$.  Let $(U,C)$ be a finitary \pgp\ with
$U\in \langle \L \rangle$: we want to show that $(U,C) \in
\langle\pgpairs(\L)\rangle$.  Combining technical results from
\cite{EsikWeil TCS} (namely Propositions 4.5 and 4.16), we may assume
that there exist preclones $S^{(1)}$,\dots, $S^{(n)}$ in \L\ such that
$U = \phi(T')$ for some morphism $\phi\colon T' \rightarrow U$ where
$T'$ is a sub-preclone of $\prod_i S^{(i)}$.  Since \L\ consists of
finitely generated preclones, let $A^{(i)}$ be a finite set of
generators of $S^{(i)}$.  Let $B$ be a finite subset of $T'$ such that
$\phi(B) = C$ and for each $i$, let $B^{(i)}$ be the projection of $B$
onto the $i$-th component.  Finally, let $T$ be the sub-preclone of
$T'$ generated by $B$.  Then $(U,C) = \phi(T,B)$ and $(T,B)$ is a
sub-\pgp\ of $\prod_i(S^{(i)},A^{(i)} \cup B^{(i)})$.  This
establishes that $(U,C) \in \langle\pgpairs(\L)\rangle$.

Let now \K\ be a full class of finitary \pgs\ and let $\W =
\langle\precl(\K)\rangle$.  We verify that $\langle\K\rangle =
\pgpairs(\W)$, which implies that $\langle\K\rangle$ is full.  Indeed,
since \K\ is full, we have $\K = \pgpairs(\precl(\K))$ and hence
$\langle\K\rangle = \langle \pgpairs(\precl(\K)) \rangle \subseteq
\langle \pgpairs(\W)\rangle$.  The first part of the proof establishes
that $\pgpairs(\W)$ is a pseudovariety, so $\langle\K\rangle \subseteq
\pgpairs(\W)$.  Moreover, the discussion in the previous paragraph,
applied to $\L = \precl(\K)$, shows that $\pgpairs(\W) \subseteq
\langle \K\rangle$.  The expected equality follows.

Applying this result to $\K = \V$, a full pseudovariety of \pgs, and
to $\W = \langle\precl(\V)\rangle$, shows that $\V = \pgpairs(\W)$, as
announced.  Finally, if $\V = \pgpairs(\W')$ for some other
pseudovariety of preclones $\W'$, then \W\ and $\W'$ have the same
finitely generated elements, and hence must be equal by
Proposition~\ref{cor 48}.  This concludes the proof of the
proposition.
\eop

Before we discuss varieties of tree languages, let us define quotients
of tree languages.  Let $L \subseteq \Sigma M_k$, let $k_1,k_2$ be
integers with $k_1 + k_2 \le k$ and let $u\in \Sigma M_{k_1+1+k_2}$.
The \textit{left quotient} of $L$ by $(u,k_1,k_2)$ is the subset of $\Sigma 
M_{k - k_1 - k_2}$
$$(u,k_1,k_2)\inv L = \{f \in \Sigma 
M_{k - k_1 - k_2} \mid u \cdot (\k_1 \oplus f \oplus \k_2) \in L\}.$$
If $v \in \Sigma M_{n,k}$, then the \textit{right quotient} of $L$ by 
$v$ is
$$L v\inv = \{f \in \Sigma M_n \mid f\cdot v \in L\}.$$

\begin{remark}\label{remark syntactic class}
    With the above notation, $(u,k_1,v,k_2)$ is an $L$-context of an
    element $f$ if and only if $f \in \big((u,k_1,k_2)\inv L\big)
    v\inv = (u,k_1,k_2)\inv \big(L (\k_1 \oplus v \oplus \k_2)\inv\big)$.
    
    Moreover, if $(u,k_1,v,k_2)$ and $(u',k_1,v',k_2)$ are contexts
    such that $u \sim_L u'$ and $v\sim_L v'$, then
    $\big((u,k_1,k_2)\inv L\big) v\inv =  \big((u',k_1,k_2)\inv L\big)
    {v'}\inv$.
\end{remark}

We say that a morphism $\phi\colon \Sigma M\rightarrow \Sigma' M$ is a
\textit{literal morphism} if $\phi(\Sigma) \subseteq \Sigma'$.  A
\textit{variety of tree languages} (resp.  a \textit{literal variety
of tree languages}) is a collection $\calV =
(\calV_{\Sigma,k})_{\Sigma,k}$, where $\Sigma$ runs over all ranked
alphabets, $k$ runs over non-negative integers, such that each
$\calV_{\Sigma,k}$ is a Boolean algebra of recognizable languages in
$\Sigma M_k$, and $\calV$ is closed under quotients and under inverse
morphisms (resp.  inverse literal morphisms) between free preclones. 
In particular, every variety of tree languages is a literal variety.

If \V\ is a pseudovariety of preclones (resp.  \pgs), we let $\var(\V)
= (\calV_{\Sigma,k})$ be such that $\calV_{\Sigma,k}$ is the class of
languages in $\Sigma M_k$ with syntactic preclone (resp.  \pgp) in \V.
If $\calV$ is a variety (resp.  literal variety) of tree languages,
let $\psv(\calV)$ be the class of finitary preclones (resp.  finitary
\pgs) which only accept languages in $\calV$.  The following result
was proved in \cite{EsikWeil TCS}.

\begin{thm}\label{thm eilenberg}
    The mappings $\var$ and $\psv$ are mutually inverse lattice 
    isomorphisms between the lattice of pseudovarieties of preclones 
    (resp. \pgs) and the lattice of varieties (resp. literal 
    varieties) of tree languages.
\end{thm}

We note the following corollary of Theorem~\ref{thm eilenberg},
which will be used in the sequel.

\begin{cor}\label{eilenberg full}
    Let $\calV$ be a literal variety and let $\V$ be the corresponding
    pseudovariety of \pgs.  Then \V\ is full if and only if $\calV$ is
    a variety.
\end{cor}    

\begin{expl}\label{variety of T exists}
    Let $\langle T_\exists\rangle$ be the pseudovariety of preclones
    generated by $T_\exists$, and let $\calV$ be the corresponding
    tree language variety.  Then a language $L \subseteq \Sigma M_k$
    is in $\calV_{\Sigma,k}$ if and only if $L$ is a Boolean
    combination of languages of the form $\Sigma'M_k$, $\Sigma'
    \subseteq \Sigma$, see Example~\ref{lg Texists} and \cite[Section
    5.2.1]{EsikWeil TCS}.
\end{expl}

More complex examples are discussed in \cite[Section 5.2]{EsikWeil
TCS}, and the main results of this article provide further examples.

\section{Logically defined tree languages}\label{sec logically}

Let $\Sigma$ be a ranked alphabet.  We will define tree languages by
means of logical formulas.  We consider the \textit{atomic formulas}
of the following form
$$P_\sigma(x),\ x < y,\ \Suc_i(x,y),\ \root(x),\ \maxf_{i,j}(x),\
\leftf_j(x) \hbox{ and }\rightf_j(x),$$
where $\sigma \in \Sigma$, $i,j$ are positive integers, $i$ is less
than or equal to the maximal rank of a letter in $\Sigma$, and $x, y$
are first-order variables.  If $k\geq 0$, subsets of $\Sigma M_{k}$
will be defined by \textit{formulas of rank $k$}, composed using
atomic formulas with $j\in[k]$, the Boolean constants $\false$ and
$\true$, the Boolean connectives and a family of generalized
quantifiers called \textit{Lindstr\"om quantifiers}, defined in
Section~\ref{sec Lind quantifiers} below.  As usual, each quantifier
binds a first-order variable (within the scope of the quantifier), and
variables that are not bound are called \textit{free}.  A formula
without free variables is called a \textit{sentence}.  We denote by
$\Lin$ the logic defined in this fashion.

When a $\Lin$-formula is interpreted on a tree $t\in\Sigma M_{k}$,
first-order variables are interpreted as nodes in $\NV(t)$ --- and we
assume $t\ne\1$, so that $\NV(t)$ is non-empty.  Then $P_{\sigma}(x)$
holds if $x$ is labeled $\sigma$ ($\sigma\in\Sigma$), $x < y$ holds if
$y$ is a proper descendant of $x$, and $\Suc_{i}(x,y)$ holds if $y$ is
the $i$-th successor of $x$.  Moreover, $\root(x)$ holds if $x$ is the
root of $t$ and $\maxf_{i,j}(x)$ holds if the $i$-th successor of $x$
is labeled by $v_j$, the $j$-th variable.  Finally, $\leftf_j(x)$
(resp.  $\rightf_j(x)$) holds if the index of the highest (resp.
least) numbered variable labeling a leaf to the left (resp.  right) of
the frontier of the subtree rooted at $x$ is $j$, see Figure~\ref{fig
leftj}.  The interpretation of Lindstr\"om quantifiers is described in
Section~\ref{sec Lind quantifiers}.
\begin{figure}[ht]
    \centering
    \begin{picture}(40,43)(0,-43)
	\drawline[AHnb=0](20.0,0.0)(0,-40)
	\drawline[AHnb=0](20,0)(40,-40)
	\drawline[AHnb=0](0,-40)(40,-40)
	\drawline[AHnb=0](20,-20)(10,-40)
	\drawline[AHnb=0](20,-20)(30,-40)
	\node[Nfill=y,fillcolor=Black,Nw=1.0,Nh=1.0,Nmr=1.0](n0)(2.0,-40){}
	\node[Nfill=y,fillcolor=Black,Nw=1.0,Nh=1.0,Nmr=1.0](n0)(8.0,-40){}
	\node[Nfill=y,fillcolor=Black,Nw=1.0,Nh=1.0,Nmr=1.0](n0)(32.0,-40){}
	\node[Nfill=y,fillcolor=Black,Nw=1.0,Nh=1.0,Nmr=1.0](n0)(38.0,-40){}
	\put(22,-20){$x$}
	\put(1,-43){$v_1$}
	\put(7,-43){$v_h$}
	\put(31,-43){$v_j$}
	\put(37,-43){$v_k$}
\end{picture}
\caption{$\leftf_h(x) \land \rightf_j(x)$}
\label{fig leftj}
\end{figure}

Recall that formally, an \textit{interpretation} is a mapping
$\lambda$ from the set of free variables of a formula $\phi$ (or from
a set containing the free variables of $\phi$) to the set $\NV(t)$ of
$\Sigma$-labeled nodes of a tree $t$.  If $t$ satisfies $\phi$ with
this interpretation, we say that $(t,\lambda)$ \textit{satisfies}
$\phi$ and we write $(t,\lambda) \models \phi$.  If $\phi$ is a
sentence, we simply write $t \models \phi$.

\begin{remark}
    In $\Lin$-formulas, first-order variables are never interpreted as
    one of the $v_i$-labeled leaves.  In fact, as far as logical
    constructs go, these particular leaves are not considered as
    proper nodes of the tree, but rather as place markers -- which
    explains the fact that they are \textit{labeled} by their position
    in the left-to-right order, and may be relabeled appropriately
    when trees are composed.
    
    When we deal with traditional trees, that is, trees in $\Sigma
    M_0$, this peculiarity disappears, and we observe that in that
    case, our atomic formulas ($P_\sigma$, $<$, $\root$ and the
    $\Suc_i$) are the atomic formulas of the usal logic on rooted
    ranked trees \cite{ThomasHdBook}.
\end{remark}    
    
\begin{expl}\label{atomic formulas on words}
    Let $A$ be an unranked alphabet, viewed as a ranked alphabet as in
    Example~\ref{words as trees}.  Then $AM_1$ is equal to the set
    $A^*v_1$, and is isomorphic to the free monoid $A^*$.  In this
    situation, the boundary of the trees in $AM$ consist of a single
    node, labeled $v_1$, that is $\leftf_1(x)$ and $\rightf_1(x)$
    always evaluate to $\false$.  Thus the relevant atomic formulas
    are $P_a(x)$ ($a\in A$), $x < y$, $\Suc_1(x,y)$, $\root(x)$ and
    $\maxf_{1,1}(x)$.  Note that in this case, $\root(x)$ is the
    predicate usually denoted by $\minf(x)$ (or $x = \minf$) and
    $\maxf_{1,1}(x)$ is the predicate $\maxf(x)$ (or $x = \maxf$).
    That is, we have the same atomic formulas as in B\"uchi's
    classical sequential calculus
    \cite{PinLogicSurvey,ThomasHdBook,Straubing}.  The condition
    $t\ne\1$ imposed to interpret formulas, is equivalent to the fact
    that logical formulas are not interpreted on the empty word.
\end{expl}    
    
Next to the atomic formulas defined above, we also use the following
shorthand notation.  Let $k>0$ and let $\leftf_0(x)$ be the formula of
rank $k$ $\leftf_0(x) = \bigwedge_{j \in [k]} \neg \leftf_j(x)$.  Then
$\leftf_0(x)$ holds if no leaf situated to the left of the frontier of
the subtree rooted at $x$, is labeled by a variable.  We observe that
for different values of $k$, we get different formulas $\leftf_0(x)$,
and our notation assumes that $k$ is clear from the context.
    
Similarly, if $k$ is clear from the context, we define
$\rightf_{k+1}(x)$ to be the formula of rank $k$ $\rightf_{k+1}(x) =
\bigwedge_{j \in [k]} \neg \rightf_j(x)$.  Its meaning is that no leaf
situated to the right of the frontier of the subtree rooted at $x$, is
labeled by a variable.

\subsection{Lindstr\"om quantifiers}\label{sec Lind quantifiers}

Before we give formal definitions, we discuss an important example.
   
\begin{expl}\label{ex introductory}
    Let us consider the first order formula $\exists x\cdot
    \phi(x)$, where $\phi$ is a formula with free variables in a set
    $Y \cup \{x\}$ ($x \not\in Y$).  Let $\lambda\colon Y\rightarrow
    \NV(t)$.  Recall that $(t,\lambda) \models \exists x \cdot \phi$ if
    there exists a node $v$ in $\NV(t)$ such that $(t,[\lambda; x\mapsto
    v]) \models \phi$.  For convenience, let $\lambda_v$ denote the
    interpretation $[\lambda; x\mapsto v]$.  We can express the
    satisfaction of $\exists x \cdot \phi$ in the following, more
    generalizable fashion: we label each node $v\in \NV(t)$ with $1$
    if $(t,\lambda_v) \models \phi$, with $0$ otherwise (the variable
    labeled nodes are left unchanged).  If $\bar t_\lambda$ denotes
    the resulting Boolean-labeled tree, then $(t,\lambda) \models
    \exists x \cdot \phi$ if and only if $\bar t_\lambda$ belongs to
    the set of trees with at least one 1 label.
    
    To be formally accurate, the nodes of $\bar t_\lambda$ must be
    labeled by a ranked alphabet, that is, we need to have, for each
    rank $n$, a letter $1_n$ and a letter $0_n$.  The definition of
    Lindstr\"om quantifiers below generalizes this example.
\end{expl}    

Let $\Delta$ be a ranked alphabet containing letters of rank $n$ for
each $n$ such that $\Sigma_{n}\ne\emptyset$ and let
$\langle\phi_{\delta}\rangle_{\delta\in \Delta}$ be a family of rank
$k$ formulas on $\Sigma$-trees.  We say that this family is
\emph{deterministic with respect to} a first-order variable $x$ if for
each tree $t\in\Sigma M_{k}$, for each integer $n$, and for each
interpretation $\lambda$ of the free variables in the $\phi_{\delta}$
mapping $x$ to a rank $n$ node of $t$, then $(t,\lambda)$ satisfies
exactly one of the $\phi_{\delta}$, $\delta\in\Delta_{n}$.  Whenever
needed, we will also assume that $x$ is not bound in any of the
$\phi_{\delta}$.

\begin{expl}\label{first ex deterministic}
    If $\Delta = \Sigma$, a very simple example of such a family is
    given by letting $\phi_\delta(x) = P_\delta(x)$ for each
    $\delta\in \Delta$.
\end{expl}
    
\begin{expl}\label{ex boolean alphabet}
    Another natural example is given over a ranked Boolean alphabet
    $\Delta$, that is, an alphabet such that whenever $\Delta_n
    \ne\emptyset$, then $\Delta_n = \{1_n, 0_n\}$.  If for each such
    $n$, $\phi_{0_n}$ is logically equivalent to $\neg\phi_{1_n}$,
    then $\langle \phi_\delta\rangle_{\delta\in \Delta}$ is
    deterministic with respect to any first order variable $x$.
    
    In later examples, when dealing with ranked Boolean alphabets, we
    will write $\phi_n$ instead of $\phi_{1_n}$ and we will assume
    that $\phi_{0_n} = \neg\phi_n$.  Then a deterministic family will
    simply be written $\langle \phi_n\rangle_n$.
\end{expl}    

With this notion, we define (\textit{simple}) \textit{Lindstr\"om
quantifiers}, a definition adap\-ted from
\cite{Lindstrom,EbbinghausFlum} to the case of finite trees.  Let
$K\subseteq \Delta M_{k}$ be a language of rank $k$ trees and let
$\langle\phi_{\delta}\rangle_{\delta\in \Delta}$ be a family of rank
$k$ formulas which is deterministic with respect to $x$.  Then the
quantified formula $Q_{K}x \cdot \langle\phi_{\delta}\rangle_{\delta
\in\Delta}$, where the quantifier $Q_K$ binds the variable $x$, is
interpreted in the following manner.

Given a tree $t\in\Sigma M_{k}$ and an interpretation $\lambda$ of the
free variables in the $\phi_{\delta}$ except for $x$, we construct a
tree $\bar{t}_{\lambda}\in\Delta M_{k}$ as follows: $t$ and $\bar
t_{\lambda}$ have the same underlying tree structure with the same
variable-labeled nodes, that is, the same set of nodes and the
same relations $<$, $\Suc_{i}$, $\root$, $\maxf_{i,j}$, $\leftf_j$ and
$\rightf_j)$.  Moreover, for each rank $n$ node $v$ of $t$ (for some
$n$), let $\lambda_{v}$ be the interpretation $[\lambda, x\mapsto v]$:
then the node $v$ in $\bar t_{\lambda}$ is labeled by the unique
element $\delta\in\Delta_{n}$ such that $(t,\lambda_{v})$ satisfies
$\phi_{\delta}$.  The tree $\bar t_\lambda$ is called the
\textit{characteristic tree} determined by $t$, $\lambda$ and the
formulas $\phi_\delta$.  If the $\phi_\delta$ have no free variable
other than $x$, we write $\bar t$ for $\bar t_{\lambda}$.  Finally, we
say that $(t,\lambda)$ \textit{satisfies} $Q_{K}x \cdot \langle
\phi_{\delta} \rangle_{\delta \in \Delta}$ if $\bar t_{\lambda}\in K$.

\begin{remark}\label{remark 25}
    With the above notation, we note that $(t, \lambda_v) \models
    \phi_\delta(x)$ if and only if $(\bar t_\lambda, [x\mapsto v])
    \models P_\delta(x)$.  Since $x$ is the only free variable in
    $P_\delta(x)$, this is also equivalent to $(\bar t_\lambda,
    \lambda_v) \models P_\delta(x)$.
\end{remark}   

\begin{expl}\label{ex define K}
    Suppose that $\Delta = \Sigma$ and $\phi_\delta = P_\delta(x)$
    as in Example~\ref{first ex deterministic}.  If $t\in \Sigma M_k$,
    it is easily verified that the trees $\bar t$ and $t$ are equal.
    In particular, if $K \subseteq \Delta M_k$ is a language of
    $k$-ary trees, then $t$ satisfies $Q_K x\cdot \langle \phi_\delta
    \rangle_{\delta \in \Delta}$ if and only if $t\in K$.
\end{expl}

\begin{expl}\label{example word lg}
    Let $A$ be an unranked alphabet, seen as a ranked alphabet as
    usual, suppose that $\Delta_k = \emptyset$ for all $k\ne 1$ and
    let $K\subseteq \Delta M_1$.  Then $K$ can also be seen as a word
    language since $\Delta M_1$ is isomorphic to the free monoid
    $\Delta_1^*$, and the logic $\Lin$ is analogous to the
    logic for word languages studied by \'Esik and Larsen in
    \cite{EsikLarsen} (the latter does not include $\minf$ and 
    $\maxf$ among its atomic formulas).
\end{expl}

In the next examples, $\Delta$ is a ranked \emph{Boolean} alphabet
such that $\Delta_n$ is non-empty whenever $\Sigma_n$ is, and
$\langle\phi_n \rangle_n$ is a family of formulas which is
deterministic with respect to a first order variable $x$, see
Example~\ref{ex boolean alphabet}.

\begin{expl}\label{example exists}
    Let $K = K_{k}(\exists)$ denote the set of all trees in $\Delta
    M_k$ containing at least one node labeled $1_{n}$ (for some $n$),
    see Example~\ref{lg Texists}.  Then the Lindstr\"om quantifier
    $Q_K$ is a generalization of the existential quantifier, as
    indicated in Example~\ref{ex introductory}.
    
    More precisely, $(t,\lambda)$ satisfies $Q_{K}x\cdot\langle
    \phi_{n} \rangle_{n}$ if and only if there exists a node $v \in
    \NV(t)$ such that $(t,\lambda_v)$ satisfies $\phi_{n}$, where $n$
    is the rank of $v$ and $\lambda_v$ is the interpretation
    $[\lambda,x\mapsto v]$.
       
    Let finally $A$ be an unranked alphabet, viewed as a ranked
    alphabet as in Example~\ref{words as trees}, and suppose that $k =
    1$.  Then $(t,\lambda) \models Q_{K}x\cdot \langle \phi_n
    \rangle_n$ (where $t$ is viewed as a tree in $AM_1$) if and only
    if $(t,\lambda) \models \exists x\cdot\phi_{1}(x)$ (where $t$ is
    viewed as a non-empty word in $A^*$).
\end{expl}

\begin{expl}\label{example exists mod p}
    In the same manner as in Example~\ref{example exists}, if $p \geq
    1$, $r < p$ and $K = K(\exists^r_p)$ denotes the set of those
    trees in $\Delta M_k$ such that the number of nodes labeled
    $1_{n}$ (for some $n$) is congruent to $r$ modulo $p$ (see
    Example~\ref{lg Texists}), then the Lindstr\"om quantifier $Q_{K}$
    is a generalization of a modular quantifier.
    
    More precisely, $(t,\lambda)$ satisfies $Q_{K}x \cdot
    \langle\phi_n\rangle_n$ if and only if, for some $n$, the number
    of nodes $v\in \NV(t)$ such that $(t,\lambda_v)$ satisfies
    $\phi_{n}(x)$ (where $n$ is the rank of $v$) is congruent to $r$
    mod $p$.
    
    If $A$ is an unranked alphabet, then $(t,\lambda) \models
    Q_{K}x\cdot \langle \phi_n \rangle_n$ (where $t$ is viewed as a
    tree in $AM_1$) if and only if $(t,\lambda) \models \exists^r_p
    x\cdot\phi_{1}(x)$ (where $t$ is viewed as a non-empty word in $A^*$).
\end{expl}

\begin{expl}\label{example exists path}
    Let $K = K_k(\exists_{\path})$ be the set of all trees in $\Delta
    M_k$ such that all the nodes along at least one path from the root
    to a leaf are labeled $1_{n}$ (for appropriate values of $n$).

    Then $(t,\lambda)$ satisfies $Q_{K}x \cdot \langle \phi_n
    \rangle_n$ if and only if there exists a root-to-leaf path such
    that, for every node $v\in \NV(t)$ along this path, $(t,\lambda_v)
    \models \phi_{n}(x)$ (where $n$ is the rank of $v$).
    
    If $A$ is an unranked alphabet, then $(t,\lambda) \models
    Q_{K}x\cdot \langle \phi_n \rangle_n$
    (where $t$ is viewed as a tree in $AM_1$) if and only if
    $(t,\lambda) \models \forall x\cdot\phi_{1}(x)$ (where $t$ is
    viewed as a non-empty word in $A^*$).
\end{expl}

\begin{expl}\label{example exists next}
    Let $K_k(\forall_\next)$ be the set of all trees in $\Delta M_k$
    such that the children of the root are labeled $1_{n}$ (for the
    appropriate $n$).

    Then $(t,\lambda)$ satisfies $Q_{K}x \cdot \langle \phi_n \rangle_n$
    if and only if, for every child $v$ of the root, $(t,\lambda_v)
    \models \phi_{n}(x)$ (where $n$ is the rank of $v$).
    
    If $A$ is an unranked alphabet, then $(t,\lambda) \models
    Q_{K}x\cdot \langle \phi_n \rangle_n$ (where $t$ is viewed as a
    tree in $AM_1$) if and only if $(t,\lambda) \models \exists x\cdot
    (\Suc(1,x) \land \phi_{1}(x))$, --- more formally, $(t,\lambda)
    \models \exists x\cdot \big(\big(\forall y\cdot (\min(y)
    \rightarrow \Suc(y,x))\big) \land \phi_{1}(x)\big)$ --- (where $t$
    is viewed as a non-empty word in $A^*$).
    
    Other next modalities can be expressed likewise, \textit{e.g.},
    requesting that at least one (resp.  an even number, etc.)  of the
    children of the root satisfies the appropriate $\phi_n$.
\end{expl}

\subsection{The language associated with a $\Lin$-formula}

Let $\phi$ be a $\Lin$-sentence of rank $k$ over $\Sigma$.  We denote
by $L_\phi$ the set of trees in $\Sigma M_k$ that satisfy $\phi$, and
we say that $L_\phi$ is \textit{defined} by the formula $\phi$.

For a class $\cK$ of tree languages, we let $\Lin(\cK)$ denote the
fragment of $\Lin$ consisting of the formulas in which all
Lindstr\"om quantifiers are of the form $Q_K$ with $K \in \cK$.  If
$\phi$ is a $\Lin(\cK)$ sentence, we say that $L_\phi$ is
\textit{$\Lin(\cK)$-definable}, and we let $\Linlg(\cK)$ denote the
class of $\Lin(\cK)$-definable tree languages.

\begin{expl}\label{expl-fo+mod2}
    Let $\cK_{\exists}$ be the class of all the languages of the form
    $K_k(\exists)$ on a Boolean ranked alphabet.  In view of the
    discussion in Example~\ref{example exists}, it is reasonable to
    say that $\Linlg(\cK_{\exists})$ is exactly the class of
    $\FO$-definable tree languages.  Examples~\ref{example exists}
    and~\ref{example exists mod p} show that if $\cK_{\exists,\mod}$
    is the class of all languages of the form $K_k(\exists)$ or
    $K_k(\exists_p^r)$, then $\Linlg(\cK_{\exists,\mod})$ is the class
    of $(\FO+\MOD)$-definable tree languages.
\end{expl}

It will be useful to associate a tree language also with the
$\Lin$-formulas that contain free variables (as is done in
\cite[Section II-2]{Straubing} for word languages).  Let $Z$ be a
finite set.  We extend $\Sigma$ to the ranked alphabet $\Sigma_{Z}$,
whose set of letters of rank $m$ ($m \geq 0$) is
$\Sigma_{m}\times\cP(Z)$.  We identify each $\sigma\in \Sigma$ with
the pair $(\sigma,\emptyset) \in \Sigma_Z$.  An element $z\in Z$ is
said to \textit{occur} in $t\in \Sigma_Z M$ at node $v$ if the label
of $v$ is of the form $(\sigma,Z')$ and $z\in Z'$.  If each $z \in Z$
occurs exactly once in $t \in \Sigma_Z M_k$, then $t$ is called a
\textit{$Z$-structure of rank $k$} over $\Sigma$.  We note that a
$Z$-structure uniquely determines a tree $t \in \Sigma M$ and a
mapping $\lambda\colon Z \to \NV(t)$.  Conversely, any such pair
$(t,\lambda)$ determines a unique $Z$-structure, written
$\str(t,\lambda)$.  Now let $\phi$ be a rank $k$ $\Lin$-formula with
free variables in a set $Y$.  Let $\str(t,\lambda)$ be a $Z$-structure
with $Z \subseteq Y$.  If $\mu\colon Y\setminus Z \longrightarrow
NV(t)$, we write $(\str(t,\lambda),\mu) \models \phi$ if $(t,
[\lambda;\mu]) \models \phi$, where $[\lambda;\mu]$ is the map from
$Y$ to $\NV(t)$ determined by $\lambda$ and $\mu$.  If $Z = Y$, we
write simply $\str(t,\lambda) \models \phi$ and we say that
$\str(t,\lambda)$ \textit{satisfies} $\phi$.  We let $L_\phi$ be the
set of $Y$-structures satisfying $\phi$.

\begin{expl}\label{expl-atomic}
    Let $\sigma \in \Sigma_m$, and let $\phi$ be the rank $k$ formula
    $\phi = P_\sigma(x)$.  Let $Y$ be a set containing $x$.  Then
    $L_\phi$ is the collection of all $Y$-structures of rank $k$ over
    $\Sigma$ such that some (necessarily unique) node has a label of
    the form $(\sigma,Y')$ with $x \in Y'$.  It is immediate to
    observe that any two trees of equal rank in $\Sigma M$ have the
    same contexts in $L_\phi$, that is, the restriction of the
    syntactic congruence of $L_\phi$ to $\Sigma M$ is the universal
    relation.  The same holds for any atomic formula $\phi$.
\end{expl}

\subsection{Properties of the operator $\Linlg$}

We now explore the properties of the operator $\Linlg$ on families of
languages.

\subsubsection{$\Linlg$ is a closure operator}\label{sec closure 
operator}

\begin{thm}\label{thm-closure}
    $\Linlg$ is a closure operator on classes of languages. That is, 
    for all language classes $\cK$ and $\cK'$, the following holds.
   
    \begin{itemize}
	\item[(1)] $\cK \subseteq \Linlg(\cK)$;
	
	\item[(2)] if $\cK \subseteq \cK'$ then $\Linlg(\cK) \subseteq
	\Linlg(\cK')$;
	
	\item[(3)] $\Linlg(\Linlg(\cK)) \subseteq \Linlg(\cK)$.
    \end{itemize}
\end{thm}

Item (1) follows immediately from Example~\ref{ex define K}, and Item
(2) is immediate from the definition.  The rest of Section~\ref{sec
closure operator} is devoted to the proof of Item (3).

Let $\phi$ be a $\Lin(\Linlg(\cK))$-formula of rank $k$ over $\Sigma$.
We argue by induction on the structure of $\phi$ to show that there is
an equivalent formula $\hat{\phi}$ of $\Lin(\cK)$, that is, a formula
with the same free variables as $\phi$ and such that $L_\phi =
L_{\hat\phi}$ in $\Sigma_YM_k$ for any finite set $Y$ containing the
free variables of $\phi$. This will be sufficient to prove 
Theorem~\ref{thm-closure}.

If $\phi$ is an atomic formula, we let $\hat{\phi} = \phi$, since
$\phi$ is also a $\Lin(\cK)$-formula.  If $\phi = \phi_1 \vee \phi_2$
(resp.  $\phi = \neg\phi_1$), we let $\hat{\phi} = \hat{\phi}_1 \vee
\hat{\phi}_2$ (resp.  $\hat{\phi} = \neg{\hat\phi}_1$).  The equivalence
of $\phi$ and $\hat\phi$ is easily verified.

The last case occurs when $\phi$ is of the form $\phi = Q_K x\cdot
\langle \phi_\delta \rangle_{\delta \in \Delta}$, where $K \in
\Linlg(\cK)$ and the $\phi_\delta$ form a family of rank $k$ formulas
of $\Lin(\Linlg(\cK))$ over $\Sigma$ that is deterministic with
respect to $x$.  In particular, $K = L_\psi$ where $\psi$ is a rank
$k$ $\Lin(\cK)$-sentence over $\Delta$.

By induction, for each $\delta\in \Delta$, there exists a
$\Lin(\cK)$-formula $\hat\phi_\delta$ equivalent to $\phi_\delta$, so
that $\phi$ is equivalent to $Q_K x\cdot \langle \hat\phi_\delta
\rangle_{\delta \in \Delta}$.  Thus, we may assume that the
$\phi_\delta$ are $\Lin(\cK)$-formulas.

Before we proceed with the end of the proof, we establish a technical
fact.  If $\chi$ is a formula and $p,q$ are variables, we denote by
$\chi[q/p]$ the formula obtained from $\chi$ by substituting the
variable $q$ for all free occurrences of $p$.  (Bound occurrences of
$q$ in $\chi$ are renamed as usual.)

Let $\chi$ be a rank $k$ formula over $\Delta$.  We then define
$\tilde\chi$ to be the rank $k$ formula over $\Sigma$ obtained from
$\chi$ by replacing each subformula of the form $P_\delta(z)$, where
$z$ is any first-order variable, by the formula $\phi_\delta[z/x]$.
Since the quantifiers in $\chi$ also occur in $\tilde \chi$, and the
quantifiers in $\tilde\chi$ occur either in $\chi$ or in the
$\phi_\delta$, it is clear that $\chi$ is a $\Lin(\cK)$-formula if and
only if $\tilde\chi$ is one.  In the sequel, we
assume that neither $x$ nor any free variable of one of the
$\phi_\delta$, is free in $\chi$, and that no free variable has bound
occurrences in the formulas under consideration.  Let us then assume
that $Y$ (the finite set containing the free variables of $\phi$ and
not containing $x$) also contains the free variables of $\chi$.

\begin{fact}\label{fact closure}
    With the notation above, let $t\in \Sigma M_k$, let $\lambda\colon
    Y \to \NV(t)$ be a function, and let $\bar{t}_\lambda \in \Delta
    M_k$ be the characteristic tree determined by $t$, $\lambda$ and
    the formulas $\phi_\delta$. Then we have
    $$(t,\lambda)\models \tilde\chi \enspace \Longleftrightarrow
    \enspace (\bar{t}_\lambda, \lambda) \models \chi.$$
\end{fact}    

\proof
We argue by induction on the structure of $\chi$.  Suppose first that
$\chi = P_\delta(z)$.  Then $\tilde\chi = \phi_\delta[z/x]$.  Let
$\mu$ be the restriction of $\lambda$ to $Y \setminus \{z\}$ and let
$\bar t_\mu$ be the characteristic tree determined by $t$, $\mu$ and
the $\phi_\delta[z/x]$.  A node $v$ is labeled $\delta$ in $\bar
t_\lambda$ if and only if $(t,\lambda_v) = (t, [\lambda; x\mapsto v])
\models \phi_\delta$.  Since $z$ does not occur in $\phi_\delta$, this
is equivalent to $(t, [\mu; z \mapsto v]) \models \phi_\delta[z/x]$,
and hence to the labeling of $v$ by $\delta$ in $\bar t_\mu$.  Thus
$\bar t_\lambda = \bar t_\mu$.  Then we have:
\begin{eqnarray*}
    (t,\lambda)\models \tilde\chi = \phi_\delta[z/x]
    &\Longleftrightarrow & (t,[\mu; z\mapsto \lambda(z)]) \models
    \phi_\delta[z/x] \textrm{ by definition of $\mu$}\\
    &\Longleftrightarrow & (\bar{t}_\mu,[z \mapsto \lambda(z)]) =
    (\bar{t}_\lambda,[z \mapsto \lambda(z)])\models P_\delta(z)\\
    &\Longleftrightarrow &
    (\bar{t}_\lambda, \lambda)\models P_\delta(z) = \chi.
\end{eqnarray*}
If $\chi$ is another atomic formula (namely, $z_1 < z_2$,
$\Suc_i(z_1,z_2)$, $\root(x)$, $\maxf_{i,j}(z)$, $\leftf_j(z)$ or
$\rightf_j(z)$ with $z,z_1,z_2 \in Y \cup \{x\}$), then $\tilde\chi =
\chi$.  Since $t$ and $\bar t_\lambda$ have the same variable-labeled
nodes and they differ only in the labeling of their nodes, and since
$\chi$ does not depend on that labeling, we have in each case
$$(t,\lambda)\models \tilde\chi \enspace \Longleftrightarrow
\enspace (\bar{t}_\lambda,\lambda)\models \chi.$$
We have now established our claim for atomic formulas. 

The induction step is immediate if $\chi$ is of the form $\chi =
\chi_1 \vee \chi_2$ or $\chi = \neg \chi_1$.  We now assume that
$\chi= Q_Lz \cdot \langle \chi_\omega \rangle_{\omega \in \Omega}$
where $L\subseteq \Omega M_k$, $z\not\in Y\cup\{x\}$, and $\langle
\chi_\omega \rangle_{\omega \in \Omega}$ is a family of rank $k$
formulas over $\Delta$ with free variables in $Y \cup \{z\}$, which
is deterministic with respect to $z$.

By construction $\tilde\chi = Q_Lz \cdot \langle \tilde\chi_\omega
\rangle_{\omega \in \Omega}$, and by induction hypothesis, for each
node $w \in \NV(t)$ and for each $\omega \in \Omega$, we have
$$(t,[\lambda;z \mapsto w])\models \tilde\chi_\omega \enspace
\Longleftrightarrow \enspace (\bar{t}_\lambda,[\lambda; z \mapsto w])
\models \chi_\omega.$$
It follows in particular that $\langle \tilde\chi_\omega
\rangle_{\omega \in \Omega}$ is deterministic with respect to $z$.
Moreover, the characteristic tree determined by $t$, $\lambda$ and
$\langle \tilde\chi_\omega \rangle_{\omega \in \Omega}$ is the same as
that determined by $\bar t_\lambda$, $\lambda$ and $\langle \chi_\omega
\rangle_{\omega \in \Omega}$.  Thus we have
$$(t,\lambda)\models \tilde\chi \enspace \Longleftrightarrow \enspace
(\bar{t}_\lambda, \lambda) \models \chi,$$
which concludes the induction and the proof.
\eop

We now return to the proof of Theorem~\ref{thm-closure}.  Recall that
$\phi = Q_K x\cdot \langle \phi_\delta \rangle_{\delta \in \Delta}$
and $K = L_\chi \subseteq \Delta M_k$ for some rank $k$
$\Lin(\cK)$-sentence $\chi$ over $\Delta$ (without free variables).
We want to construct a formula in $\Lin(\cK)$ that is equivalent to
$\phi$ and we claim that $\tilde\chi$ is such a formula.

Indeed, let $t\in\Sigma M_k$, let $\lambda$ be a mapping
$\lambda\colon Y \mapsto \NV(t)$, and let $\bar t_\lambda \in \Delta
M_k$ be the characteristic tree determined by $t$, $\lambda$ and the
$\phi_\delta$.  By definition, $(t,\lambda) \models \phi$ if and only
if $\bar t_\lambda \in K$, that is, $\bar t_\lambda \models \psi$, or
equivalently, $(\bar t_\lambda,\lambda) \models \psi$.  It was
established in Fact~\ref{fact closure} that this is equivalent to
$(t,\lambda) \models \tilde\psi$, which concludes the proof.
\cqfd

\subsubsection{Closure properties of $\Linlg(\cK)$}\label{sec closure}

The objective of this section is to prove the closure properties
summarized in Theorem~\ref{thm-literalvar} below.

\begin{thm}\label{thm-literalvar} 
    $\Linlg(\cK)$ is closed under Boolean operations and inverse
    literal morphisms.  Moreover, $\Linlg(\cK)$ is closed under left
    (resp.  right) quotients if and only if any left (resp.  right)
    quotient of a language in $\cK$ belongs to $\Linlg(\cK)$.
\end{thm}

We now prove Theorem~\ref{thm-literalvar}, by considering separately
each closure property.
 
\paragraph{Boolean operations}
The fact that $\Linlg(\cK)$ is closed under the Boolean operations
follows directly from the fact that $\Lin(\cK)$-formulas are closed
under disjunction and negation.
\cqfd

\paragraph{Inverse literal morphisms}
Let $h\colon \Sigma'\rightarrow \Sigma$ be a rank-preserving mapping,
and let us also denote by $h$ the induced morphism $h\colon
(\Sigma'M,\Sigma') \rightarrow (\Sigma M,\Sigma)$.  Note that if $t$
is a tree, then $h(t)$ differs from $t$ only in the labeling of the
nodes in $\NV(t)$.  Let $\phi$ be a rank $k$ $\Lin(\cK)$-formula
over $\Sigma$ with free variables in a finite set $Y$.  We show by
structural induction on $\phi$ that there exists a rank $k$
$\Lin(\cK)$-formula $\phi'$ over $\Sigma'$, with the same free
variables as $\phi$, and such that $(t,\lambda) \models \phi'$ if and
only if $(h(t),\lambda) \models \phi$ for any tree $t\in \Sigma'M_k$
and any interpretation $\lambda\colon Y\rightarrow \NV(t)$.

If $\phi = P_\sigma(x)$ for some $\sigma\in\Sigma$, we let $\phi' =
\bigvee P_{\sigma'}(x)$, where the disjunction runs over the letters
$\sigma' \in \Sigma'$ such that $h(\sigma') = \sigma$.  If $\phi$ is
another type of atomic formula, then $\phi$ does not depend on the
labeling of the tree, and it suffices to choose $\phi' = \phi$.

The inductive step for the Boolean connectives is equally natural: if
$\phi = \phi_1 \vee \phi_2$ (resp.  $\phi = \neg\phi_1$), then we let
$\phi' = \phi'_1 \vee \phi'_2$ (resp.  $\phi' = \neg\phi'_1$).

Suppose finally that $\phi$ is of the form $Q_K x\cdot \langle
\phi_{\delta} \rangle_{\delta \in \Delta}$.  By induction, there exist
formulas $\phi'_\delta$ over $\Sigma'$ such that, for each $\delta$,
$(t,\lambda_v) \models \phi'_\delta$ if and only if $(h(t),\lambda_v)
\models \phi_\delta$ for any tree $t\in \Sigma'M_k$, node $v$ in $t$
and mapping $\lambda\colon Y \rightarrow \NV(t)$.  It follows that the
characteristic tree determined by $t$, $\lambda$ and
$\langle\phi'_{\delta}\rangle_{\delta \in \Delta}$, and the
characteristic tree determined by $h(t)$, $\lambda$ and
$\langle\phi_\delta\rangle_{\delta \in \Delta}$ coincide.  As a
result, we have $(t,\lambda) \models \phi'$ if and only if
$(h(t),\lambda) \models \phi$.
\cqfd

\paragraph{Left quotients}
We now assume that any left quotient of a language in $\cK$ belongs to
$\Linlg(\cK)$.  Let $k_1, k_2, \ell$ be non-negative integers and let
$k = k_1 + \ell + k_2$.  Let also $\phi$ be a rank $k$
$\Lin(\cK)$-formula over $\Sigma$ with free variables in a finite set
$Y$, and let $U = \str(u,\mu)$ be a $Z$-structure of rank $k_1+1+k_2$
for some $Z \subseteq Y$. (Without loss of generality, we may assume 
that $u \ne \1$.) Let $X = Y\setminus Z$.  We prove by
structural induction on $\phi$ that there exists a rank $\ell$
$\Lin(\cK)$-formula $\phi'$ over $\Sigma$, with free variables in $X$
and such that, for every tree $t\in \Sigma M_\ell$ and every mapping
$\lambda\colon X \rightarrow \NV(t)$ (see Figure~\ref{fig left 
quotients}), we have
$$(t,\lambda) \models \phi' \enspace \Longleftrightarrow \enspace 
(U\cdot (\k_1\oplus t\oplus \k_2), \lambda) \models \phi.$$

If $\phi$ is a formula without free variables ($X = Y = Z = 
\emptyset$, $U = u \in \Sigma M_{k_1+1+k_2})$, this shows that 
$L_{\phi'} = (u,k_1,k_2)\inv L_\phi$, and hence that $\Linlg(\cK)$ 
is closed under left quotients. 

We now proceed with the proof.  We first observe that $\NV(t)$ may be
viewed as a subset of $\NV(U\cdot (\k_1\oplus t\oplus \k_2))$: more
precisely, the latter set is equal to the disjoint union of $\NV(t)$
and $\NV(U) = \NV(u)$.
\begin{figure}[ht]
    \centering
    \begin{picture}(75,35)(0,-35)
	\drawline[AHnb=0](15.0,0.0)(0,-20)
	\drawline[AHnb=0](15,0)(30,-20)
	\drawline[AHnb=0](0,-20)(30,-20)
	\node[Nfill=y,fillcolor=Black,Nw=1.0,Nh=1.0,Nmr=1.0](n0)(15.0,-20){}
	\drawline[AHnb=0](15.0,-20.0)(7,-35)
	\drawline[AHnb=0](15,-20)(23,-35)
	\drawline[AHnb=0](7,-35)(23,-35)
	\put(13.5,-12){$U$}
	\put(7,-18.5){$k_1$}
	\put(21,-18.5){$k_2$}
	\put(14,-30){$t$}
	\put(30,-12){a $Z$-structure}
	\put(30,-30){variables in $X$ are interpreted here}
\end{picture}
\caption{$S = U\cdot (\k_1\oplus t\oplus \k_2)$}
\label{fig left quotients}
\end{figure}
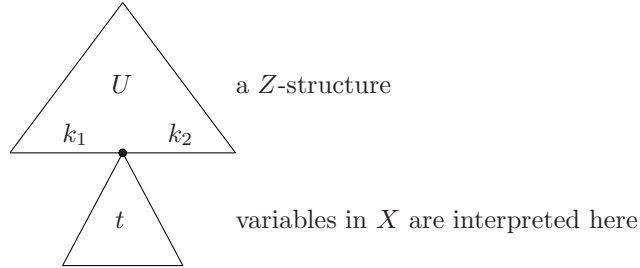

If $\phi$ is equal to $P_\sigma(x)$, we let $\phi' = \phi$ if $x
\not\in Z$, and $\phi' = \true$ (resp.  $\false$) if $x \in Z$ and $x$
occurs at a node of $U$ for which the first component of the label
is (resp.  is not) $\sigma$.  That is, if $x\in Z$ and $U$ satisfies
(resp.  does not satisfy) $\phi$.

Let now $\phi = \leftf_j(x)$ (resp.  $\rightf_j(x)$, $\maxf_{i,j}(x)$)
with $1\le j \le k$.  If $x \not\in Z$, we let $\phi' =
\leftf_{j-k_1}(x)$ if $k_1 \leq j \leq k_1 + \ell$ (resp.
$\rightf_{j-k_1}(x)$ if $k_1 < j \leq k_1 + \ell + 1$,
$\maxf_{i,j-k_1}(x)$ if $k_1 < j \le k_1 + \ell$), and $\phi' =
\false$ otherwise.  If $x\in Z$, then it does not depend on $t$ and
$\lambda$ whether $(U\cdot (\k_1\oplus t\oplus \k_2), \lambda) \models
\phi$ or not, and we let $\phi' = \true$ or $\false$ accordingly.

If $\phi = \root(x)$ and $x\not\in Z$, we let $\phi' = \false$.  If
$x\in Z$, we let $\phi' = \true$ or $\false$ depending on whether $U
\models \phi$.

Now consider the case where $\phi = (x<y)$.  If $x,y \not\in Z$, we
let $\phi' = \phi$.  If $x \not\in Z$ and $y\in Z$, we let $\phi' =
\false$.  If $x \in Z$ but $y\not\in Z$, we let $\phi' = \true$ or
$\false$ depending whether the node of $U$ where $x$ occurs in an
ancestor of the $(k_1+1)$-st variable leaf.  Finally, if $x,y \in Z$,
we let $\phi' = \true$ or $\false$ depending whether $U\models \phi$.

The last case of an atomic formula occurs if $\phi$ is of the form
$\Suc_i(x,y)$.  If $x,y\not\in Z$, we let $\phi' = \phi$.  If $x,y\in
Z$, we let $\phi' = \true$ or $\false$, depending on whether $U
\models \phi$.  If $x\not\in Z$ and $y\in Z$, we let $\phi' = \false$.
Finally, if $x\in Z$ and $y\not\in Z$, let $w$ be the node of $U$
where $x$ occurs.  If the $i$-th successor of $w$ is the ($k_1+1$)-st
variable leaf, we let $\phi' = \root(y)$; otherwise we let $\phi' =
\false$.

As usual, if $\phi = \phi_1 \vee \phi_2$ (resp.  $\phi = \neg\phi_1$),
then we let $\phi' = \phi'_1 \vee \phi'_2$ (resp.  $\phi' =
\neg\phi'_1$).

We now consider the case where $\phi = Q_K x\cdot \langle \phi_\delta
\rangle_{\delta \in \Delta}$.  We may assume that $x\not\in Y$.  Let
$S = U\cdot (\k_1 \oplus t \oplus \k_2)$, let $s = u\cdot (\k_1 \oplus
t \oplus \k_2)$ and let $\lambda\colon X \rightarrow \NV(t)$.  Let
$\bar s_\lambda$ be the characteristic tree determined by $s$,
$[\lambda;\mu]$ and $\langle \phi_\delta \rangle_{\delta \in \Delta}$.
Then, for any $v\in \NV(s)$ and for any $\delta$, we have
\begin{eqnarray*}
    (\bar s_\lambda, [x\mapsto v]) \models P_\delta(x)
    &\Longleftrightarrow& (s, [\lambda; \mu; x\mapsto v]) \models
    \phi_\delta \\
    &\Longleftrightarrow& (S, [\lambda; x\mapsto v]) \models
    \phi_\delta.
\end{eqnarray*}
Moreover,
$$(s,[\lambda;\mu]) \models \phi \enspace\Longleftrightarrow\enspace
(S,\lambda) \models \phi \enspace \Longleftrightarrow \enspace \bar
s_\lambda \in K.$$
 
For each $\delta\in \Delta$, let $\phi'_\delta$ be the formula
associated with $\phi_\delta$ and $U$ by the induction hypothesis.
Let $\bar t_\lambda$ be the characteristic tree determined by $t$,
$\lambda$ and $\langle \phi'_\delta \rangle_{\delta\in \Delta}$.
Then, for any node $v\in \NV(t)$, we have
\begin{eqnarray*}
(S,[\lambda; x\mapsto v]) \models \phi_\delta
&\Longleftrightarrow& (t,[\lambda; x\mapsto v]) \models \phi'_\delta\\
&\Longleftrightarrow& (\bar t_\lambda,[x\mapsto v]) \models P_\delta(x),
\end{eqnarray*}
and hence $\bar s_\lambda$ is of the form $\bar s_\lambda = \hat u\cdot
(\k_1 \oplus \bar t_\lambda \oplus \k_2)$ for some tree $\hat u$ which
differs from $u$ only in the labeling of the nodes in $\NV(u)$.

For each $v\in \NV(u)$, we let $U^{(v)}$ be the
($Z\cup\{x\}$)-structure obtained from $U$ by adding $x$ to the second
component of the label of $v$.  Then, for each $\delta\in\Delta$, we
let $\psi_{\delta,v}$ be the formula associated with $\phi_\delta$ and
$U^{(v)}$ by the induction hypothesis. Then we have
\begin{eqnarray*}
(t,\lambda) \models \psi_{\delta,v} &\Longleftrightarrow&
(U^{(v)}\cdot(\k_1\oplus t \oplus \k_2),\lambda) \models
\phi_\delta\\
&\Longleftrightarrow& (S,[\lambda; x\mapsto v]) \models \phi_\delta \\
&\Longleftrightarrow& (\bar s_\lambda, [x\mapsto v]) \models P_\delta(x).
\end{eqnarray*}

Now, for each mapping $\alpha\colon \NV(u)\rightarrow \Delta$, let
$\hat u_\alpha$ be the
tree obtained from $u$ by relabeling each node $v\in \NV(u)$ with
$\alpha(v)$.  Let also $\psi_\alpha$ be the conjunction of the
$\psi_{\alpha(v),v}$ when $v$ runs over $\NV(u)$.  Then
$$(t,\lambda) \models \psi_\alpha \enspace\Longleftrightarrow\enspace 
\bar s_\lambda = \hat u_\alpha \cdot (\k_1 \oplus \bar t_\lambda \oplus 
\k_2).$$
Finally, let
$$\phi'' = \bigvee_\alpha \Big(\psi_\alpha \land Q_{(\hat
u_\alpha,k_1,k_2)\inv K} \langle \phi'_\delta\rangle_\delta\Big),$$
where the disjunction runs over all mappings $\alpha\colon
\NV(u)\rightarrow \Delta$.  Then the above discussion establishes that
$(t,\lambda)$ satisfies $\phi''$ if and only if $(U\cdot(\k_1\oplus
t\oplus \k_2),\lambda)$ satisfies $\phi$.  Moreover, since each $(\hat
u_\alpha,k_1,k_2)\inv K$ is in $\Linlg(\cK)$, the formula $\phi''$ is a
$\Lin(\Linlg(\cK))$-formula, and by Theorem~\ref{thm-closure},
$\phi''$ is equivalent to some $\Lin(\cK)$-formula $\phi'$, which
concludes this proof.
\cqfd

\paragraph{Right quotients}
The proof concerning the closure under right quotients is similar.  We
assume that every right quotient of a language in $\cK$ belongs to
$\Linlg(\cK)$.  Let $k\ge 0$ and let $\phi$ be a rank $k$
$\Lin(\cK)$-formula over $\Sigma$ with free variables in a finite set
$Y$.  Let $n\ge 1$ and $Z \subseteq Y$, and let $U = U_1 \oplus \cdots
\oplus U_n \in \Sigma_Z M_{n,k}$ where each $U_i$ is a $Z_i$-structure
of rank $k_i$, $U_i = \str(u_i,\mu_i)$, $k = \sum_i k_i$ and the $Z_i$
form a partition of $Z$.  Let $u = \oplus_i u_i$, $\mu =
[\mu_1,\ldots,\mu_n]$ and $X = Y \setminus Z$.

\begin{figure}[ht]
    \centering
    \begin{picture}(90,26)(0,-31)
	\drawline[AHnb=0](20.0,0.0)(5,-20)
	\drawline[AHnb=0](20,0)(35,-20)
	\drawline[AHnb=0](5,-20)(35,-20)
	\put(19,-13){$t$}
	\drawline[AHnb=0](5.0,-21)(0,-31)
	\drawline[AHnb=0](35,-21)(40,-31)
	\drawline[AHnb=0](0.0,-31)(40,-31)
	\drawline[AHnb=0](5,-21)(35,-21)
	\put(19.5,-27){$U$}
	\put(45,-13){variables in $X$ are interpreted here}
	\put(45,-27){a $\oplus$-sum of $Z_i$-structures}
\end{picture}
\caption{$S = t\cdot U$}
\label{fig right quotients}
\end{figure}
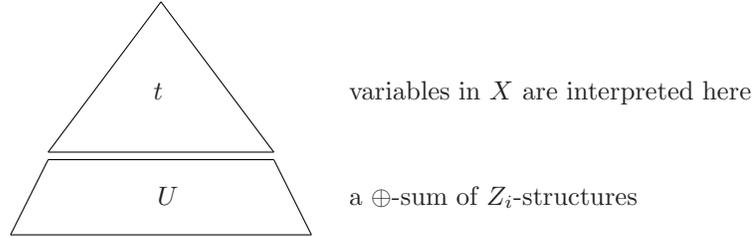

We show by structural induction on $\phi$ that there exists a rank $n$
$\Lin(\cK)$-formula $\phi'$ with free variables in $X$ such that, for
every tree $t\in \Sigma M_n$ and every mapping $\lambda\colon X
\rightarrow \NV(t)$ (see Figure~\ref{fig right quotients}), we have
$$(t,\lambda) \models \phi' \enspace\Longleftrightarrow\enspace
(t\cdot U,\lambda) \models \phi.$$

We first consider the case of atomic formulas.  If $\phi =
P_\sigma(x)$, we let $\phi' = \phi$ if $x\not\in Z$ and $\phi' =
\true$ or $\false$ if $x\in Z$, depending on whether $U$ satisfies
$\phi$.

If $\phi = \root(x)$, we let $\phi' = \phi$ if $x\not\in Z$, and
$\false$ if $x\in Z$.

If $\phi = \maxf_{i,j}(x)$ and $x\not\in Z$, we let $\phi' = \phi$ if
$j = k_1+\cdots+k_{h-1} +1$ for some $h$ such that $U_h = \1$, and
$\phi' = \false$ otherwise.  If $x\in Z$ and $k_1+\cdots+k_{h-1} \le j
\le k_1+\cdots+k_h$, we let $\phi' = \true$ or $\false$ depending on
whether $U_h \models \maxf_{i,j-(k_1+\cdots+k_{h-1})}(x)$.

Suppose now that $\phi = (x < y)$.  If $x,y \not\in Z$, we let $\phi'
= \phi$.  If $x,y\in Z$, we let $\phi' = \true$ or $\false$ depending
on whether one of the $U_j$ satisfies $\phi$.  If $x\in Z$ and
$y\not\in Z$, we let $\phi' = \false$.  Finally, if $x\not\in Z$ and
$y\in Z$, let $1\le j \le n$ be such that $y \in Z_j$ (i.e. $y$ occurs
in $U_j$).  Then we let $\phi' = \bigvee_{i<j}\leftf_i(x) \land
\bigvee_{j<h} \rightf_h(x)$.

The situation is similar if $\phi = \Suc_i(x,y)$.  If $x,y \not\in Z$,
we let $\phi' = \phi$.  If $x,y\in Z$, we let $\phi' = \true$ or
$\false$ depending on whether one of the $U_j$ satisfies $\phi$.  If
$x\in Z$ and $y\not\in Z$, we let $\phi' = \false$.  Finally, if
$x\not\in Z$ and $y\in Z$, let $j$ be such that $y\in Z_j$.  If $y$
does not occur at the root of $U_j$, we let $\phi'= \false$.  If $y$
does occur at the root of $U_j$, we let $\phi' = \maxf_{i,j}(x)$.

Finally, suppose that $\phi = \leftf_j(x)$ (resp.  $\rightf_j(x)$).
If $x\in Z$, let $i$ be such that $x \in Z_i$.  Then we let $\phi' =
\true$ or $\false$ according to whether $k_1+\cdots+k_{i-1} \le j \le
k_1+\cdots+k_i$ and $U_i$ satisfies
$\leftf_{j-(k_1+\cdots+k_{i-1})}(x)$ (resp.
$\rightf_{j-(k_1+\cdots+k_{i-1})}(x)$).  If $x\not\in Z$, we let $\phi'
= \leftf_{h}(x)$ if $j = \sum_{i\le h} k_i$ (resp.  $j = 1 +
\sum_{i\le h} k_i$) for some $h$, and $\phi' = \false$ if $j$ is not
of that form.

If $\phi = \phi_1 \vee \phi_2$ (resp.  $\phi = \neg\phi_1$), then we
let $\phi' = \phi'_1 \vee \phi'_2$ (resp.  $\phi' = \neg\phi'_1$), and
we now assume that $\phi = Q_K x\cdot \langle \phi_\delta
\rangle_{\delta \in \Delta}$, with $x\not\in Y$.  Let $S = t\cdot U$,
let $s = t\cdot u$ and let $\lambda\colon X \rightarrow \NV(t)$.  Let
$\bar s_\lambda$ be the characteristic tree determined by $s$,
$[\lambda; \mu]$ and $\langle \phi_\delta \rangle_{\delta \in
\Delta}$.  For each $\delta\in \Delta$, let $\phi'_\delta$ be the
formula associated with $\phi_\delta$ and $U$ by the induction
hypothesis, and let $\bar t_\lambda$ be the characteristic tree
determined by $t$, $\lambda$ and $\langle \phi'_\delta
\rangle_{\delta\in \Delta}$.  Then the tree $\bar s_\lambda$ is of the
form $\bar s_\lambda = \bar t_\lambda \cdot \hat u$ for some tree $\hat
u$ which differs from $u$ only in the labeling of the nodes in
$\NV(u)$.

We continue as in the left quotient case.  For each $v\in \NV(u)$, we
let $U^{(v)}$ be the structure obtained from $U$ by adding $x$ to the
second component of the label of $v$ and for each $\delta\in \Delta$,
we let $\psi_{\delta,v}$ be the formula associated with $\phi_\delta$
and $U^{(v)}$ by the induction hypothesis.  As above, we verify that
if $\hat u_\alpha$ is the relabeling of $u$ determined by the mapping
$\alpha\colon \NV(u) \longrightarrow \Delta$, and if $\psi_\alpha$ is 
the conjunction of the $\psi_{\alpha(v),v}$ (over the nodes $v\in 
NV(u)$), then
$$(t,\lambda) \models \psi_\alpha \enspace \Longleftrightarrow
\enspace \bar s_\lambda = \bar t_\lambda \cdot \hat u_\alpha.$$
We then let
$$\phi'' = \bigvee_\alpha \Big(\psi_\alpha \land Q_{K \hat
u_\alpha\inv} \langle \phi'_\delta\rangle_\delta\Big),$$
where the disjunction runs over all mappings $\alpha\colon
\NV(u)\rightarrow \Delta$, and we note that $(t,\lambda)$ satisfies
$\phi''$ if and only if $(t\cdot U,\lambda)$ satisfies $\phi$.  Since
each $K\hat u_\alpha\inv$ is in $\Linlg(\cK)$, the formula $\phi''$ is
a $\Lin(\Linlg(\cK))$-formula, and hence is equivalent to a
$\Lin(\cK)$-formula $\phi'$, which concludes the proof.
\cqfd

\subsection{Logics admitting relativization}

We say that a fragment $\L$ of $\Lin$ \textit{admits relativization}
if Properties $R1$ and $R2$ below hold.

\begin{figure}[ht]
    \centering
    \begin{picture}(75,39)(0,-39)
	\drawline[AHnb=0](15,0)(0,-30)
	\drawline[AHnb=0](15,0)(30,-30)
	\drawline[AHnb=0](0,-30)(9,-30)
	\drawline[AHnb=0](9,-30)(15,-18)
	\drawline[AHnb=0](15,-18)(21,-30)
	\drawline[AHnb=0](21,-30)(30,-30)
	\node[Nfill=y,fillcolor=Black,Nw=1.0,Nh=1.0,Nmr=1.0](n0)(15.0,-18){}
	\drawline[AHnb=0](15,-21)(8,-35)
	\drawline[AHnb=0](15,-21)(22,-35)
	\drawline[AHnb=0](8,-35)(22,-35)
	\put(14,-12){$r$}
	\put(2,-33){$k_1$}
	\put(25,-33){$k_2$}
	\put(17,-18){$v$}
	\put(6,-39){$k-(k_1+k_2)$}
	\put(14,-31){$s$}
	\drawline[AHnb=0](60,0)(45,-30)
	\drawline[AHnb=0](60,0)(75,-30)
	\drawline[AHnb=0](45,-30)(54,-30)
	\drawline[AHnb=0](54,-30)(60,-18)
	\drawline[AHnb=0](60,-18)(66,-30)
	\drawline[AHnb=0](66,-30)(75,-30)
	\node[Nfill=y,fillcolor=Black,Nw=1.0,Nh=1.0,Nmr=1.0](n0)(60.0,-18){}
	\drawline[AHnb=0](60,-21)(53,-35)
	\drawline[AHnb=0](60,-21)(67,-35)
	\drawline[AHnb=0](53,-35)(67,-35)
	\put(52,-24){$r$}
	\put(47,-33){$k_1$}
	\put(70,-33){$k_2$}
	\put(62,-20){$w$}
	\put(59,-39){$\ell$}
	\put(59,-31){$s$}
	\put(57,-12){$v$}
	\drawline[AHnb=0](58,-13)(54,-18)
	\drawline[AHnb=0](58,-13)(60,-18)
	\put(57.5,-17){$\scriptstyle i$}
	\drawline[AHnb=0](58,-13)(66,-18)
\end{picture}
\caption{The factorizations of $t$ in Properties $R1$ and $R2$ 
respectively}
\label{fig R1R2}
\end{figure}
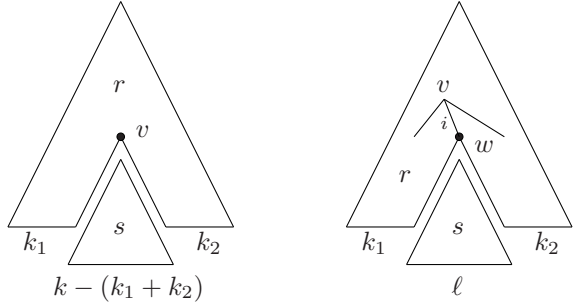

\paragraph{Property $R1$} For all integers $k_1,k_2\ge 0$ and $k \ge
k_1 + k_2$, for each $\L$-sentence $\phi$ of rank $k_1 + 1 + k_2$ over
an alphabet $\Sigma$ and for each first-order variable $x$ without
occurrence in $\phi$, there exists an $\L$-formula $\phi[\not > x]$ of
rank $k$ in the free variable $x$ with the following property.  For
each tree $t \in \Sigma M_k$ and for each node $v\in \NV(t)$, $(t, x
\mapsto v)$ satisfies $\phi[\not > x]$ if and only if
\begin{itemize}
    \item if $s$ is the subtree of $t$ with root $v$, then $t$ is of
    the form $t = r \cdot (\k_1 \oplus s \oplus \k_2)$ (see 
    Figure~\ref{fig R1R2}), and
    
    \item $r \models \phi$.
\end{itemize}

\paragraph{Property $R2$} For all ranked alphabet $\Sigma$, integer $i
\geq 1$ less than or equal to the maximal rank of a letter in $\Sigma$
and integers $k_1, k_2, \ell \ge 0$, for each rank $\ell$
$\L$-sentence $\phi$ and for each first-order variable $x$ without
occurrence in $\phi$, there exists an $\L$-formula $\phi[\geq xi]$ of
rank $k_1 + \ell + k_2$ over $\Sigma$ in the free variable $x$ with
the following property.  For each tree $t \in \Sigma M_{k_1 + \ell +
k_2}$ and for each node $v\in \NV(t)$, $(t, x \mapsto v)$ satisfies
$\phi[\geq xi]$ if and only if
\begin{itemize}
    \item the rank of $v$ is greater than or equal to $i$, and its
    $i$-th child, $w$, has rank $\ell$
    
    \item if $s$ is the subtree of $t$ with root $w$, then $t$ is of
    the form $t = r \cdot (\k_1 \oplus s \oplus \k_2)$, and $s \models
    \phi$ (see Figure~\ref{fig R1R2}).
\end{itemize}

\begin{prop}\label{prop relativization conditions}
    Let $\cK$ be a class of tree languages containing $\cK_\exists$
    and closed under the following operations.  Let $k_1, k_2, \ell
    \ge 0$, let $\Delta$ be a ranked alphabet and let $E$ be a
    disjoint ranked alphabet such that $\card(E_n) = 1$ if
    $\Delta_n\ne\emptyset$, and $\card(E_n) = 0$ otherwise: if $K
    \subseteq \Delta M_{k_1+1+k_2}$ belongs to $\cK$, then $K \cdot
    (\k_1 \oplus EM_\ell \oplus \k_2) \in \Linlg(\cK)$; if $K
    \subseteq \Delta M_\ell$ belongs to $\cK$, then $EM_{k_1+1+k_2}
    \cdot (\k_1 \oplus K \oplus \k_2) \in \Linlg(\cK)$.
    
    Then $\Lin(\cK)$ admits relativization.
\end{prop}    

\proof
We first consider Property $R1$.  Let $k_1, k_2 \ge 0$, let $k \ge k_1
+ k_2$ and let $\phi$ be a $\Lin(\cK)$ formula over $\Sigma$, of rank
$k_1 + 1 + k_2$, without any occurrence of $x$.  We show by structural
induction on $\phi$ that there exists a rank $k$ $\Lin(\cK)$-formula
$\phi[\not >x]$ where $x$ is a free variable and such that, for any
tree $t\in \Sigma M_k$, the following holds: if $v\in \NV(t)$ and $t =
r\cdot (\k_1 \oplus s \oplus \k_2)$ with the tree $s$ rooted at $v$,
and if $\lambda\colon Y \rightarrow \NV(r)$ is an interpretation (where
$Y$ is a set containing the free variables of $\phi$ and not
containing $x$), then
$$(t,[\lambda; x\mapsto v]) \models \phi[\not >x] \enspace
\Longleftrightarrow \enspace
(r,\lambda) \models \phi.$$
If $\phi = \leftf_j(y)$ with $j > k_1$, we let $\phi[\not>x] =
\leftf_{j+\ell}(y)$, where $\ell = k - (k_1 + k_2)$.  If $\phi =
\rightf_j(y)$ with $j > k_1+1$, we let $\phi[\not>x] =
\rightf_{j+\ell}(y)$.  If $\phi = \max_{i,j}(y)$ with $j > k_1 + 1$,
we let $\phi[\not>x] = \maxf_{i,j+\ell-1}(y)$.  And if $\phi =
\max_{i,k_1+1}(y)$, we let $\phi[\not>x] = \Suc_{i}(y,x)$.

For all other atomic formulas, we let $\phi[\not>x] = \phi$.  It is
elementary to verify that these choices guarantee the expected
equivalence.  Similarly, if $\phi = \phi_1 \lor \phi_2$ (resp.  $\phi
= \neg\phi_1$), we let $\phi[\not>x] = \phi_1[\not>x] \lor
\phi_2[\not>x]$ (resp.  $\phi[\not>x] = \neg\phi_1[\not>x]$).

Let us now assume that $\phi = Q_K y \cdot \langle \phi_\delta
\rangle_{\delta\in \Delta}$ where $K \subseteq \Delta M_{k_1 + 1 +
k_2}$ is in $\cK$, $y \not\in Y \cup \{x\}$ and the $\phi_\delta$ are
deterministic with respect to $y$.  Let $E$ be a ranked alphabet
disjoint from $\Delta$, with a single rank $n$ element $\epsilon_n$
for each $n$ such that $\Delta_n \ne \emptyset$; and let $\Delta' =
\Delta\cup E$.  Let $L = K \cdot (\k_1 \oplus E M_\ell \oplus \k_2)$;
then $L\in \Linlg(\cK)$ by assumption.  For each $\delta\in \Delta$,
we let $\psi_\delta = \neg(y \ge x) \land \phi_\delta[\not>x]$; and we
let $\psi_{\epsilon} = (y \ge x)$ for each $\epsilon \in 
E$.%
\footnote{To justify this choice of $\psi_\delta$, we need to verify
that $y=x$ is expressible: it is equivalent to $\forall z \wedge_{i =
1}^n \Suc_i(x,z) \leftrightarrow \Suc_i(y,z) \wedge \Suc_i(z,x)
\leftrightarrow \Suc_i(z,y)$ where $n$ denotes the maximal rank of a
letter in $\Sigma$.  The presence of a universal quantifier is
acceptable since we have assumed that $\cK$ contains $\cK_\exists$.}
We note
that the $\psi_\delta$ have their free variables in $Y \cup \{x,y\}$.
Using the induction hypothesis, one verifies that $\langle \psi_\delta
\rangle_{\delta \in \Delta'}$ is deterministic with respect to $y$,
and we let $\psi = Q_L y \cdot \langle \psi_\delta \rangle_{\delta \in
\Delta'}$.  Then $\psi$ is a $\Lin(\Linlg(\cK))$-formula, and by
Theorem~\ref{thm-closure}, there exists an equivalent
$\Lin(\cK)$-formula $\psi'$.

By the induction hypothesis, for every $w$ in $\NV(r)$ and $\delta \in
\Delta$, $(t,[\lambda;x \mapsto v,y \mapsto w]) \models \psi_\delta$
if and only if $(r,[\lambda,y \mapsto w]) \models \varphi_\delta$.
Also, $(t,[\lambda;x \mapsto v,y \mapsto w]) \models (y > x)$ for all
$w \in \NV(s)$.  Thus, the characteristic tree determined by $t$,
$[\lambda; x \mapsto v]$ and the $\psi_\delta$ is of the form $\hat{r}
\cdot (\k_1 \oplus \hat{s} \oplus \k_2)$, where $\hat{r}$ is the
characteristic tree determined by $r$, $\lambda$ and the
$\varphi_\delta$, and where each $w \in \NV(\hat{s})$ is labeled in
$E$.  Thus, letting $\varphi[\not > x] = \psi'$, we have the desired
equivalence.

%

Let us now consider Property $R2$.  Let $i \ge 1$, $k_1, k_2, \ell \ge
0$, let $k = k_1 + \ell + k_2$ and let $\phi$ be a rank $\ell$
$\Lin(\cK)$-formula over $\Sigma$ without any occurrence of $x$.  We
show by structural induction on $\phi$ that there exists a rank $k$
$\Lin(\cK)$-formula $\phi[\ge xi]$ where $x$ is a free variable and
such that, for any tree $t\in \Sigma M_k$, the following holds: if
$v\in \NV(t)$, then
$$(t,[\lambda; x\mapsto v]) \models \phi[\ge xi] \enspace
\Longleftrightarrow \enspace
\cases{& $v$ has rank at least $i$, \cr
       & $t$ factors as $t = r\cdot (\k_1 \oplus s \oplus \k_2)$ \cr
       & $(s,\lambda) \models \phi$,}$$ 
where $s$ is the subtree of $t$ rooted at the $i$-th successor of $v$
and $\lambda\colon Y \rightarrow \NV(s)$.

If $\phi = \leftf_j(y)$ (resp.  $\rightf_j(y)$, $\maxf_{h,j}(y)$), we
let $\phi[\ge xi] = \leftf_{k_1+j}(y)$ (resp.  $\rightf_{k_1+j}(y)$,
$\maxf_{h,k_1+j}(y)$).  If $\phi = \root(y)$, we let $\phi[\ge xi] =
\Suc_i(x,y)$.  For all other atomic formulas, we let $\phi[\ge xi] =
\phi$.  If $\phi = \phi_1 \lor \phi_2$ (resp.  $\phi = \neg\phi_1$),
we take $\phi[\ge xi] = \phi_1[\ge xi] \lor \phi_2[\ge xi]$ (resp.
$\phi[\ge xi] = \neg\phi_1[\ge xi]$).  Again, it is elementary to
verify that these choices guarantee the expected equivalence.

Let us now assume that $\phi = Q_K y \cdot \langle \phi_\delta
\rangle_{\delta\in \Delta}$ where $K \subseteq \Delta M_\ell$ is in
$\cK$, $y \not\in Y \cup \{x\}$ and the $\phi_\delta$ are
deterministic with respect to $y$.  Let $E$ and $\Delta'$ be as in the
first part of the proof, and let $L = E M_{k_1+1+k_2} \cdot (\k_1
\oplus K \oplus \k_2)$; then $L\in \Linlg(\cK)$ by assumption.

For each $n \ge 0$ such that $\Sigma_n \ne \emptyset$, let $\chi_n$ be
the formula $\Suc_i(x,z) \land (z \leq y)$ (independent of $n$), and
let $\chi = Q_{K_k(\exists)} z \cdot \langle \chi_n\rangle$.  By
assumption, $\chi$ is a $\Lin(\cK)$-formula.  Moreover, $(t,[x\mapsto
v;y\mapsto w])$ satisfies $\chi$ if and only if $v$ has rank at least
$i$ and $w$ is a descendant of the $i$-th child of $v$.

For each $\epsilon \in E$, let $\psi_\epsilon = \neg\chi$, and for
each $\delta\in \Delta$, let $\psi_\delta = \phi_\delta[\ge xi] \land
\chi$.  By induction, the $\psi_\delta$ ($\delta\in \Delta'$) are
$\Lin(\cK)$-formulas with free variables in $Y \cup \{x,y\}$.  Using
the induction hypothesis again, one verifies that $\langle \psi_\delta
\rangle_{\delta \in \Delta'}$ is deterministic with respect to $y$,
and we let $\psi = Q_L y \cdot \langle \psi_\delta \rangle_{\delta \in
\Delta'}$. Then $\psi$ is a $\Lin(\Linlg(\cK))$-formula, and by
Theorem~\ref{thm-closure}, there exists an equivalent
$\Lin(\cK)$-formula $\psi'$.

It follows as above that $(t,[\lambda; x \mapsto v]) \models \psi'$ if
and only if the rank of $v$ is at least $i$ and $t$ factors as $t = r
\cdot (\k_1 \oplus s \oplus \k_2)$ with $(s,\lambda) \models \varphi$,
where $s$ is the subtree of $t$ with root $v$.
%
%
%
\eop

This can be applied to the classes $\cK_\exists$ and
$\cK_{\exists,\mod}$ discussed in Example~\ref{expl-fo+mod2}.

\begin{cor}\label{cor relativization}
    The logics $\Lin(\cK_\exists)$ and $\Lin(\cK_{\exists,\mod})$
    admit relativization.
\end{cor}    

\proof
Let $\Delta$ be a ranked Boolean alphabet, let $E$ be a disjoint
ranked alphabet as in the statement of Proposition~\ref{prop
relativization conditions}, and let $\Delta' = \Delta \cup E$.  Denote
by $\epsilon_n$ the element of rank $n$ in $E$, if it exists.  Let
$k_1, k_2, \ell \ge 0$, and $k = k_1+\ell+k_2$.

Let $K = K_{k_1+1+k_2}(\exists)$ and $L = K \cdot (\k_1 \oplus EM_\ell
\oplus \k_2)$.  Define, for each $m\ge 0$ such that $\Sigma_m \ne \emptyset$
\begin{eqnarray*}
    \phi_m &=& P_{1_m}(x),\\
    \phi &=& Q_{K_k(\exists)} x \cdot \langle \phi_n\rangle_n,\\
    \chi_m &=& (x \le y) \land \neg P_{\epsilon_m}(y),\\
    \chi &=& Q_{K_k(\exists)} y \cdot \langle \chi_n\rangle_n,\\
    \omega_m &=& \neg(x \le y) \land P_{\epsilon_m}(y),\\
    \omega &=& Q_{K_k(\exists)} y \cdot \langle \omega_n\rangle_n,\\
    \psi_m &=& \leftf_{k_1}(x) \land \rightf_{k_1+\ell+1}(x) \land
    \neg\chi \land \neg\omega \enspace \textrm{and}\\
    \psi &=& Q_{K_k(\exists)} x \cdot \langle \psi_n\rangle_n.
\end{eqnarray*}
Then a tree $t\in \Delta'M_k$ satisfies $\phi$ if and only if a letter
of the form $1_n$ occurs at least once in $t$; $(t, [x\mapsto v])$
satisfies $\chi$ (resp.  $\omega$) if some descendant (resp.
non-descendant) of $v$ has its label in $\Delta$ (resp.  in $E$); and
$t$ satisfies $\psi$ if and only if $t$ can be factored as $t = r\cdot
(\k_1 \oplus s \oplus \k_2)$ with all the nodes in $\NV(s)$ labeled in
$E$ and all the nodes in $\NV(r)$ labeled in $\Delta$.  It is immediate
that $L$ is defined by the $\Lin(\cK_\exists)$-formula $\phi \land \psi$.

Now let $K = K_\ell(\exists)$ and $L = EM_{k_1+1+k_2} \cdot (\k_1
\oplus K \oplus \k_2)$.  Define, for each $m\ge 0$ such that $\Sigma_m
\ne \emptyset$
\begin{eqnarray*}
    \chi_m &=& \neg(x < y) \land \neg P_{\epsilon_m}(y),\\
    \chi &=& Q_{K_k(\exists)} y \cdot \langle \chi_n\rangle_n,\\
    \omega_m &=& (x < y) \land  P_{\epsilon_m}(y),\\
    \omega &=& Q_{K_k(\exists)} y \cdot \langle \omega_n\rangle_n,\\
    \psi_m &=& \leftf_{k_1}(x) \land \rightf_{k_1+\ell+1}(x) \land
    \neg\chi \land \neg\omega \enspace \textrm{and}\\
    \psi &=& Q_{K_k(\exists)} x \cdot \langle \psi_n\rangle_n.
\end{eqnarray*}
Then if $t\in \Delta'M_k$, we have $(t, [x\mapsto v]) \models \omega$
(resp.  $\chi$) if some proper descendant (resp.
non proper-descendant) of $v$ has its label in $E$ (resp.  in
$\Delta$); and $t$ satisfies $\psi$ if and only if $t$ can be factored
as $t = r\cdot (\k_1 \oplus s \oplus \k_2)$ with all the nodes in
$\NV(r)$ labeled in $E$ and all the nodes in $\NV(s)$ labeled in $E$.
It is immediate that $L$ is defined by the $\Lin(\cK_\exists)$-formula
$\phi \land \psi$.

The proof that $\Lin(\cK_{\exists,\mod})$ admits relativization is 
similar. 
\eop

\section{Algebraic characterization of logically defined tree languages}
\label{sec alg charact}

\subsection{The block product of preclones}

In this section, we introduce our main algebraic tool, the block
product of preclones and of \pgs.  This is a generalization of an
operation on monoids that was introduced by Rhodes and Tilson
\cite{RhodesTilson}, as a two-sided generalization of the more
classical wreath product.

Let us first (attempt to) briefly summarize the spirit of the block
product of monoids, which was introduced \cite{RhodesTilson} in
relation with the description of bimachines (Eilenberg
\cite{Eilenberg}).  Let $T$ be a finite monoid and let $\tau\colon A^*
\rightarrow T$ be a morphism.  The associated bimachine $\mathcal T$
represents the simultaneous operations of left-to-right and
right-to-left processing of a string in $A^*$ by $\tau$: if $a_1\cdots
a_n \in A^*$, the $i$-th component of this processing is the triple
$(\tau(a_1\cdots a_{i-1}),a_i,\tau(a_{i+1}\cdots a_n)) \in T\times
A\times T$, and the output of $\mathcal T$ is the product of these
components, namely the following string in $(T\times A\times T)^*$:
$$(1,a_1,\tau(a_2\cdots a_n))(\tau(a_1),a_2,\tau(a_3\cdots a_n))\ 
\cdots\ (\tau(a_1\cdots a_{n-1}),a_n,1).$$
The idea of the block product is to capture the (cascade) product of 
this bimachine with an ordinary automaton, that is, to use the output 
of the bimachine as input for another automaton $\mathcal S$ 
operating on alphabet $T\times A\times T$. This translates to a 
monoid morphism $\sigma\colon (T\times A\times T)^*\rightarrow S$ 
into a finite monoid $S$ (the transition monoid of $\mathcal S$) -- 
which is entirely determined by the images of the triples $(t,a,t') 
\in T\times A\times T$. For each $a\in A$, let us denote by $f_a$ the 
map $f_a(t,t') = \sigma(t,a,t')$. Then the composed machine output, 
on input $a_1\cdots a_n$ is
$$f_{a_1}(1,\tau(a_2\cdots a_n))f_{a_2}(\tau(a_1),\tau(a_3\cdots
a_n))\ \cdots\ f_{a_n}(\tau(a_1\cdots a_{n-1}),1).$$
Note that $f_a(t,t')$ is the $\sigma$-image (the $S$-value) of the
effect of letter $a$ in bimachine $\mathcal T$, when $a$ is in a
left-right context whose $T$-values are $t$ and $t'$.  The map $f_a$
itself records the effect of letter $a$ in all possible contexts.

The general definition of the block product of monoids is an
abstraction of these ideas: $S \block T$ is the set of pairs $(f,t)
\in S^{T\times T}\times T$ and the product $(f_1,t_1)\cdots (f_n,t_n)$
is equal to $(g,t_1\cdots t_n)$, with
$$g(t,t') = f_1(1,t_2\cdots t_n)f_2(t_1,t_3\cdots t_n)\ \cdots\
f_{n-1}(t_1\cdots t_{n-2},t_n)f_n(t_1\cdots t_{n-1},1).$$
This operation on monoids proved to be useful to decompose morphisms
\cite{RhodesTilson,mps1,mps2} and to explain the connection between
first-order logic and aperiodic monoids (see \cite{Straubing}).  We
now extend these ideas to preclones.  The resulting definition is more
complex as our contexts are not just left-right pairs (see the
definition of contexts in Section~\ref{sec syntactic}) and we need to
take into account the rank of elements.  In particular, this leads to
the definition of a sequence of block products $S \block_k T$ ($k \geq
0$).

Formally, let $S$ and $T$ be preclones.  We define preclones $S
\block_k T$ for each $k \geq 0$.  Recall (Section~\ref{sec syntactic})
that, for each $k,n \geq 0$, $I_{k,n}$ denotes the set of $n$-ary
contexts in $T_{k}$.  The set of rank $n$ elements of $S \block_k T$
is defined to be
$$(S \block_k T)_n = S_n^{I_{k,n}} \times T_n, \quad n \geq 0.$$

The identity $\1$ is the pair $(F_\1, \1)$, where $F_\1(C) = \1$, for
all $C \in I_{k,1}$.  As for the composition operation, let $(F,f) \in
(S \block_k T)_n$, and let $(G_i, g_i)\in (S \block_k T)_{m_i}$ for
each $i\in [n]$.  Let $g = g_1\oplus \cdots \oplus g_n \in T_{n,m}$,
where $m = \sum_{i=1}^n m_i$.  Then we let
$$(F,f) \cdot ((G_1,g_1)\oplus \cdots \oplus (G_n,g_n)) = (Q, f\cdot
g),$$
an element of $(S \block_k T)_m = S_m^{I_{k,m}} \times T_m$, where
$Q\colon I_{k,m} \rightarrow S_m$ is described as follows.

\begin{figure}[ht]
    \centering
    \begin{picture}(110,70)(0,-70)
	\drawline[AHnb=0](15.0,0.0)(0,-30)
	\drawline[AHnb=0](15,0)(30,-30)
	\drawline[AHnb=0](0,-30)(30,-30)
	\node[Nfill=y,fillcolor=Black,Nw=1.0,Nh=1.0,Nmr=1.0](n0)(15.0,-30){}
	\put(14,-17){$u$}
	\put(7,-28){$k_1$}
	\put(21,-28){$k_2$}
	\drawline[dash={2.0 1.0}{0.0},AHnb=0](15.0,-30.0)(5,-45)
	\drawline[dash={2.0 1.0}{0.0},AHnb=0](15,-30)(25,-45)
	\drawline[dash={2.0 1.0}{0.0},AHnb=0](5,-45)(25,-45)
	\drawline[AHnb=0](5.0,-46)(1.5,-56)
	\drawline[AHnb=0](25,-46)(29.5,-56)
	\drawline[AHnb=0](5.0,-46)(25,-46)
	\drawline[AHnb=0](1.5,-56)(29.5,-56)
	\put(14,-52){$v$}
	\put(14,-44){$m$}
	\put(14,-60){$\ell$}
	\drawline[AHnb=0](55.0,0.0)(40,-30)
	\drawline[AHnb=0](55,0)(70,-30)
	\drawline[AHnb=0](40,-30)(70,-30)
	\node[Nfill=y,fillcolor=Black,Nw=1.0,Nh=1.0,Nmr=1.0](n1)(55.0,-30){}
	\put(54,-17){$u$}
	\put(47,-28){$k_1$}
	\put(61,-28){$k_2$}
	\drawline[AHnb=0](45.0,-46)(48,-46)
	\drawline[AHnb=0](45.0,-46)(41.5,-56)
	\drawline[AHnb=0](41.5,-56)(51.5,-56)
	\drawline[AHnb=0](48,-46)(51.5,-56)
	\drawline[AHnb=0](65,-46)(68.5,-56)
	\drawline[AHnb=0](62,-46)(58.5,-56)
	\drawline[AHnb=0](62,-46)(65,-46)
	\drawline[AHnb=0](58.5,-56)(68.5,-56)
	\put(53,-52){$\cdots$}
	\put(45,-52){$\bar v_1$}	
	\put(45,-44){$m_1$}	
	\put(45,-60){$\ell_1$}	
	\put(62,-52){$\bar v_n$}
	\put(62,-44){$m_n$}	
	\put(62,-60){$\ell_n$}	
	\drawline[AHnb=0](95.0,0.0)(80,-30)
	\drawline[AHnb=0](95,0)(110,-30)
	\drawline[AHnb=0](80,-30)(110,-30)
	\node[Nfill=y,fillcolor=Black,Nw=1.0,Nh=1.0,Nmr=1.0](n2)(95.0,-30){}
	\drawline[AHnb=0](95,-30)(83,-45)
	\drawline[AHnb=0](95,-30)(107,-45)
	\drawline[AHnb=0](83,-45)(107,-45)
	\node[Nfill=y,fillcolor=Black,Nw=1.0,Nh=1.0,Nmr=1.0](n3)(85.0,-45){}
	\node[Nfill=y,fillcolor=Black,Nw=1.0,Nh=1.0,Nmr=1.0](n4)(100.0,-45){}
	\node[Nfill=y,fillcolor=Black,Nw=1.0,Nh=1.0,Nmr=1.0](n3)(105.0,-45){}
	\drawline[AHnb=0](85,-45)(83,-53)
	\drawline[AHnb=0](85,-45)(87,-53)
	\drawline[AHnb=0](83,-53)(87,-53)
	\drawline[AHnb=0](83,-54)(87,-54)
	\drawline[AHnb=0](82,-60)(88,-60)
	\drawline[AHnb=0](83,-54)(82,-60)
	\drawline[AHnb=0](87,-54)(88,-60)
	\drawline[AHnb=0](100,-45)(98,-53)
	\drawline[AHnb=0](100,-45)(102,-53)
	\drawline[AHnb=0](98,-53)(102,-53)
	\drawline[AHnb=0](98,-54)(102,-54)
	\drawline[AHnb=0](97,-60)(103,-60)
	\drawline[AHnb=0](98,-54)(97,-60)
	\drawline[AHnb=0](102,-54)(103,-60)
	\put(94,-17){$u$}
	\put(87,-28){$k_1$}
	\put(101,-28){$k_2$}
	\put(94,-39){$f$}
	\put(79,-51){$g_1$}
	\put(83.5,-63){$\bar v_1$}
	\put(91,-58){$\cdots$}
	\put(90,-51){$g_{n-1}$}
	\put(97,-63){$\bar v_{n-1}$}
	\drawline[dash={1.0 1.0}{0.0},AHnb=0](105,-45)(106,-60)
	\drawline[dash={1.0 1.0}{0.0},AHnb=0](105,-45)(110,-60)
	\drawline[dash={1.0 1.0}{0.0},AHnb=0](106,-60)(110,-60)
	\drawline[AHnb=0](106,-61)(110,-61)
	\drawline[AHnb=0](105,-67)(111,-67)
	\drawline[AHnb=0](106,-61)(105,-67)
	\drawline[AHnb=0](110,-61)(111,-67)
	\put(107,-70){$\bar v_n$}
\end{picture}
\caption{Two views of $(u,k_1,k_2,v) \in I_{k,m}$, and the context
$C_n$}
\label{fig def block product}
\end{figure}
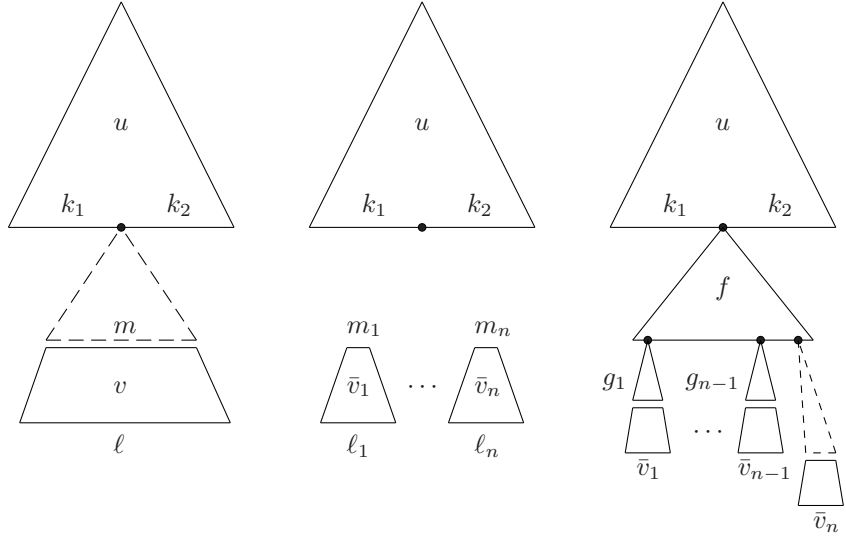

For each $(u,k_1,v,k_2) \in I_{k,m}$, we have $v = v_1 \os \cdots \os
v_m \in T_{m,\ell}$, where $\ell = k - (k_1 + k_2)$.  Let $\bar v_1$
be the $\oplus$-sum of the first $m_1$ $v_j$'s, $\bar v_1 = v_1 \oplus
\cdots \oplus v_{m_1}$, let $\bar v_2$ be the $\oplus$-sum of the next
$m_2$ $v_j$'s, etc, until $\bar v_n = v_{m-m_n+1} \oplus \cdots \oplus
v_m$ is the $\oplus$-sum of the last $m_n$ $v_j$'s, see
Figure~\ref{fig def block product}.  In particular, $v =
\bigoplus_{i=1}^n\bar v_i$.  For each $i \in [n]$, let $\ell_i$ be the
total rank of $\bar v_i$, so that $\bar v_i \in T_{m_i,\ell_i}$ and
$\sum_i\ell_i = \ell$.

For each $i\in [n]$, we observe that $g_i\cdot \bar v_i \in
T_{\ell_i}$, and we let
$$c_i = u \cdot \Big(\k_1 \oplus f\cdot(g_1\cdot\bar v_1 \oplus \cdots
\oplus g_{i-1}\cdot\bar v_{i-1} \oplus \1 \oplus g_{i+1}\cdot\bar
v_{i+1} \oplus \cdots \oplus g_n\cdot\bar v_n) \oplus \k_2\Big).$$
We note that $u\cdot (\k_1 \oplus f
\cdot g \cdot v \oplus \k_2) = c_i \cdot (\p_1\oplus g_i \cdot \bar v_i\oplus
\p_2)$,
where $p_1 = k_1 + \sum_{j<i}\ell_j$ and $p_2 = \sum_{j>i}\ell_j
+ k_2$.  Then $c_i$ is an element of $T$ with rank $p_1 + 1 + p_2 =
k_1 + k_2 + \ell - \ell_i +1 = k - \ell_i +1$.  (Of course, the
integers $p_1$ and $p_2$ depend on $i$ even though our notation does
not show it.)

In particular, $C_i = (c_i, p_1, \bar v_i, p_2)$ is a context in
$I_{k,m_i}$, see Figure~\ref{fig def block product}.  We are finally
ready to define $Q$:
$$ Q(u,k_1,v,k_2) = F(u,k_1,g \cdot v,k_2) \cdot (G_1(C_1) \oplus
\cdots \oplus G_n(C_n)).$$

\begin{lem}
    The above definition satisfies the axioms of preclones.
\end{lem}

\proof
Let us first verify the axioms concerning the identity element.  Let
$(G,g) \in (S\block_k T)_m$ and let $(Q,g) = (F_\1,\1)\cdot (G,g)$.
Let $(u,k_1,v,k_2)\in I_{k,m}$.  With reference to the notation in the
definition above, we have $n = 1$ and $C_1 = (u,k_1,v,k_2)$.  It
follows that $Q = G$, so $(F_\1,\1)\cdot (G,g) = (G,g)$.

Let now $(F,f) \in (S\block_k T)_n$ and let $(Q,f) = (F,f) \cdot
((F_\1,\1)\oplus\cdots\oplus(F_\1,\1))$.  Let $(u,k_1,v,k_2)\in
I_{k,n}$.  Then $Q(u,k_1,v,k_2) = F(u,k_1,v,k_2) \cdot (F_\1(C_1)
\oplus \cdots \oplus F_\1(C_n))$ for some $C_1,\ldots,C_n$, and hence
$Q(u,k_1,v,k_2) = F(u,k_1,v,k_2) \cdot (\1 \oplus \cdots \oplus \1) =
F(u,k_1,v,k_2)$.  Thus $(F,f) \cdot ((F_\1,\1) \oplus \cdots \oplus
(F_\1,\1)) = (F,f)$.

Next let $(F,f) \in (S \block_k T)_n$; for $i\in [n]$ let $(G_i,g_i)
\in (S \block_k T)_{m_i}$ and let $m = \sum_{i\in[n]} m_i$; for $j\in
[m]$, let $(H_j,h_j) \in (S \block_k T)_{p_j}$ and let $p =
\sum_{j\in[m]} p_j$.

Let $g = \oplus_{i\in[n]} g_i$ and let $h = \oplus_{j\in[m]} h_j$.  We
also denote by $\bar h_1$ the $\oplus$-sum of the first $m_1$ $h_j$'s,
$\bar h_2$ the $\oplus$-sum of the next $m_2$ $h_j$'s, etc, to $\bar
h_n$ the $\oplus$-sum of the last $m_n$ $h_j$'s, so that $h =
\oplus_{i\in[n]}\bar h_i$.  The rank of $\bar h_i$ is $\sum_{j=
m_1+\cdots+m_{i-1}+1}^{m_1+\cdots+m_i}p_j$.

We need to consider $p$-ary contexts in $T_k$: let $(u,k_1,v,k_2) \in
I_{k,p}$ be such a context.  Then $v = v_1 \oplus \cdots \oplus v_p
\in T_{p,\ell}$ with $\ell = k - k_1 - k_2$.  Let $\bar v_1$ denote
the $\oplus$-sum of the first $p_1$ $v_i$'s, $\bar v_2$ the
$\oplus$-sum of the next $p_2$ $v_i$'s, etc to $\bar v_m$ the
$\oplus$-sum of the last $p_m$ $v_i$'s.  For $i\in [n]$, we also
denote by $\bar{\bar v}_i$ the $\oplus$-sum of the $\bar v_j$ where
$h_j$ is part of the summation defining $\bar h_i$.  That is,
$\bar{\bar v}_1 = \bar v_1\oplus\cdots\oplus \bar v_{m_1}$,\ldots,
$\bar{\bar v}_n = \bar v_{m-m_n+1}\oplus\cdots\oplus \bar v_m$.

We first consider the product
\begin{eqnarray*}
    && \Big((F,f) \cdot \big((G_1,g_1)\oplus\cdots \oplus
    (G_n,g_n)\big)\Big) \cdot \big((H_1,h_1) \oplus \cdots \oplus
    (H_m,h_m)\big) \\
    &=& (Q,f\cdot g) \cdot \big((H_1,h_1) \oplus \cdots \oplus
    (H_m,h_m)\big) \\
    &=& (R, f\cdot g\cdot h).
\end{eqnarray*}  
Then we have $R(u,k_1,v,k_2) = Q(u,k_1,h\cdot v,k_2) \cdot
(H_1(B_1)\oplus \cdots \oplus H_m(B_m))$, with $B_j =
(b_j,p_1',\bar{v}_j, p_2')$ where $b_j$ ($j\in [m]$) is
$$u \cdot (\k_1 \oplus f
\cdot g \cdot (h_1 \cdot \bar{v}_1 \oplus \cdots \oplus h_{j-1}\cdot
\bar{v}_{j-1} \oplus \1 \oplus h_{j+1}\cdot \bar{v}_{j+1} \oplus
\cdots \oplus h_{m}\cdot
\bar{v}_{m}) \oplus \k_2),$$
$p_1' = k_1 + \sum_{s=1}^{j-1} \rank(\bar v_s)$ and $p'_2 = k_2 + 
\sum_{s=j+1}^{m} \rank(\bar v_s)$, so that
$$u \cdot (\k_1 \oplus f\cdot g\cdot h \cdot v \oplus \k_2) = b_j
\cdot (\p'_1 \oplus h_j\cdot \bar v_j \oplus \p'_2).$$
%
%
%
\begin{figure}[ht]
    \centering
    \begin{picture}(110,92)(0,-92)
	\drawline[AHnb=0](15.0,0.0)(0,-30)
	\drawline[AHnb=0](15,0)(30,-30)
	\drawline[AHnb=0](0,-30)(30,-30)
	\node[Nfill=y,fillcolor=Black,Nw=1.0,Nh=1.0,Nmr=1.0](n0)(15.0,-30){}
	\put(14,-17){$u$}
	\put(7,-28){$k_1$}
	\put(21,-28){$k_2$}
	\drawline[AHnb=0](15.0,-30.0)(8,-43)
	\drawline[AHnb=0](15,-30)(23,-43)
	\drawline[AHnb=0](8,-43)(23,-43)
	\put(14,-39){$f$}
	\drawline[AHnb=0](8,-44)(6,-53)
	\drawline[AHnb=0](23,-44)(25,-53)
	\drawline[AHnb=0](8,-44)(23,-44)
	\drawline[AHnb=0](6,-53)(25,-53)
	\node[Nfill=y,fillcolor=Black,Nw=1.0,Nh=1.0,Nmr=1.0](n3)(23.5,-53){}
	\put(14,-49){$g$}
	\drawline[AHnb=0](6,-54)(4,-63)
	\drawline[AHnb=0](22,-54)(23,-63)
	\drawline[AHnb=0](6,-54)(22,-54)
	\drawline[AHnb=0](4,-63)(23,-63)
	\put(5.5,-59){$\scriptstyle h_1\oplus\cdots\oplus h_{m-1}$}
	\drawline[AHnb=0](4,-64)(2,-73)
	\drawline[AHnb=0](23,-64)(24,-73)
	\drawline[AHnb=0](4,-64)(23,-64)
	\drawline[AHnb=0](2,-73)(24,-73)
	\put(5.5,-69){$\scriptstyle \bar v_1\oplus\cdots\oplus \bar v_{m-1}$}
	\drawline[dash={1.0 1.0}{0.0},AHnb=0](23.5,-53)(26,-73)
	\drawline[dash={1.0 1.0}{0.0},AHnb=0](23.5,-53)(29,-73)
	\drawline[dash={1.0 1.0}{0.0},AHnb=0](26,-73)(29,-73)
	\drawline[AHnb=0](26,-74)(27,-83)
	\drawline[AHnb=0](29,-74)(32,-83)
	\drawline[AHnb=0](26,-74)(29,-74)
	\drawline[AHnb=0](27,-83)(32,-83)
	\put(27,-81){$\bar v_{m}$}
%
%
%
	\drawline[AHnb=0](55.0,0.0)(40,-30)
	\drawline[AHnb=0](55,0)(70,-30)
	\drawline[AHnb=0](40,-30)(70,-30)
	\node[Nfill=y,fillcolor=Black,Nw=1.0,Nh=1.0,Nmr=1.0](n1)(55.0,-30){}
	\put(54,-17){$u$}
	\put(47,-28){$k_1$}
	\put(61,-28){$k_2$}
	\drawline[AHnb=0](55,-30)(48,-43)
	\drawline[AHnb=0](55,-30)(63,-43)
	\drawline[AHnb=0](48,-43)(63,-43)
	\node[Nfill=y,fillcolor=Black,Nw=1.0,Nh=1.0,Nmr=1.0](n4)(61.5,-43){}
	\put(54,-39){$f$}
	\drawline[AHnb=0](48,-44)(47.45,-46.5)
	\drawline[AHnb=0](46.67,-50)(46,-53)
	\drawline[AHnb=0](60,-44)(61,-53)
	\drawline[AHnb=0](48,-44)(60,-44)
	\drawline[AHnb=0](46,-53)(61,-53)
	\put(44,-49){$\scriptstyle g_1\oplus\cdots\oplus g_{n-1}$}
	\drawline[AHnb=0](46,-54)(45.45,-56.5)
	\drawline[AHnb=0](44.67,-60)(44,-63)
	\drawline[AHnb=0](61,-54)(62,-63)
	\drawline[AHnb=0](46,-54)(61,-54)
	\drawline[AHnb=0](44,-63)(62,-63)
	\put(44,-59){$\scriptstyle \bar h_1\oplus\cdots\oplus \bar 
	h_{n-1}$}
	\drawline[AHnb=0](44,-64)(42,-73)
	\drawline[AHnb=0](62,-64)(63,-73)
	\drawline[AHnb=0](44,-64)(62,-64)
	\drawline[AHnb=0](42,-73)(63,-73)
	\put(44.5,-69){$\scriptstyle \bar{\bar v}_1\oplus\cdots\oplus 
	\bar{\bar v}_{n-1}$}
	\drawline[dash={1.0 1.0}{0.0},AHnb=0](61.5,-43)(66,-73)
	\drawline[dash={1.0 1.0}{0.0},AHnb=0](61.5,-43)(69,-73)
	\drawline[dash={1.0 1.0}{0.0},AHnb=0](66,-73)(69,-73)
	\drawline[AHnb=0](66,-74)(67,-83)
	\drawline[AHnb=0](69,-74)(72,-83)
	\drawline[AHnb=0](66,-74)(69,-74)
	\drawline[AHnb=0](67,-83)(72,-83)
	\put(67,-81){$\bar h_{n}$}
	\drawline[AHnb=0](67,-84)(68,-93)
	\drawline[AHnb=0](72,-84)(75,-93)
	\drawline[AHnb=0](67,-84)(72,-84)
	\drawline[AHnb=0](68,-93)(75,-93)
	\put(69,-90){$\bar{\bar v}_{n}$}
%
%
%
%
%
	\drawline[AHnb=0](95.0,0.0)(80,-30)
	\drawline[AHnb=0](95,0)(110,-30)
	\drawline[AHnb=0](80,-30)(110,-30)
	\node[Nfill=y,fillcolor=Black,Nw=1.0,Nh=1.0,Nmr=1.0](n2)(95.0,-30){}
	\put(94,-17){$u$}
	\put(87,-28){$k_1$}
	\put(101,-28){$k_2$}
	\drawline[AHnb=0](95,-30)(88,-43)
	\drawline[AHnb=0](95,-30)(103,-43)
	\drawline[AHnb=0](88,-43)(103,-43)
	\node[Nfill=y,fillcolor=Black,Nw=1.0,Nh=1.0,Nmr=1.0](n5)(101.5,-43){}
	\put(94,-39){$f$}
	\drawline[AHnb=0](88,-44)(87.45,-46.5)
	\drawline[AHnb=0](86.67,-50)(86,-53)
	\drawline[AHnb=0](100,-44)(101,-53)
	\drawline[AHnb=0](88,-44)(100,-44)
	\drawline[AHnb=0](86,-53)(101,-53)
	\put(84,-49){$\scriptstyle g_1\oplus\cdots\oplus g_{n-1}$}
	\drawline[AHnb=0](86,-54)(85.45,-56.5)
	\drawline[AHnb=0](84.67,-60)(84,-63)
	\drawline[AHnb=0](101,-54)(102,-63)
	\drawline[AHnb=0](86,-54)(101,-54)
	\drawline[AHnb=0](84,-63)(102,-63)
	\put(84,-59){$\scriptstyle \bar h_1\oplus\cdots\oplus \bar 
	h_{n-1}$}
	\drawline[AHnb=0](84,-64)(82,-73)
	\drawline[AHnb=0](102,-64)(103,-73)
	\drawline[AHnb=0](84,-64)(102,-64)
	\drawline[AHnb=0](82,-73)(103,-73)
	\put(84.5,-69){$\scriptstyle \bar{\bar v}_1\oplus\cdots\oplus 
	\bar{\bar v}_{n-1}$}
	\drawline[dash={1.0 1.0}{0.0},AHnb=0](101.5,-43)(106,-73)
	\drawline[dash={1.0 1.0}{0.0},AHnb=0](101.5,-43)(109,-73)
	\drawline[dash={1.0 1.0}{0.0},AHnb=0](106,-73)(109,-73)
	\drawline[AHnb=0](106,-74)(107,-83)
	\drawline[AHnb=0](109,-74)(112,-83)
	\drawline[AHnb=0](106,-74)(109,-74)
	\drawline[AHnb=0](107,-83)(112,-83)
	\put(107,-80){$\bar{\bar v}_{n}$}
\end{picture}
\caption{The contexts $B_m$, $C_n$ and $D_n$}
\label{fig associativity}
\end{figure}
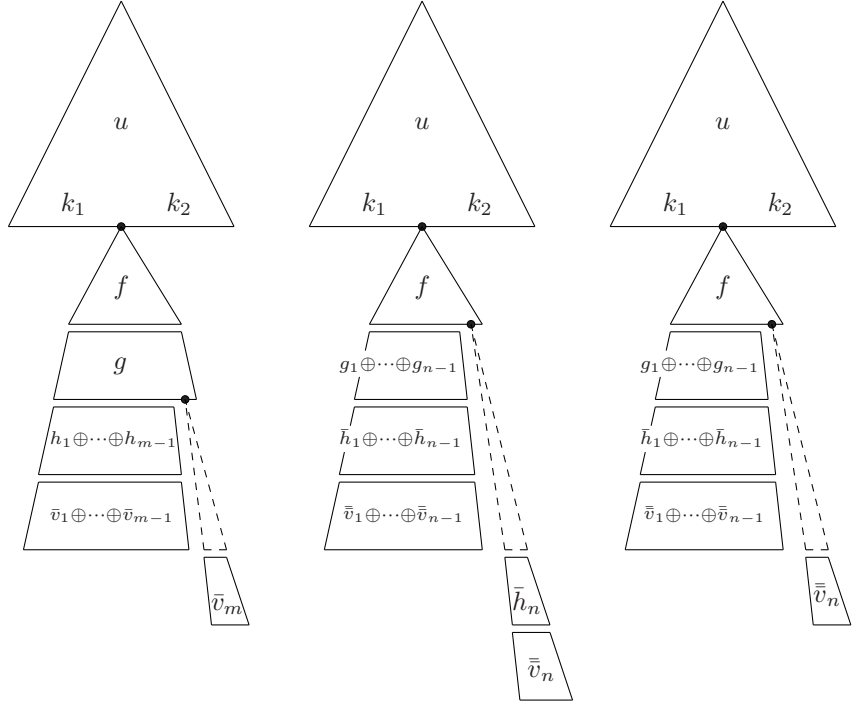 

Moreover, $Q(u,k_1,h\cdot v,k_2) = F(u,k_1,g\cdot h\cdot v,k_2) \cdot
(G_1(C_1) \oplus \cdots \oplus G_n(C_n))$, where $C_i$ ($i\in [n]$) is
the context $(c_i,r_1,\bar h_i\cdot \bar{\bar v}_i,r_2)$ with $c_i$ 
equal to
$$u \cdot \big(\k_1 \oplus f \cdot (g_1 \cdot \bar{h}_1 \cdot \bar{\bar v}_1 \oplus
\cdots \oplus g_{i-1} \cdot \bar{h}_{i-1} \cdot \bar{\bar v}_{i-1} 
\oplus \1 \oplus
 g_{i+1} \cdot \bar{h}_{i+1} \cdot \bar{\bar v}_{i+1} \oplus 
\cdots \oplus  g_{n} \cdot \bar{h}_{n} \cdot \bar{\bar v}_{n})\oplus 
\k_2\big),$$
$r_1 = k_1 + \sum_{s = 1}^{i-1}\rank(\bar{\bar v}_s)$, $r_2 = k_2 +
\sum_{s=i+1}^n\rank(\bar{\bar v}_s)$, so that
$$u \cdot (\k_1 \oplus
f\cdot g\cdot h \cdot v \oplus \k_2) = c_i \cdot (\r_1 \oplus g_i
\cdot \bar h_i \cdot \bar{\bar v}_i \oplus \r_2).$$
See Figure~\ref{fig associativity}.  Thus
%
%
%
$$R(u,k_1,v,k_2) = F(u,k_1,g\cdot h\cdot v,k_2) \cdot
\Big(\bigoplus_{i=1}^n G_i(C_i)\Big) \cdot \Big(\bigoplus_{j=1}^m
H_j(B_j)\Big).$$
We compare this result with the product
\begin{eqnarray*}
    && (F,f) \cdot \Big(\big((G_1,g_1)\oplus\cdots \oplus
    (G_n,g_n)\big) \cdot \big((H_1,h_1) \oplus \cdots \oplus
    (H_m,h_m)\big)\Big) \\
    &=& (F,f) \cdot \big((Q'_1,g_1\cdot \bar h_1) \oplus \cdots \oplus
    (Q'_n,g_n\cdot \bar h_n)\big) \\
    &=& (R', f\cdot g\cdot h).
\end{eqnarray*}  
Then we have $R'(u,k_1,v,k_2) = F(u,k_1,g\cdot h\cdot v,k_2) \cdot
(Q'_1(D_1)\oplus \cdots \oplus Q'_n(D_n))$, where $D_i$ ($i\in [n]$)
is the context $(c_i,r_1,\bar{\bar v}_i,r_2)$, where $c_i, r_1, 
r_2$ are defined above.
%

Next we compute $Q'_1(D_1)$: we have
$$Q'_1(D_1) = Q'_1(c_1,r_1,\bar{\bar v}_1,r_2) = G(c_1,r_1,\bar h_1
\cdot \bar{\bar v}_1,r_2) \cdot (H_1(E_{1}) \oplus \cdots \oplus
H_{m_1}(E_{m_1}))$$
where $E_{j}$ ($j\in [m_1]$) is the context $(e_{j},r'_{1,j},\bar
v_j,r'_{2,j})$,
\begin{eqnarray*} 
e_j & = & u \cdot (\k_1 \oplus f \cdot (g_1 \cdot (h_1 \cdot \bar{v}_1 \oplus 
\cdots \oplus h_{j-1}\cdot \bar{v}_{j-1} \oplus \1 \\
&&\oplus  h_{j+1}\cdot \bar{v}_{j+1} \oplus \cdots \oplus h_{m_1} \cdot
\bar{v}_{m_1}) \oplus g_2 \cdot \bar{h}_2 \cdot \bar{\bar{v}}_2 \oplus \cdots
\oplus g_n \cdot \bar{h}_n \cdot \bar{\bar{v}}_n) \oplus \k_2)
\end{eqnarray*} 
and $r'_{1,j}$ and $r'_{2,j}$ are
appropriate integers so that $c_1\cdot(\r_1 \oplus
g_1\cdot\bar h_1\cdot\bar{\bar v}_1 \oplus \r_2) = e_j \cdot (\r_{1,j} \oplus
h_j\cdot \bar v_j \oplus \r_{2,j})$, see Figure~\ref{fig associativity}.

We observe now that $e_j = b_j$ and $E_j = B_j$ for $j\in [m_1]$.  So
we have $Q'_1(D_1) = G_1(C_1) \cdot \bigoplus_{j = 1}^{m_1}H_j(B_j)$.

Similarly, for each $i\in[n]$, we have
$$Q'_i(D_i) = G_i(C_i) \cdot \bigoplus_{j =
1}^{m_i}H_{m_1+\cdots+m_{i-1}+j}(B_{m_1+\cdots+m_{i-1}+j}),$$
and we have verified that $R = R'$.
\eop

We also define block products of \pgs.  If $(S,A)$ and $(T,B)$ are
\pgs\ and $k \geq 0$, we define $(S,A) \block_k (T,B)$ to be the
sub-\pgp\ of $S \block_k T$ generated by those pairs $(F,g)$ such
that for some $n \ge 0$, $g \in B_n$ and $F(c) \in A_n$ for each $c\in
I_{k,n}$.

\begin{figure}[ht]
    \centering
    \begin{picture}(25,20)(0,-20)
	\gasset{Nframe=n,Nadjust=w,Nh=6,Nmr=0}
	\node(S)(0,-2){$\Sigma M$}
	\node(B)(25,-2){$BM$}
	\node(U)(0,-20){$U$}
	\node(T)(25,-20){$T$}
	\drawedge(S,B){$\pi$}
	\drawedge(B,T){$\beta$}
	\drawedge(S,T){$\tau$}
	\drawedge[ELside=r](S,U){$\phi$}
	\drawedge(U,T){$$}
\end{picture}
\caption{The morphisms $\phi$, $\pi$ and $\beta$}
\label{diagram beta}
\end{figure}
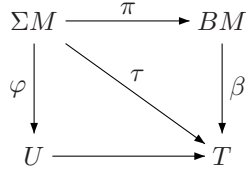

Let $(S,A)$ and $(T,B)$ be \pgs\ and let $\alpha\colon AM \rightarrow
S$ and $\beta\colon BM \rightarrow T$ be the natural morphisms, so
that $\alpha(a) = a$ and $\beta(b) = b$ for all $a\in A$ and $b\in B$.
Let $(U,\Sigma) = (S,A) \block_k (T,B)$, and let $\phi\colon \Sigma M
\rightarrow U \subseteq S \block_k T$ be the natural morphism.  By
definition, each $\sigma \in \Sigma_n$ ($n \ge 0$) is a pair $\sigma =
(F_\sigma,b_\sigma)$ with $b_\sigma \in B_n$ and $F_\sigma \in
A_n^{I_{k,n}}$.  Let $\pi\colon \Sigma M \rightarrow BM$ be the
morphism induced by the second component projection from $\Sigma$ to
$B$, and let $\tau = \beta\circ\pi\colon \Sigma M \rightarrow T$, see
Figure~\ref{diagram beta}.  We now describe a way of computing
$\phi(t)$ for a tree $t \in \Sigma M_n$, say $\phi(t) = (Q_t,
\tau(t))$.

Let $D = (u,k_1,v,k_2) \in I_{k,n}$.  We define the tree $\bar t_D$ by
relabeling the nodes of $t$ in $\NV(t)$ with elements of $A$ as
follows.  Let $x$ be a node of $t$ of rank $m\ge 0$, let $g$ be the
subtree of $t$ whose root is $x$, and let $h_1,\ldots,h_m$ be the
subtrees whose roots are the children of $x$, see Figure~\ref{fig
fact32}.  Let $h = h_1 \oplus \cdots \oplus h_m$, let $\sigma \in
\Sigma_m$ be the label of $x$ in $t$ and let $r_2 \ge 0$ be such that
$g = \sigma \cdot h \in \Sigma M_{r_2}$ and $h \in \Sigma M_{m,r_2}$.
\begin{figure}[ht]
    \centering
    \begin{picture}(110,80)(0,-80)
	\drawline[AHnb=0](15,0)(0,-30)
	\drawline[AHnb=0](15,0)(30,-30)
	\drawline[AHnb=0](0,-30)(9,-30)
	\drawline[AHnb=0](9,-30)(15,-18)
	\drawline[AHnb=0](15,-18)(21,-30)
	\drawline[AHnb=0](21,-30)(30,-30)
	\node[Nfill=y,fillcolor=Black,Nw=1.0,Nh=1.0,Nmr=1.0](n0)(15.0,-18){}
	\drawline[AHnb=0](15,-21)(8,-35)
	\drawline[AHnb=0](15,-21)(22,-35)
	\drawline[AHnb=0](8,-35)(22,-35)
	\put(14,-12){$f$}
	\put(2,-33){$r_1$}
	\put(25,-33){$r_3$}
	\put(17,-18){$x$}
	\put(13,-38){$r_2$}
	\put(14,-31){$g$}
%
%
%
	\put(54,1){$\sigma$}
	\drawline[AHnb=0](55,-2)(45,-10)
	\drawline[AHnb=0](55,-2)(65,-10)
	\drawline[AHnb=0](45,-10)(40,-25)
	\drawline[AHnb=0](45,-10)(50,-25)
	\drawline[AHnb=0](40,-25)(50,-25)
	\put(43.5,-21){$h_1$}
	\put(53,-21){$\cdots$}
	\drawline[AHnb=0](65,-10)(60,-25)
	\drawline[AHnb=0](65,-10)(70,-25)
	\drawline[AHnb=0](60,-25)(70,-25)
	\put(63,-21){$h_m$}
%
%
%
%
	\drawline[AHnb=0](95.0,0.0)(80,-30)
	\drawline[AHnb=0](95,0)(110,-30)
	\drawline[AHnb=0](80,-30)(110,-30)
	\node[Nfill=y,fillcolor=Black,Nw=1.0,Nh=1.0,Nmr=1.0](n2)(95.0,-30){}
	\put(94,-17){$u$}
	\put(87,-33){$k_1$}
	\put(101,-33){$k_2$}
	\drawline[AHnb=0](95,-30)(85,-45)
	\drawline[AHnb=0](95,-30)(105,-45)
	\drawline[AHnb=0](85,-45)(105,-45)
	\node[Nfill=y,fillcolor=Black,Nw=1.0,Nh=1.0,Nmr=1.0](n5)(95,-45){}
	\put(92,-39){$\tau(f)$}
	\put(89,-43.5){$r_1$}
	\put(99,-43.5){$r_3$}
	\drawline[AHnb=0](85,-46)(94,-46)
	\drawline[AHnb=0](79,-55)(91,-55)
	\drawline[AHnb=0](85,-46)(79,-55)
	\drawline[AHnb=0](94,-46)(91,-55)
	\put(85,-52){$\bar v_1$}
	\put(84,-58){$p_1$}
	\drawline[AHnb=0](105,-46)(96,-46)
	\drawline[AHnb=0](111,-55)(99,-55)
	\drawline[AHnb=0](105,-46)(111,-55)
	\drawline[AHnb=0](96,-46)(99,-55)
	\put(102,-52){$\bar v_3$}
	\put(103,-58){$p_3$}
	\drawline[dash={1.0 1.0}{0.0},AHnb=0](95,-45)(90,-60)
	\drawline[dash={1.0 1.0}{0.0},AHnb=0](95,-45)(100,-60)
	\drawline[dash={1.0 1.0}{0.0},AHnb=0](90,-60)(100,-60)
	\drawline[AHnb=0](90,-61)(100,-61)
	\drawline[AHnb=0](89,-70)(101,-70)
	\drawline[AHnb=0](90,-61)(89,-70)
	\drawline[AHnb=0](100,-61)(101,-70)
	\put(92,-66.5){$\tau(h)$}
	\drawline[AHnb=0](89,-71)(101,-71)
	\drawline[AHnb=0](88,-80)(102,-80)
	\drawline[AHnb=0](89,-71)(88,-80)
	\drawline[AHnb=0](101,-71)(102,-80)
	\put(94,-76.5){$\bar v_2$}
\end{picture}
\caption{The trees $t$ and $g = \sigma\cdot h$, and the context $C$}
\label{fig fact32}
\end{figure}
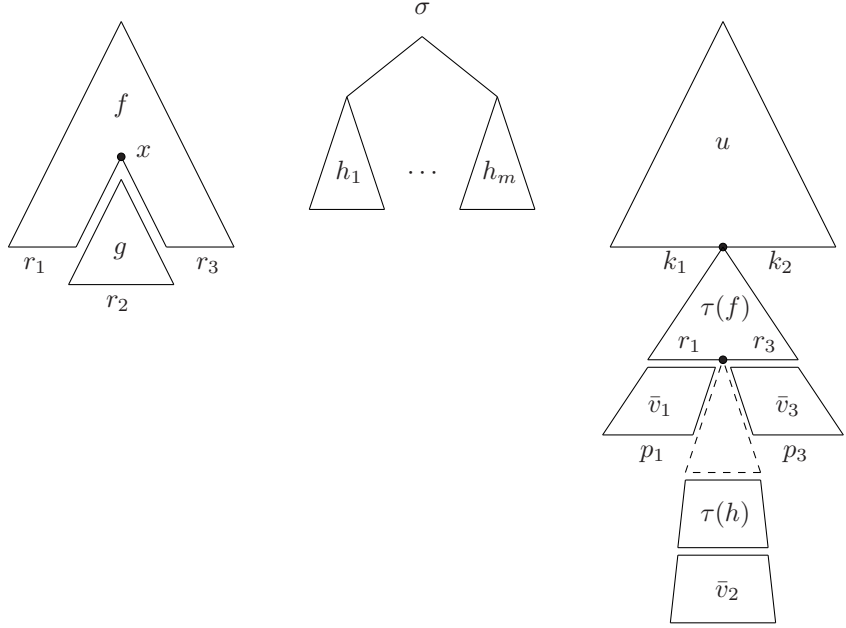 

Let us write $t = f\cdot (\r_1 \oplus g \oplus \r_3)$, where $r_1$ and
$r_3$ are integers such that the node $x$ is now labeled by a variable
in $f$ (that is, $\leftf_{r_1}(x)$ and $\rightf_{n - r_3+1}(x)$ in
$t$, and $n = r_1 + r_2 + r_3$).  Let $\bar v_1$ be the $\oplus$-sum
of the first $r_1$ $v_j$'s, $\bar v_2$ the $\oplus$-sum of the next
$r_2$ $v_j$'s and $\bar v_3$ the $\oplus$-sum of the last $r_3$
$v_j$'s.  Then we have $\bar v_1 \in T_{r_1,p_1}$, $\bar v_2 \in
T_{r_2,p_2}$ and $\bar v_3 \in T_{r_3,p_3}$ for some $p_1, p_2, p_3
\ge 0$ (and $k = k_1 + p_1 + p_2 + p_3 + k_2$).  Let then $c = u \cdot
(\k_1 \oplus \tau(f)\cdot (\bar v_1 \oplus \1 \oplus \bar v_3) \oplus
\k_2)$, so that $C = (c, k_1 + p_1, \tau(h)\cdot \bar v_2, p_3 + k_2)
\in I_{k,m}$, see Figure~\ref{fig fact32}.  We finally label the node
$x$ in $\bar t_D$ by $F_\sigma(C)$.

The resulting tree $\bar t_D$ is an element of $AM_n$.  We now show the
following fact.

\begin{fact} \label{prop-tree}
    With the notation above, $\phi(t) = (Q_t, \tau(t))$ where $Q_t(D) =
    \alpha(\bar t_D)$ for each context $D \in I_{k,n}$.
\end{fact} 
 
\proof 
The proof is by structural induction on $t$.  If $t = \1$, then $\bar
t_D = \1$ for each $D$, and $\phi(t) = (F_\1,\1)$, so the announced
result holds.

If $t$ consists of a single node, then $t = \sigma\in \Sigma_0$ and
$\phi(t) = (F_\sigma, b_\sigma)$.  Now let $D = (u,k_1,\0,k_2) \in
I_{k,0}$.  With the notation above, we have $g = \sigma = t$, $h =
\0$, $f = \1$, and $p_i = r_j = 0$.  In particular, $c = u$.  It
follows that $C = D$ and $\bar t_D = F_\sigma(D)$.  Moreover, since
$F_\sigma(D) \in A$, we have $\alpha(\bar t_D) = F_\sigma(D)$.  This
concludes the verification of the equality for one-node trees.

Let us now assume that $t \in \Sigma M_n$ ($n\ge 0$) has more than 
one node, let $\sigma \in
\Sigma_m$ be the label of the root of $t$, and let $s^{(1)},\ldots,
s^{(m)}$ be the subtrees of $t$ attached to the children of the root.
Let also $s = s^{(1)} \oplus \cdots \oplus s^{(m)}$, so that $t =
\sigma \cdot s$.  Let $D = (u,k_1,w,k_2) \in I_{k,n}$.  By induction,
we have
$$Q_t(D) = F_\sigma(u, k_1, \tau(s)\cdot w, k_2) \cdot
\Big(\alpha(\bar{s}^{(1)}_{C_1}) \oplus \cdots \oplus
\alpha(\bar{s}^{(m)}_{C_m})\Big),$$
where $w = \bar w_1 \oplus \cdots \oplus \bar w_n$, $C_i =
(c_i,q_1,\bar w_i,q_2)$,
$$c_i = u \cdot (\k_1 \oplus \tau(\sigma \cdot (s^{(1)} \oplus \cdots
\oplus s^{(i-1)} \oplus \1 \oplus s^{(i+1)} \oplus \cdots \oplus s^{(m)}))
\oplus \k_2)$$
and $q_1$ and $q_2$ are appropriate integers (which depend on $i$)
such that
$$u\cdot (\k_1 \oplus \tau(t) \cdot w \oplus \k_2) = c_i
\cdot (\q_1 \oplus \tau(s^{(i)})\cdot \bar w_i \oplus \q_2).$$
We compare this value with $\alpha(\bar t_D)$.  If $a$ is the label of
the root of $\bar t_D$ and if $d_1,\ldots, d_m$ are the
subtrees of $\bar t_D$ attached to the children of the root, then
$\alpha(\bar t_D) = \alpha(a) \cdot \bigoplus_i \alpha(d_i)$.  We
first discuss the value of $a$.  With reference to the notation in the
definition of the labels of $\bar t_D$ above, since $t = \sigma\cdot
s$, the integers $p_1, p_3, r_1, r_3$ are all equal to $0$ and
$\bar v_1 = \bar v_3 = \0$.  In particular, $a = 
F_\sigma(u,k_1,\tau(s) \cdot w, k_2)$.  Thus $\alpha(a) =
\alpha(F_\sigma(u,k_1,\tau(s)\cdot w,k_2)) =
F_\sigma(u,k_1,\tau(s)\cdot w,k_2)$.

To conclude, we need only to verify that $d_i = \bar{s}^{(i)}_{C_i}$
for each $i\in [m]$, that is, each node $x$ of $t$ in $\NV(s^{(i)})$, has
the same label in $d_i$ and in $\bar{s}^{(i)}_{C_i}$. But it is easy 
to see that the label of $x$ in both $\bar t_D$ and 
$\bar{s}^{(i)}_{C_i}$ is of the form $F_\rho(C)$ where $\rho\in 
\Sigma$ is the label of $x$ in $s^{(i)}$ and $C$ is appropriate.
%
%
\eop

\subsection{Closed pseudovarieties}

We say that a pseudovariety $\V$ of preclones is \textit{closed} if
every block product $S \block_k T$ with $S, T\in \V$ and $k\ge 0$
belongs to \V. Closed pseudovarieties of \pgs\ are defined similarly. 
Since the intersection of a family of closed pseudovarieties is 
closed, there exists a least closed pseudovariety containing any 
given class $\K$ of finitary preclones (resp. finitary \pgs).

We now give a technical result on closed pseudovarieties, that will be
used in the proof of our main result.  We consider the situation where
$S,T,T'$ are preclones and $T$ is a sub-preclone of $T'$.  Then the
elements of $S \block_k T'$ whose second component belongs to $T$,
form a sub-preclone of $S \block_k T'$ which we denote by
$S\block_k^{T'} T$.

\begin{prop}\label{prop technique}
    Let $\V$ be a closed pseudovariety of preclones.  Let $S,T \in \V$
    and let $T'$ be a finitary preclone such that $T$ is a
    sub-preclone of $T'$.  For each $k \geq 0$, the product $S
    \block_k^{T'} T$ belongs to $\V$.
\end{prop} 

Before we prove Proposition~\ref{prop technique}, we verify a
technical lemma.  In this lemma, we use the notation in the
proposition.  In particular, we need to consider contexts in both
$T_k$ and $T'_k$.  We denote by $I_{k,n}$ (resp.  $I'_{k,n}$) the set
of $n$-ary contexts in $T_k$ (resp.  $T'_k$).

\begin{lem}\label{lem-gen-block}
    Let $S$, $T$, $T'$ and $k$ be as in Proposition~\ref{prop
    technique} and let $C \in I'_{k,n}$.  There exists a morphism
    $\alpha^C \colon S \block_k^{T'} T \rightarrow S \block_n T$ such
    that, if $(F,f) \in (S \block_k^{T'} T)_n$, then $\alpha^C(F,f)$
    is of the form $\alpha^C(F,f) = (F^C,f)$ with $F^C(\1,0,\n,0) =
    F(C)$.
\end{lem} 

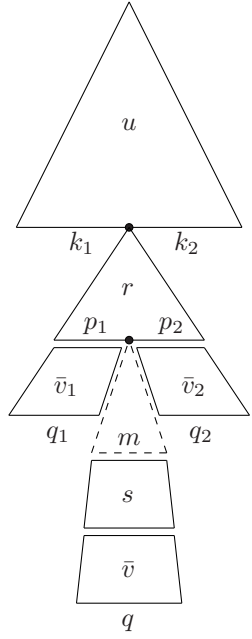
\begin{figure}[ht]
    \centering
    \begin{picture}(30,82)(0,-82)
	\drawline[AHnb=0](15.0,0.0)(0,-30)
	\drawline[AHnb=0](15,0)(30,-30)
	\drawline[AHnb=0](0,-30)(30,-30)
	\node[Nfill=y,fillcolor=Black,Nw=1.0,Nh=1.0,Nmr=1.0](n2)(15.0,-30){}
	\put(14,-17){$u$}
	\put(7,-33){$k_1$}
	\put(21,-33){$k_2$}
	\drawline[AHnb=0](15,-30)(5,-45)
	\drawline[AHnb=0](15,-30)(25,-45)
	\drawline[AHnb=0](5,-45)(25,-45)
	\node[Nfill=y,fillcolor=Black,Nw=1.0,Nh=1.0,Nmr=1.0](n5)(15,-45){}
	\put(14,-39){$r$}
	\put(9,-43.5){$p_1$}
	\put(19,-43.5){$p_2$}
	\drawline[AHnb=0](5,-46)(14,-46)
	\drawline[AHnb=0](-1,-55)(11,-55)
	\drawline[AHnb=0](5,-46)(-1,-55)
	\drawline[AHnb=0](14,-46)(11,-55)
	\put(5,-52){$\bar v_1$}
	\put(4,-58){$q_1$}
	\drawline[AHnb=0](25,-46)(16,-46)
	\drawline[AHnb=0](31,-55)(19,-55)
	\drawline[AHnb=0](25,-46)(31,-55)
	\drawline[AHnb=0](16,-46)(19,-55)
	\put(22,-52){$\bar v_2$}
	\put(23,-58){$q_2$}
	\drawline[dash={1.0 1.0}{0.0},AHnb=0](15,-45)(10,-60)
	\drawline[dash={1.0 1.0}{0.0},AHnb=0](15,-45)(20,-60)
	\drawline[dash={1.0 1.0}{0.0},AHnb=0](10,-60)(20,-60)
	\put(13.5,-59){$m$}
	\drawline[AHnb=0](10,-61)(20,-61)
	\drawline[AHnb=0](9,-70)(21,-70)
	\drawline[AHnb=0](10,-61)(9,-70)
	\drawline[AHnb=0](20,-61)(21,-70)
	\put(14,-66.5){$s$}
	\drawline[AHnb=0](9,-71)(21,-71)
	\drawline[AHnb=0](8,-80)(22,-80)
	\drawline[AHnb=0](9,-71)(8,-80)
	\drawline[AHnb=0](21,-71)(22,-80)
	\put(14,-76.5){$\bar v$}
	\put(14,-83){$q$}
\end{picture}
\caption{$F^C(D)$ is the image by $F$ of the context represented here}
\label{fig lemma34}
\end{figure}

\proof
Let $C = (u,k_1,v,k_2) \in I'_{k,n}$ and let $(F,f) \in (S
\block_k^{T'} T)_m$, $m \ge 0$.  We first define a mapping $F^C\colon
I_{n,m} \rightarrow S_m$.  Let $D = (r,p_1,s,p_2) \in I_{n,m}$.  By
definition, $p_1 + p_2 \le n$, and we let $\bar v_1$ be the
$\oplus$-sum of the first $p_1$ $v_i$'s, $\bar v_2$ be the
$\oplus$-sum of the last $p_2$ $v_i$'s, and $\bar v$ be the
$\oplus$-sum of the middle $n-p_1-p_2$ $v_i$'s.  In particular, there
exist integers $q_1,q,q_2$ such that $\bar v_1 \in T_{p_1,q_1}$, $\bar
v \in T_{n-p_1-p_2,q}$, $\bar v_2 \in T_{p_2,q_2}$ and $k_1 + p_1 +
q_1 + q + q_2 + p_2 + k_2 = k$, see Figure~\ref{fig lemma34}.  We let
$$F^C(D) = F\big(u\cdot (\k_1 \oplus r\cdot (\bar v_1 \oplus \1 \oplus
\bar v_2) \oplus \k_2), k_1+q_1, s\cdot \bar v, q_2+k_2\big).$$
The verification that $F^C(\1,0,\n,0) = F(C)$ is straightforward, and
we need to show that $\alpha^C \colon (F,f) \mapsto (F^C,f)$ defines a
morphism of preclones.

Let $(F,f) \in (S \block_k^{T'} T)_m$ and let $(G_i,g_i) \in
(S\block_k^{T'} T)_{h_i}$ ($i \in [m]$).  For convenience, we let
$g$ be the $\oplus$-sum of the $g_i$, so $g \in T_{m,h}$ with $h =
\sum_i h_i$.  Let $(Q,f\cdot g) = (F,f) \cdot \bigoplus_i (G_i,g_i)$,
so that $\alpha^C((F,f) \cdot \bigoplus_i (G_i,g_i)) = (Q^C, f\cdot
g)$.  Moreover, let $(R,f\cdot g) = (F^C,f) \cdot \bigoplus_i
(G_i^C,g_i)$.  We need to verify that $Q^C = R$.

Let $D = (r,p_1,s,p_2) \in I_{h,n}$.  For each $i \in [m]$, let
$\bar s_i$ be the appropriate $\oplus$-sum of $s_j$'s such that $g
\cdot s = g_1 \cdot \bar s_1 \oplus \cdots \oplus g_m \cdot \bar s_m$.
Then $R(D) = F^C(r,p_1,g\cdot s,p_2) \cdot \bigoplus_i G_i^C(D_i)$,
where $D_i = (r_i,\ell_1,\bar s_i, \ell_2)$, $r_i$ is
$$r\cdot (\p_1 \oplus f \cdot (g_1 \cdot \bar{s}_1 \oplus \cdots
\oplus g_{i-1}\cdot \bar s_{i-1} \oplus \1 \oplus g_{i+1}\cdot \bar
s_{i+1} \oplus \cdots \oplus g_m \cdot \bar{s}_m) \oplus \p_2),$$
and $\ell_1, \ell_2$ are such that $r\cdot (\p_1 \oplus f\cdot g \cdot
s \oplus \p_2) = r_i \cdot (\ell_1 \oplus g_i\cdot\bar s_i \oplus
\ell_2)$.

Thus $R(D)$ is the composition of
$$F\big(u\cdot(\k_1 \oplus r \cdot (\bar v_1 \oplus \1 \oplus 
\bar v_2) \oplus \k_2), k_1 + p_1, g\cdot s\cdot \bar v, 
p_2+k_2\big)$$
with
$$\bigoplus_{i=1}^m G_i\big(u\cdot(\k_1 \oplus r_i \cdot (\bar v^i_1
\oplus \1 \oplus \bar v^i_2) \oplus \k_2), k_1 + \ell_1, g\cdot s\cdot
\bar v^i, \ell_2 + k_2\big),$$
where $v = \bar v_1^i \oplus \bar v^i \oplus \bar v_2^i$ is the
appropriate grouping.
Observe that
$$u\cdot(\k_1 \oplus r_i \cdot (\bar v^i_1 \oplus \1
\oplus \bar v^i_2) \oplus \k_2)$$
is equal to
$$u \cdot (\k_1 \oplus \p_1 \oplus f \cdot (g_1 \cdot \bar{s}_1 \oplus
\cdots \oplus g_{i-1} \cdot \bar{s}_{i-1} \oplus \1 \oplus g_{i+1}
\cdot \bar{s}_{i+1}\oplus \cdots \oplus g_m \cdot \bar{s}_m)
\oplus \p_2 \oplus \k_2).$$

Now, $Q^C(D) = Q(u \cdot (\k_1 \oplus r \cdot (\bar v_1 \oplus \1
\oplus \bar v_2) \oplus \k_2), k_1 + p_1, s\cdot \bar v, p_2 + k_2)$. 
By definition of the block product, this is equal to the composition 
of
$$F\big(u \cdot (\k_1 \oplus r \cdot (\bar v_1 \oplus \1
\oplus \bar v_2) \oplus \k_2), k_1 + p_1, g\cdot s\cdot \bar v, p_2 + 
k_2\big)$$
with
$$\bigoplus_{i=1}^m G_i(r'_i,z_1,\bar s_i\cdot \bar v'_i,z_2),$$
where $\bar v = \bigoplus_i \bar v'_i$ is the appropriate grouping,
\begin{eqnarray*}
r'_i & = & u \cdot (\k_1 \oplus r \cdot (\bar{v}_1 \oplus f \cdot (g_1
\cdot \bar{s}_1 \cdot \bar{\bar{v}}_1 \oplus \cdots \oplus g_{i-1}
\cdot \bar{s}_{i-1} \cdot \bar{\bar{v}}_{i-1} \oplus \1\\
&& \oplus g_{i+1} \cdot \bar{s}_{i+1} \cdot \bar{\bar{v}}_{i+1} \oplus
\cdots \oplus g_m \cdot \bar{s}_m \cdot \bar{\bar{v}}_m) \oplus
\bar{v}_2) \oplus \k_2)
\end{eqnarray*}
where $v = \bar{\bar{v}}_1 \oplus \cdots  \oplus \bar{\bar{v}}_m$ and 
$z_1$ and $z_2$ are appropriate integers.

It is now a straightforward verification that $R(D) = Q^C(D)$, which
concludes the proof.
\eop

\rm \trivlist \item[\hskip \labelsep{\bf Proof of
Proposition~\ref{prop technique}.}]
By Proposition~\ref{423 TCS}, it suffices to verify that distinct
elements of equal rank in $S \block_k^{T'} T$ can be separated by a
morphism into an element of \V.

So we consider $(F_1,f_1)$ and $(F_2,f_2)$, rank $n$ elements of $S
\block_k^{T'} T$.  If $f_1 \ne f_2$, the second component projection
is a morphism into $T \in \V$ which separates the two elements.  If
$f_1 = f_2$, then $F_1 \ne F_2$ and we let $C \in I'_{k,n}$ such that
$F_1(C) \ne F_2(C)$.  Then the morphism $\alpha^C$ in
Lemma~\ref{lem-gen-block} is a morphism into $S \block_n T \in \V$
which separates the two elements.
\cqfd

If $(S,A)$ and $(T',B')$ are \pgs\ and $(T,B)$ is a sub-\pgp\ of
$(T',B')$, recall that the block product $(S,A) \block_k (T',B')$ is
generated by a ranked alphabet $\Sigma'$ such that $\Sigma'_m =
A_m^{I'_{k,m}} \times B'_m$.  We let $(S,A) \block_k^{(T',B')} (T,B)$
be the sub-\pgp\ of $(S,A) \block_k (T',B')$ generated by the subset
$\Sigma$ of $\Sigma'$ such that $\Sigma_m = A_m^{I'_{k,m}} \times B_m$
for each $m$.

\begin{prop}\label{prop technique pg}
    Let $\V$ be a closed pseudovariety of \pgs.  Let $(S,A)$ and
    $(T,B)$ be \pgs\ in \V\ and let $(T',B')$ be a finitary \pgp\ such
    that $(T,B)$ is a sub-\pgp\ of $(T',B')$.  For each $k \geq 0$,
    the product $(S,A) \block_k^{(T',B')} (T,B)$ belongs to $\V$.
\end{prop} 

\proof
We note that the morphism $\alpha^C$ in the proof of
Lemma~\ref{lem-gen-block}, maps each generator of $(S,A)
\block_k^{(T',B')} (T,B)$ to a generator of $(S,A) \block_n (T,B)$, so
$\alpha^C$ is also a morphism of \pgs\ between these block products.

The same scheme as in the proof of Proposition~\ref{prop technique}
can then be applied, using Proposition~\ref{424 TCS} instead of
Proposition~\ref{423 TCS}.
\eop

We conclude with a result on full pseudovarieties (see
Section~\ref{sec var}).

\begin{prop}\label{prop full closure}
    Let \W\ be a pseudovariety of preclones, let $\V = \pgpairs(\W)$
    and let $\widehat\W$ and $\widehat\V$ be the closure of \W\ and
    \V\ respectively.  Then \V\ and $\widehat\V$ are full and
    $\widehat\V = \pgpairs(\widehat\W)$.
\end{prop}

\proof
In view of Proposition~\ref{psv pgs or preclones}, it suffices to show
that $\widehat\V = \pgpairs(\widehat\W)$.  We first verify that if
$(S,A) \in \widehat\V$, then $S\in \widehat\W$, by induction on the
construction of $(S,A)$ from elements of \V\ by means of block
products.  If $(S,A)\in\V$, then $S\in\W$ by definition and hence,
$S\in\widehat\W$.  Now suppose that $(S,A) < (S^{(1)},A^{(1)})
\block_k (S^{(2)},A^{(2)})$ and $S^{(1)},S^{(2)} \in \widehat\W$.  By
definition of the block product of \pgs, $S < S^{(1)} \block_k
S^{(2)}$ and hence $S \in \widehat\W$.

Now we show that if $(S,A)$ is a finitary \pgp\ with $S\in\widehat\W$,
then $(S,A)\in\widehat\V$.  The proof is by induction on the
construction of $S$ from elements of \W\ by means of block products.
If $S \in \W$, then $(S,A) \in \pgpairs(\W) = \V$ by definition.  Now
suppose that $S = S^{(1)} \block_k S^{(2)}$ and $\pgpairs(S^{(i)})
\subseteq \widehat\V$.  Let $B^{(2)}$ be the projection of $A$ onto
$S^{(2)}$.  Each element of $A$ is of the form $(F,b)$ with $b\in
B^{(2)}$.  Let $B^{(1)}$ be the union of the ranges of the first
components of elements of $A$.  Then, if $T^{(i)}$ is the sub-preclone
of $S^{(i)}$ generated by $B^{(i)}$, we have $(S,A) \subseteq
(T^{(1)}, A^{(1)}) \block_k (T^{(2)}, A^{(2)})$.  It follows that
$(S,A) \in \widehat\V$.
\eop

\subsection{Characterizing $\Linlg(\calK)$}

Our main result is:

\begin{thm}\label{thm-main} 
    Let $\cK$ be a class of recognizable tree languages such that each
    quotient of a language in $\cK$ belongs to $\Linlg(\cK)$ and such
    that $\Lin(\cK)$ admits relativization.  Then a language is in
    $\Linlg(\cK)$ if and only if its syntactic \pgp\ belongs to the
    least closed pseudovariety of \pgs\ containing the syntactic \pgs\
    of the languages in $\cK$.
\end{thm}

The proof of Theorem~\ref{thm-main} is based on
Propositions~\ref{prop-main1} and \ref{prop-main2} below.
Applications are considered in the next section.

\begin{prop}\label{prop-main1}
    Let $\cK$ be a class of recognizable tree languages such that
    $\Lin(\cK)$ admits relativization and let $(S,A)$ and $(T,B)$ be
    \pgs\ such that every language recognizable by $(S,A)$ or $(T,B)$
    belongs to $\Linlg(\cK)$.  Then every language recognizable by a
    block product of $(S,A)$ and $(T,B)$ also belongs to
    $\Linlg(\cK)$.
\end{prop}

\proof
Let $k\ge 0$, $(U,\Sigma) = (S,A) \block_k (T,B)$, and $\phi\colon
(\Sigma M,\Sigma) \rightarrow (U,\Sigma)$ be the morphism induced by
the identity map of $\Sigma$.  Let $L$ be a tree language recognized
by a morphism $\phi'\colon (\Sigma' M,\Sigma') \rightarrow
(U,\Sigma)$.  Since $\phi$ is onto and $(\Sigma' M,\Sigma')$ is free,
there exists a morphism $\psi\colon (\Sigma' M,\Sigma') \rightarrow
(\Sigma M,\Sigma)$ such that $\phi' = \phi\circ\psi$.  In particular,
$L = {\phi'}\inv(\phi'(L)) = \psi\inv(\phi\inv(\phi'(L)))$.  In view
of Theorem~\ref{thm-literalvar}, it suffices to show that every
language recognized by $\phi$ lies in $\Linlg(\cK)$.  This in turn
reduces to showing that $\phi\inv(F,g) \in \Linlg(\cK)$ for each
$(F,g) \in U$.

We use the information obtained in Fact~\ref{prop-tree} on the
computation of $\phi(t)$.  As in that statement, we let $\alpha\colon
AM \rightarrow S$ and $\beta\colon BM \rightarrow T$ be the natural
morphisms, we let $\pi\colon \Sigma M \rightarrow BM$ be the morphism
induced by the second coordinate projection from $\Sigma$ to $B$, and
we let $\tau = \beta \circ \pi \colon \Sigma M \rightarrow T$.  For
each $\sigma \in \Sigma$, we let $\phi(\sigma) = (F_\sigma,
b_\sigma)$.

Let $(F,g) \in U_n$.  For each $t\in \Sigma M_n$, we have $\phi(t) =
(Q_t, \tau(t))$, where $Q_t \in S_n^{I_{n,k}}$ is described in
Fact~\ref{prop-tree}.  We note that $\tau\inv(g)$ is recognized by
$(T,B)$, and hence is in $\Linlg(\cK)$.  We denote by $\chi_g$ a rank
$n$ $\Lin(\cK)$-sentence defining $\tau\inv(g)$.

Recall (from Fact~\ref{prop-tree}) that $Q_t(D) = \alpha(\bar t_D)$ 
for all $D \in I_{k,n}$.
For each $s\in S$, the tree language $K_s = \alpha\inv(s)$ is
recognized by $(S,A)$, and hence $K_s \in \Linlg(\cK)$. It now
suffices to show that, for each $s \in S_n$ and $D \in I_{k,n}$,
there exists a rank $n$ $\Lin(\cK)$-sentence $\psi_{s,D}$ defining
the language
$$\{ t \in \Sigma M_n \mid \bar t_D \in K_s \}.$$
Indeed, since $I_{k,n}$ is finite, it will follow that $\phi\inv(F,g)$
is defined by the conjunction of $\chi_g$ and the $\psi_{s,D}$ ($D \in
I_{k,n}$ and $F(D) = s$).  We construct the sentence $\psi_{s,D}$ in
the form $\psi_{s,D} = Q_{K_s} z \cdot \langle \psi_a \rangle_{a\in
A}$ (where $\psi_a$ is a rank $n$ $\Lin(\cK)$-formula on $\Sigma$
depending on $a$ and $D$).  (The formula $\psi_{s,D}$ is actually a
$\Lin(\Linlg(\cK))$-formula but this is sufficient for our purpose in
view of Theorem~\ref{thm-closure}~(3).)

Let $a\in A_m$ and $D = (u,k_1,v,k_2) \in I_{k,n}$.  For each $0 \le i
< j \le n+1$, let $\bar v_1 = \oplus_{q=1}^i v_q$, $\bar v_2 =
\oplus_{q = i+1}^{j-1} v_q$ and $\bar v_3 = \oplus_{q = j}^n v_q$;
and let $p_1$, $p_2$, $p_3$ be the ranks of $\bar v_1$, $\bar v_2$ and
$\bar v_3$ respectively.  For each such $i,j$, for $m\ge 0$, $\sigma
\in \Sigma_m$, $c\in T_{k_1+p_1+1+p_3+k_2}$, $c_1\oplus\cdots \oplus
c_m \in T_{m,p_2}$, we let
$$\psi' = P_\sigma(z) \land \leftf_i(z) \land \rightf_j(z) \land
\chi_c[\not>z] \land \bigwedge_{\ell \in [m]}\chi_{c_\ell}[\ge x\ell]$$
and we let $\psi_a(z)$ be the (finite) disjunction of the $\psi'(z)$
when
$$F_\sigma(u \cdot (\k_1 \oplus c \cdot( \bar{v}_1 \oplus \1 \oplus
\bar{v}_3) \oplus \k_2), k_1 + p_1, (c_1 \oplus \cdots \oplus c_m)
\cdot \bar{v}_2, p_3 + k_2) = a.$$
It is elementary to verify that $(t, z \mapsto x) \models \psi_a$ if
and only if node $x$ is labeled $a$ in $\bar t_D$.  Thus $t$ satisfies
$\psi_{s,D}$ if and only if $\bar t_D \in K_s$, if and only if $Q(t) =
\alpha(\bar t_D) \in \alpha(K_s) = \{s\}$, which concludes the proof.
\eop

\begin{prop}\label{prop-main2}
    Let $Y$ be a set of first order variables and let $y \not\in Y$. 
    Let $\langle \phi_\delta \rangle_{\delta \in \Delta}$ be a family
    of rank $k$ $\Lin$-formulas over $\Sigma$, with free variables in
    $Y \cup \{y\}$, deterministic with respect to $y$, let $K
    \subseteq \Delta M_k$ be a tree language and let $\phi = Q_K
    y\cdot \langle \phi_\delta \rangle_{\delta \in \Delta}$.
    
    Let $(S,A)$ be a \pgp\ recognizing $K$, and let $(T,B)$ be a \pgp\
    recognizing simultaneously the languages $L_{\phi_\delta}\subseteq
    \Sigma_{Y\cup\{y\}} M_k$ ($\delta\in \Delta$).  Then the language
    $L_\phi$ (a subset of $\Sigma_Y M_k$) is recognized by
    $(S,A)\block_k (T,B)$.
\end{prop}

\proof
Let $\kappa\colon (\Delta M, \Delta) \rightarrow (S,A)$ be an onto
morphism recognizing $K$, and let $\tau\colon (\Sigma_{Y\cup\{y\}}
M,\Sigma_{Y\cup\{y\}}) \rightarrow (T,B)$ be an onto morphism
recognizing each of the $L_{\phi_\delta}$ ($\delta \in \Delta$).

We observe the following: if $t\in \Sigma_{Y\cup\{y\}} M_k$ is a
$(Y\cup\{y\})$-structure and $y$ occurs in the label of a rank $n$
node, then there exists a unique $\delta\in \Delta_n$ such that
$\tau(t) \in \tau(L_{\phi_\delta})$.  Indeed, the determinism of
$\langle\phi_\delta\rangle_\delta$ with respect to $y$ shows that $t$
lies in exactly one of the $L_{\phi_\delta}$ ($\delta \in \Delta_n)$:
by hypothesis, $\tau\inv\big(\tau(L_{\phi_\epsilon})\big) =
L_{\phi_\epsilon}$ for each $\epsilon$, so $\tau(t) \in
\tau(L_{\phi_\epsilon})$ and $\epsilon \in \Delta_n$ implies $\epsilon
= \delta$.

We consider the block product of \pgs\ $(S,A) \block_k (T,B)$.  For
each $\sigma\in \Sigma_n$ and $Z \subseteq Y$ (so that $(\sigma,Z) \in
(\Sigma_Y)_n$), we let $\gamma(\sigma,Z) = (F_{\sigma, Z},
\tau(\sigma,Z))$, where $F_{\sigma, Z}$ is defined as follows.  Let
$(u,k_1,v,k_2) \in I_{k,n}$.  If $u \cdot (\k_1 \oplus
\tau(\sigma,Z\cup\{y\}) \cdot v \oplus \k_2)$ is the $\tau$-image of
some $(Y\cup\{y\})$-structure, then we let $F_{\sigma, Z}(u,k_1,v,k_2)
= \kappa(\delta)$ where $\delta\in \Delta_n$ is uniquely determined by
the property $\tau(u \cdot (\k_1 \oplus (\sigma,Z\cup\{y\}) \cdot
v \oplus \k_2)) \in \tau(L_{\phi_\delta})$.  Otherwise, we choose
$F_{\sigma, Z}(u,k_1,v,k_2)$ arbitrarily in $A_n$.  Note that
$\gamma(\sigma,Z)$ lies in the generator set of the \pgp\ $(S,A)
\block_k (T,B)$.

Let now $t \in \Sigma_Y M_k$ and let $D = (\1,0,\k,0) \in I_{k,k}$.
Then $\gamma(t) = (Q, \tau(t))$, and $Q(D) = \alpha(\bar t_D)$, where
$\alpha\colon AM \rightarrow S$ is the natural morphism and $\bar t_D$
is described in Fact~\ref{prop-tree}.  In particular, let $x \in
\NV(t)$ be a rank $n$ node, labeled by $(\sigma,Z)$ in $t$ and let
$t$ factor as $t = f \cdot (\r_1 \oplus (\sigma,Z) \cdot h \oplus
\r_3)$ where $f$, $r_1$ and $r_3$ are such that $(\sigma,Z)\cdot h$ is
the subtree of $t$ rooted at node $x$.  The label of $x$ in $\bar
t_D$ is equal to $F_{\sigma,Z}(\tau(f),r_1,\tau(h),r_2)$, for the
computation of which we need to consider the tree $\tau(f\cdot (\r_1
\oplus (\sigma, Z\cup\{y\}) \cdot h \oplus \r_3))$, that is,
$\tau(t')$, where $t'$ is equal to $t$ with the label of $x$ changed
to $(\sigma, Z\cup \{y\})$.  Note that $t'$ is a
$(Y\cup\{y\})$-structure, so $x$ is labeled by $\kappa(\delta)$
($\delta \in \Delta_n$) in $\bar t_D$ if and only if $\tau(t') \in
L_{\phi_\delta}$.

Going back to the definition of the interpretation of Lindstr\"om
quantifiers, this shows that $t \in L_\phi$ if and only if $\bar t_D
\in K$.  As a result, $L_\phi = \gamma\inv(P)$ where $P$ consists of
the pairs $(F,f)$ such that $F(\1,0,\k,0) \in \alpha(K)$, and hence
$L_\phi$ is recognized by $(S,A) \block_k (T,B)$.
\eop 

We are now ready to complete the proof of Theorem~\ref{thm-main}.

\rm \trivlist \item[\hskip \labelsep{\bf Proof of
Theorem~\ref{thm-main}.}]
Let $\K$ be the class of syntactic \pgs\ of the elements of $\cK$, let
$\V$ be the pseudovariety of \pgs\ generated by $\K$, and let
$\widehat \V$ be the least closed pseudovariety containing \V. We
first show that if $L$ is a tree language with syntactic \pgp\ $(S,A)
\in \widehat\V$, then $L \in \Linlg(\cK)$.  In view of
Proposition~\ref{prop psv gen}, $(S,A)$ can be obtained from elements
of $\K$ by a succession of operations consisting of taking either a
sub-\pgp, a quotient, a direct product or a block product.  We let
$\sharp(S,A)$ be the least number of such operations, and we proceed
by induction on $\sharp(S,A)$.

If $\sharp(S,A) = 0$, then $(S,A) \in \K$, that is, $(S,A)$ is the
range of the syntactic morphism $\phi\colon (\Sigma
M,\Sigma)\rightarrow (S,A)$ of a language $K\subseteq \Sigma M_k$ in
$\cK$.  We want to show that every language recognized by $(S,A)$ is
in $\Linlg(\cK)$.  As in the first lines of the proof of
Proposition~\ref{prop-main1}, this reduces to showing that for each
$s\in S$, we have $\phi\inv(s) \in \Linlg(\cK)$.  Now we deduce from
Remark~\ref{remark syntactic class} that
$$\phi\inv(s) \enspace=\enspace \bigcap \big((u,k_1,k_2)\inv K\big) 
v\inv \setminus \bigcup  \big((u,k_1,k_2)\inv K\big) 
v\inv,$$
where the intersection runs over all $n$-ary contexts $(u,k_1,v,k_2)$
over $\Sigma M_k$ such that $ \big((u,k_1,k_2)\inv K\big) v\inv$ meets
$\phi\inv(s)$, and the union over the $n$-ary contexts that do not.
Moreover, by Remark~\ref{remark syntactic class} again, this union and
this intersection are finite.  It follows from
Theorem~\ref{thm-literalvar} that $\phi\inv(s)$, and hence any
language recognized by $(S,A)$ lies in $\Linlg(\cK)$.

We now suppose that $\sharp(S,A) > 0$.  If $(S,A)$ is a sub-\pgp\ or a
quotient of a \pgp\ $(T,B) \in \widehat{\V}$ with $\sharp(T,B) <
\sharp (S,A)$, Proposition~\ref{prop syntactic} establishes that $L$
is also recognized by $(T,B)$, so every such language is in
$\Linlg(\cK)$ by induction hypothesis.  If $(S,A)$ is the direct
product of \pgs\ $(T,B)$ and $(T',B')$ with lesser $\sharp$-values,
then by a standard argument, every language recognized by $(S,A)$ is a
finite union of intersections of the form $L \cap L'$, where $L$ is
recognized by $(T,B)$ and $L'$ by $(T',B')$.  In particular, such a
language is in $\Linlg(\cK)$ by Theorem~\ref{thm-literalvar}.  If on
the other hand, $(S,A)$ divides a block product of \pgs\ with lesser
$\sharp$-values, the inductive step follows directly from
Proposition~\ref{prop-main1}.

This concludes the proof that every tree language recognized by a 
\pgp\ in $\widehat\V$ is in $\Linlg(\cK)$. We now turn to the 
converse, namely showing that any tree language defined by a 
$\Lin(\cK)$-sentence has its syntactic \pgp\ in $\widehat\V$.

This is implied by the following, more precise statement: if $\phi$ is
a rank $k$ $\Lin(\cK)$-formula with free variables in a finite set
$Y$, then $L_\phi$ is recognized by a morphism $\alpha\colon
(\Sigma_YM, \Sigma_Y) \rightarrow (S,A)$ such that $\Im(\alpha) \in
\widehat\V$, where $\Im(\alpha)$ denotes the sub-\pgp\ of $(S,A)$
generated by $\alpha(\Sigma)$ (recall that $\Sigma$ is identified with
the subset $\Sigma\times\{\emptyset\}$ of $\Sigma_Y$).

We prove this statement by structural induction on $\phi$.  If $\phi$
is an atomic formula and $\alpha$ is the syntactic morphism of
$L_\phi$, then $\Im(\alpha)$ is trivial by Example~\ref{expl-atomic},
and hence lies in $\widehat\V$.  If $\phi = \phi_1 \vee \phi_2$, then
by induction hypothesis, there exist morphisms $\alpha_i\colon
(\Sigma_Y M, \Sigma_Y) \to (S_i,A_i)$ recognizing $L_{\phi_i}$ with
$\Im(\alpha_i) \in \widehat{\V}$, $i = 1,2$.  It is immediate that
$L_\phi$ is recognizable by the target tupling $\alpha =
(\alpha_1,\alpha_2)\colon (\Sigma_Y M, \Sigma_Y) \to ((S_1,A_1) \times
(S_2,A_2))$, and that $\Im(\alpha)$ is a sub-\pgp\ of the direct
product $\Im(\alpha_1) \times \Im(\alpha_2)$.  Since $\widehat{\V}$ is
a pseudovariety, it follows that $\Im(\alpha)$ is in
$\widehat{\V}$.  The case where $\phi$ is of the form $\phi = \neg
\phi_1$, is also easily treated: any morphism recognizing $L_{\phi_1}$
also recognizes its complement, namely $L_\phi$.

Finally, suppose that $\phi$ is of the form $\phi = Q_K y \cdot
\langle \phi_\delta\rangle_{\delta\in\Delta}$, where $K \subseteq
\Delta M_k$ is recognized by some $(S,A) \in \K$.  Without loss of
generality, we may assume that $y\not\in Y$.  By induction, each
$L_{\phi_\delta}$ ($\delta \in \Delta$) is recognized by a morphism
$\beta_\delta$ such that $\Im(\beta_\delta) \in \widehat\V$.  Taking
the target tupling of the $\beta_\delta$, we construct a morphism
$\beta\colon (\Sigma_{Y \cup \{y\}}M, \Sigma_{Y \cup \{y\}})
\rightarrow (T',B')$ recognizing each $L_{\phi_\delta}$ and such that
$\Im(\beta) \in \widehat\V$ (since $\Im(\beta)$ is a sub-\pgp\ of the
direct product of the $\Im(\beta_\delta)$).  By
Proposition~\ref{prop-main2} (and its proof), we find that $L_\phi$ is
recognized by a morphism $\gamma \colon (\Sigma_YM, \Sigma_Y)
\rightarrow (S,A) \block_k (T',B')$, where the composition $\pi \circ
\gamma \colon (\Sigma_YM, \Sigma_Y) \rightarrow (T',B')$ agrees with
$\beta$ (here $\pi$ is the second component projection).  In
particular, $\Im(\pi\circ\gamma)$ is contained in $\Im(\beta)$.
Thus $\Im(\gamma)$ is a sub-\pgp\ of $(S,A)
\block_k^{(T',B')}\Im(\beta)$, and hence $\Im(\gamma) \in \widehat\V$
by Proposition~\ref{prop technique pg}.
\cqfd
 
\subsection{Applications}

The first result follows directly from Theorem~\ref{thm-main} and
Proposition~\ref{prop full closure}.

\begin{thm}\label{first application}
    Let $\cK$ be a class of recognizable tree languages such that each
    quotient of a language of $\cK$ is in $\Linlg(\cK)$ and such that
    $\Lin(\cK)$ admits relativization.  Let \V\ be the least
    pseudovariety of \pgs\ containing the syntactic \pgs\ of elements
    of $\cK$ and let $\widehat\V$ be the least closed pseudovariety
    containing \V. The following holds.
    
    \begin{itemize}
	\item $\Linlg(\cK)$ is a literal variety of recognizable tree 
	languages, associated with the pseudovariety $\widehat\V$ in 
	the Eilenberg correspondence (Theorem~\ref{thm eilenberg}).
	
	\item If $\cK$ is the class of languages recognized by a  
	class \L\ of preclones, then $\Linlg(\cK)$ is a variety of 
	recognizable tree languages. Moreover, if \W\ is the 
	pseudovariety of preclones generated by \L\ and $\widehat\W$ 
	is the least closed pseudovariety of preclones containing \W, 
	then $\widehat\W$ is the pseudovariety of preclones 
	associated with $\Linlg(\cK)$.
    \end{itemize}
\end{thm}

\proof
Theorem~\ref{thm-main} shows that $\Linlg(\cK)$ consists of
recognizable languages and Theorem~\ref{thm-literalvar} then shows
that it is a literal variety.  Let \X\ be the pseudovariety of \pgs\
associated with $\Linlg(\cK)$ by Theorem~\ref{thm eilenberg}.  Then
\X\ and $\widehat\V$ contain the same syntactic \pgs\ by
Theorem~\ref{thm-main}, and Proposition~\ref{cor 48} shows
that this implies $\X = \widehat\V$.  This concludes the proof of the
first statement.

We now suppose that $\cK$ is the class of tree languages recognized by
the preclones in a class \L. By definition, $\V =
\langle\pgpairs(\L)\rangle$ and Proposition~\ref{psv pgs or preclones}
shows that \V\ is full and
$$\V = \pgpairs(\langle \precl(\V)\rangle) = \pgpairs(\langle
\L\rangle) = \pgpairs(\W).$$
Proposition~\ref{prop full closure} then shows that $\widehat\V$ is
full and $\widehat\V = \pgpairs(\widehat\W)$.  Thus $\Linlg(\cK)$ is a
variety of tree languages (Corollary~\ref{eilenberg full}) and the
corresponding pseudovariety of preclones is $\precl(\widehat\V) =
\widehat\W$.
\eop

The following statement is an important consequence of
Theorem~\ref{first application}, which motivated this work.

\begin{cor}
    $\Linlg(\cK_\exists)$ (that is, the class of $\FO$-definable tree
    languages) is a variety of tree languages and the corresponding
    variety of preclones is the least pseudovariety containing
    $T_\exists$ and closed under block product.
\end{cor}

\proof
Note that, by Example~\ref{expl-fo+mod2}, $\Linlg(\cK_\exists)$ is the
class of $\FO$-definable tree languages.  Recall that $\cK_\exists$
consists of the language $K_k(\exists) \subseteq \Delta M_k$, where
$\Delta$ is a ranked Boolean alphabet.

Now let $\calV_\exists$ be the variety of tree languages corresponding
to the pseudovariety $\langle T_\exists\rangle$ generated by
$T_\exists$.  According to Example~\ref{variety of T exists}, a
language $L \subseteq \Sigma M_k$ is in $\calV_\exists$ if and only if
it is a Boolean combination of languages of the form $\Sigma'M_k$,
$\Sigma' \subseteq \Sigma$.  Now the complement of $\Sigma'M_k$ in
$\Sigma M_k$ is the language of trees that contain at least a letter
outside $\Sigma'$: therefore this complement is the inverse
image of $K_k(\exists)$ in the literal morphism from $\Sigma M$ to
$\Delta M$ that maps $\Sigma'$ to $\{0_n \mid n\ge 0\}$ and $\Sigma
\setminus \Sigma'$ to $\{1_n \mid n\ge 0\}$.  In view of the closure
properties of $\Linlg(\cK_\exists)$ (Theorem~\ref{thm-literalvar}), it
follows that $\cK_\exists \subseteq \calV_\exists \subseteq
\Linlg(\cK_\exists)$ and hence $\Linlg(\calV_\exists) =
\Linlg(\cK_\exists)$ by Theorem~\ref{thm-closure}.

By definition, $\calV_\exists$ is a variety and as such, it is closed
under taking quotients.  The corresponding logic
$\Lin(\calV_\exists)$ is equivalent to $\Lin(\cK_\exists)$ (since
$\Linlg(\calV_\exists) = \Linlg(\cK_\exists)$) and hence it admits
relativization by Corollary~\ref{cor relativization}.  Thus we can
apply Theorem~\ref{first application} to conclude the proof.
\eop

A similar reasoning, using both $T_\exists$ and the preclones $T_p$
(see Section~\ref{sec examples} and Examples~\ref{lg Texists},
\ref{example exists mod p}, \ref{expl-fo+mod2} and Corollary~\ref{cor
relativization}), yields the following result.

\begin{cor}\label{cor FO+MOD}
    The class of $(\FO+\MOD)$-definable tree languages is a variety of
    tree languages and the corresponding variety of preclones is the
    least pseudovariety containing $T_\exists$ and the $T_{p}$ ($p\ge
    2$) and closed under block product.
\end{cor}

\section*{Conclusion}

We reduced the characterization of the expressive power of certain
naturally defined logics on trees, a chief example of which
is given by first-order sentences, to an algebraic problem.  This
algebraic problem is set in a new algebraic framework, that of
preclones, which the authors introduced in \cite{EsikWeil TCS}
precisely for the purpose of discussing tree languages.  It is worth
stating again that the notion of algebraic recognizability resulting
from this new framework coincides with the usual one: we simply gave
ourselves a richer algebraic set-up to classify recognizable tree
languages.

Our result does not yield (yet?)  a decidability result for, say,
first-order definable tree languages, but we can now look for a
solution of this problem based on the methods of algebra.  In this
process, it will probably be necessary to develop the structure theory
of preclones, to get more precise results on the block product
operation.

A positive aspect of our approach is its generality: it is not
restricted to the characterization of logics based on the use of
Lindstr\"om quantifiers, nor indeed to the characterization of logics.
Our key algebraic tool is the block product: this product was
introduced by Rhodes and Tilson \cite{RhodesTilson} for monoids, to
investigate the lattice of pseudovarieties of monoids and its
application to the theory of formal languages (of finite words), and
we adapted its definition for preclones.  The use of
wreath products instead of block products (the wreath product can be
seen as a one-sided restriction of the block product) can yield
algebraic characterizations for other natural classes of recognizable
tree languages, see \cite{ESz1}.

\medskip

Our approach also raises a number of questions.  At a technical level
first: it was shown in \cite{RhodesTilson} that for monoids, the block
product can be expressed in terms of a double semidirect product, a
two-sided generalization of the semidirect product.  It might be
convenient to have such a notion for preclones as well, and to derive
analogues of the wreath product principle and the block product
principle (general descriptions of the languages recognized by a
wreath product or a block product).  This might yield, as in the
finite word case, the characterization of the recognizing power of the
block product of two varieties, the characterization of logical
hierarchies within \FO, etc.

At a more general level, we observe that in the word language case,
the decidability of first-order definability does not stem from the
analogue of our main result, namely the fact that a language is
\FO-definable if and only if its syntactic monoid is in the least
pseudovariety containing $\{0,1\}$ and closed under block product.  It
follows rather from the characterization of that class of monoids as
the aperiodic monoids, see the theorems of McNaughton and Papert on
the equivalence of \FO-definability and star-freeness, and of
Sch\"utzenberger on the equivalence between star-freeness and
aperiodicity.  This characterization makes use in an essential way of
the notion of star-freeness and of the structure theory of finite
monoids.  The question is therefore whether we can find a useful
analogue of star-freeness for tree languages.  There were attempts in
this direction (\cite{Heuter,PT93,Potthoff95}) that established that the
more natural notions of star-freeness for trees do not coincide with
\FO-definability.  Are we missing on an important concept?  Taking the
question from a different angle, can we directly develop the relevant
fragment of a structure theory of finitary preclones, to prove
decidability of \FO-definability?

Another, more general remark is the following.  We are convinced that 
the algebraic concept of preclones is well suited for the study and the
classification of recognizable languages of finite ranked trees.
However, we are conscious that beyond its qualities (the first of
which is to allow results such as those proved in this paper), our
algebraic framework is cumbersome, and perhaps intimidating.  We 
argued in \cite{EsikWeil TCS} that several known results on the 
characterization of particular classes of tree languages can be 
expressed in a natural way in the language of preclones, --- but 
there might be an equivalent, yet lighter algebraic set-up.

This remark is related with another question.  Other algebraic
frameworks have been investigated in the literature since our results
were announced in 2003 \cite{FSTTCS}, in the (considerable) interval
it took for this paper to be written and refereed.  One of the more
promising and elegant is the notion of \textit{forest algebras},
introduced by Boja\'nczyk and Walukiewicz \cite{BojWal}, initially to
discuss languages of unranked, unordered trees.  These authors also
achieved the algebraic characterization of certain logically defined
tree languages.  More recent papers record interesting results using
forest algebras, also to discuss ranked or ordered trees (e.g.
Boja\'nczyk, S\'egoufin, Straubing, Walukiewicz
\cite{BojLICS07,BSegICALP2008,BSSLICS2008}), and it is tempting to
wonder whether this algebraic approach and ours could be unified,
since forest algebras may be seen as an unsorted version of preclones.

Possibly as a longer term project, one should consider the following.
From the point of view of applications (in the field of verification,
the investigation of distributed computation models, of game theory,
etc), being able to handle languages of infinite (ranked ordered)
trees is important.  Discussing languages of infinite words as well as
languages of finite words, was a topic of interest from the very
beginnings of automata theory (B\"uchi), automata models were proposed
quite early on, but the development of an algebraic model to handle
them (namely the notion of $\omega$-semigroups) was very slow in
coming, and was matured only in the late 1980s (through work of
Arnold, Perrin, Pin, Wilke, etc, see \cite{Perrin-Pin}).  Can an
analogous extension be developed for our preclones?  One key technical
tool in dealing with recognizable languages of infinite words is
Ramsey's theorem, and the authors are unfortunately not aware of a
relevant analogue of this theorem for trees.

\paragraph{Acknowledgements}
The authors wish to thank Szabolcs Iv\'an for his careful reading of 
preliminary versions  of this work, which helped eliminate many 
mistakes.

\thebibliography{99}

{\small

\bibitem{Almeida}
J. Almeida.  On pseudovarieties, varieties of languages, filters of
congruences, pseudoidentities and related topics, \textit{Algebra
Universalis} \textbf{27} (1990) 333-350.

\bibitem{AMSV}
J. Almeida, S. Margolis, B. Steinberg, M. Volkov.  Representation
theory of finite semigroups, semigroup radicals and formal language
theory, \url{arXiv:math/0702400v1 [math.GR]}

\bibitem{BenediktSegoufin}
M. Benedikt, L. S\'egoufin.
Regular tree languages definable in \FO\ and in $\FO_{mod}$, to appear.

\bibitem{BojLICS07}
M. Boja\'nczyk.
Two-way unary temporal logic over trees, in \textit{LICS 2007}, IEEE,
121-130.

\bibitem{BSegICALP2008}
M. Boja\'nczyk, L. S\'egoufin.  Tree Languages Defined in First-Order
Logic with One Quantifier Alternation, in (L. Aceto, I. Damg{\aa}rd,
L.A. Goldberg, M.M. Halld\'orsson, A. Ing\'olfsd\'ottir, I.
Walukiewicz eds.) \textit{ICALP 2008, part II}, Lect.  Notes in
Computer Science \textbf{5126}, Springer, 2008, 233-245.

\bibitem{BSSLICS2008}
M. Boja\'nczyk, L. S\'egoufin, H. Straubing.
Piecewise Testable Tree Languages, in \textit{LICS 2008}, IEEE,
442-451.

\bibitem{BW}
M. Boja\'nczyk, I. Walukiewicz.  Characterizing \EF\ and \EX\ tree
logics, \textit{Theoret.  Computer Science} \textbf{358} (2006)
255-272.

\bibitem{BojWal}
M. Boja\'nczyk, I. Walukiewicz.  Forest algebras, in (J. Flum, E.
Graedel, T. Wilke eds.)  \textit{Logic and Automata}, Texts in Logic
and Games, Amsterdam University Press, 2007.

\bibitem{tata}
H. Comon, M. Dauchet, R. Gilleron, F. Jacquemard, D. Lugiez, S. Tison,
M. Tommasi.  \textit{Tree Automata Techniques and Applications},
\url{http://www.grappa.univ-lille3.fr/tata}.

\bibitem{EbbinghausFlum}
H.-D. Ebbinghaus, J. Flum.
{\em Finite Model Theory}, Springer, 1995. 

\bibitem{Eilenberg}
S. Eilenberg.
{\em Automata, Languages, and Machines},
vol. A and B, Academic Press, 1974 and 1976.

\bibitem{Esik}
Z. \'Esik.
A variety theorem for trees and theories,
\textit{Publicationes Mathematicae} \textbf{54} (1999) 711-762.

\bibitem{Esik TCS 2006}
Z. \'Esik.
Characterizing CTL-like logics on finite trees,
{\em Theoret. Computer Science} \textbf{356} (2006) 136-152.

\bibitem{ESz1}
Z. \'Esik, Sz.  Iv\'an.  Some varieties of finite tree automata
related to restricted temporal logics, \textit{Fundamenta
Informatic\ae} \textbf{82} (2008) 79-103.

\bibitem{ESz2}
Z. \'Esik, Sz.  Iv\'an.  Products of tree automata with an application
to temporal logic, \textit{Fundamenta Informatic\ae} \textbf{82}
(2008) 61-78.

\bibitem{EsikLarsen}
Z. \'Esik, K. G. Larsen.  Regular languages definable by
Lindstr\"om quantifiers, {\em Theoretical Informatics and
Applications} \textbf{37} (2003) 179-241.

\bibitem{FSTTCS}
Z. \'Esik, P. Weil.  On logically defined recognizable tree languages,
in \textit{Proc.  FST TCS 2003} (P. K. Pandya, J. Radhakrishnan eds.),
Lect.  Notes in Computer Science \textbf{2914}, Springer, 2003,
195-207.

\bibitem{EsikWeil TCS}
Z. \'Esik, P. Weil.
Algebraic characterization of regular tree languages,
\textit{Theoretical Computer Science} \textbf{340} (2005) 291-321.

\bibitem{Heuter}
U. Heuter.  First-order properties of trees, star-free expressions,
and aperiodicity, \textit{Theoretical Informatics and Applications}
\textbf{25} (1991) 125-146.

\bibitem{Lindstrom}
P. Lindstr\"om.
First order predicate logic with generalized quantifiers.
\textit{Theoria} \textbf{32} (1966) 186-195.

\bibitem{Perrin-Pin}
D. Perrin, J.-E. Pin. \textit{Infinite words}, Pure and Applied
Mathematics vol.  \textbf{141}, Elsevier (2004).

\bibitem{Pin}
J.-E. Pin.
\textsl{Vari\'et\'es de langages
formels}, Masson, Paris (1984). English translation: \textsl{Varieties
of formal languages}, Plenum, New-York (1986).

\bibitem{PinLogicSurvey}
J.-E. Pin. Logic, Semigroups and Automata on Words, \textit{Annals of
Mathematics and Artificial Intelligence} \textbf{16} (1996) 343-384.

\bibitem{PinStraubingStamps}
J.-E. Pin, H. Straubing. Some results on C-varieties, \textit{Theoret.
Informatics Appl.}  \textbf{39} (2005) 239-262.

\bibitem{Podelski}
A. Podelski. A monoid approach to tree automata, in: {\em Tree automata
and languages} (M. Nivat, A. Podelski, eds.), North Holland, 1992,
41-56.

\bibitem{Potthoff95}
A. Potthoff. First order logic on finite trees, in: \textit{TAPSOFT '95}
(P.D. Mosses, M. Nielsen, M.I. Schwartzbach eds.), Lect. Notes in
Computer Science \textbf{915}, Springer, 1995, 125-139.

\bibitem{PT93}
A. Potthoff, W. Thomas. Regular tree languages without unary symbols 
are star-free, in (Z. \'Esik ed.) \textit{FCT 93},  Lect. Notes in
Computer Science \textbf{710}, Springer, 1993,  396-405.

\bibitem{RhodesTilson}
J. Rhodes, B. Tilson.  The kernel of monoid morphisms, \textit{J.
Pure and Appl.  Alg.} \textbf{62} (1989) 227-268.

\bibitem{mps1}
J. Rhodes, P. Weil.
Decomposition techniques for finite semigroups using categories,~I
{\it Journal of Pure and Applied Algebra} {\bf 62} (1989) 269-284.

\bibitem{mps2}
J. Rhodes, P. Weil.
Decomposition techniques for finite semigroups using categories,~II,
{\it Journal of Pure and Applied Algebra} {\bf 62} (1989) 285-312.

\bibitem{Salehi}
S. Salehi. Varieties of tree languages definable by syntactic monoids,
\textit{Acta Cybernetica} \textbf{17} (2005), 21-41.

\bibitem{SalehiSteinby}
S. Salehi, M. Steinby. Tree algebras and varieties of tree languages, {\em Theoret.
Comput.  Sci.} \textbf{377} (2007) 1-24.

\bibitem{Steinby1}
M. Steinby. A theory of tree language varieties, in: {\em Tree automata
and languages} (M. Nivat, A. Podelski, eds.), North Holland, 1992,
57-81.

\bibitem{Steinby2}
M. Steinby.  General varieties of tree languages, {\em Theoret.
Comput.  Sci.} \textbf{205} (1998) 1-43.

\bibitem{Straubing}
H. Straubing. {\em Finite Automata, Formal Logic, and Circuit
Complexity}, Birk\-ha\"user, 1994.

\bibitem{StraubingStamps}
H. Straubing. On logical descriptions of regular languages, in
\textit{LATIN 2002}, Lect.  Notes in Computer Science \textbf{2286},
Springer, 2002,  528-538.

\bibitem{ThomasHdBook}
W. Thomas. Languages, automata and logic, in (G. Rozenberg, A.
Salomaa, eds) \textit{Handbook of Formal Languages}, volume III, pp.
389-455,  Springer, 1997.

\bibitem{jcss}
P. Weil.
Closure of varieties under products with counter, 
{\it Journal of Computing and System Science} {\bf 45} (1992) 316-339.

\bibitem{WeilMFCS}
P. Weil.
Algebraic recognizability of languages, in:
\it MFCS 2004\rm\ (J. Fiala, V. Koubek, J. Kratochv\'\i l eds.),
Lect. Notes in Computer Science  \textbf{3153}, Springer, 2004, 149-175.

\bibitem{Wilke}
Th. Wilke. An algebraic characterization of frontier testable tree
languages, {\em Theoret. Comput. Sci.} \textbf{154} (1996) 85-106.

}

\end{document}